\documentclass[12pt,letterpaper,utf8]{article}
\usepackage{amsmath,amsthm,cancel,eurosym,geometry,graphicx,caption,color,comment,caption,pdflscape,array,mathtools,multirow,hhline,bm,bbm,enumerate,optidef,nameref,subcaption,import,placeins,booktabs}
\usepackage[normalem]{ulem}
\usepackage{thmtools,thm-restate}
\usepackage{setspace}
\usepackage[multiple]{footmisc}

\usepackage{microtype}

\usepackage[dvipsnames,table,xcdraw]{xcolor}
\usepackage{tikz}
\usetikzlibrary{calc}
\numberwithin{equation}{section}

\usepackage{enumitem}
    \setlist[itemize]{noitemsep,nolistsep}
    \setlist[enumerate,1]{noitemsep,nolistsep,label=(\arabic*)}
    \setlist[enumerate,2]{noitemsep,nolistsep,label=(\alph*)}

\usepackage[page,title,toc]{appendix}

\usepackage{amsfonts}
\usepackage{amssymb}
\usepackage[T1]{fontenc}
\usepackage[utf8]{inputenc}
\usepackage{accents}

\usepackage[nottoc,numbib]{tocbibind}
\usepackage[backend=biber,sorting=ynt,style=authoryear-comp,maxbibnames=9,citetracker=true,maxcitenames=2,uniquename=false,uniquelist=false,natbib]{biblatex}

\newrobustcmd*{\citefirstlastauthor}{\AtNextCite{\DeclareNameAlias{labelname}{given-family}}\citet*}

\makeatletter
\newcommand{\upmax}{\def\blx@maxcitenames{99}}
\newcommand{\dnmax}{\def\blx@maxcitenames{1}}
\makeatother

\theoremstyle{plain}
  \newtheorem{theorem}{Theorem}
  \newtheorem{lemma}{Lemma}
  
  \newtheorem{corollary}{Corollary}
\theoremstyle{definition}
    \newtheorem{definition}{Definition}
  \newtheorem{example}{Example}
  \newtheorem{remark}{Remark}
\theoremstyle{remark}

\AtBeginEnvironment{appendices}{\crefalias{section}{appendix}}
\AtBeginEnvironment{appendices}{\crefalias{subsection}{appendix}}
\AtBeginEnvironment{appendices}{\crefalias{subsubsection}{appendix}}

\newcolumntype{L}[1]{>{\raggedright\let\newline\\arraybackslash\hspace{0pt}}m{#1}}
\newcolumntype{C}[1]{>{\centering\let\newline\\arraybackslash\hspace{0pt}}m{#1}}
\newcolumntype{R}[1]{>{\raggedleft\let\newline\\arraybackslash\hspace{0pt}}m{#1}}

\let\phi\varphi
\usepackage{etoolbox}

\usepackage[hyperfootnotes=false]{hyperref}
\hypersetup{
	linktoc=page,
	unicode=false,          
	pdftoolbar=true,        
	pdfmenubar=true,        
	pdffitwindow=false,     
	pdfstartview={Fit},    
	pdftitle={},    
	pdfauthor={},     
	pdfcreator={},   
	pdfproducer={}, 
	pdfkeywords={}, 
	pdfnewwindow=true,      
	colorlinks=true,       
	linkcolor=MidnightBlue,          
	citecolor=MidnightBlue,        
	filecolor=black,      
	urlcolor=BrickRed,		
	anchorcolor=black,
	menucolor=black,
	runcolor=black,
}

\usepackage[nameinlink]{cleveref}
    \crefname{theorem}{theorem}{theorems}
    \Crefname{theorem}{Theorem}{Theorems}
    \Crefname{lemma}{Lemma}{Lemmata}
    \crefname{lemma}{lemma}{lemmata}
    \crefname{proposition}{proposition}{propositions}
    \Crefname{proposition}{Proposition}{Propositions}
    \crefname{definition}{definition}{definitions}
    \Crefname{definition}{Definition}{Definitions}
    \crefname{corollary}{corollary}{corollaries}
    \Crefname{corollary}{Corollary}{Corollaries}
    \Crefname{table}{Table}{Tables}
    \Crefname{tabular}{Table}{Tables}
    \Crefname{example}{Example}{Examples}
    \Crefname{remark}{Remark}{Remarks}
    \Crefname{footnote}{Footnote}{Footnotes}
    \crefname{footnote}{footnote}{footnotes}
    \newcommand{\headercref}[2]{\texorpdfstring{\Cref{#2}}{#1 \ref{#2}}}

\DeclareMathOperator{\supp}{supp}

\DeclareMathOperator*{\extremepoints}{ext}

\DeclareMathOperator*{\graph}{graph}

\DeclareMathOperator*{\deterministic}{det}

\hypersetup{
	pdftitle={Optimal Allocation with Peer Information},
	pdfauthor={Axel Niemeyer, Justus Preusser}
}

\begin{document}

\title{Optimal Allocation with Peer Information\thanks{\protect Parts of this paper previously circulated under the title ``Simple Allocation with Correlated Types.'' 
We thank Marco Battaglini, Florian Brandl, Modibo Camara, Gregorio Curello, Francesc Dilm\'{e}, Tangren Feng, Nathan Hancart, Deniz Kattwinkel, Jan Kn{\"o}pfle, Nenad Kos, Daniel Kr\"ahmer, Stephan Lauermann, Irene Lo, Andrew Mackenzie, Benny Moldovanu, Marco Ottaviani, Alessandro Pavan, Johannes Schneider, Ludvig Sinander, Francesco Squintani, Omer Tamuz, {\'E}va Tardos, Rakesh Vohra, Kai Hao Yang, and numerous seminar audiences for comments. 
Niemeyer acknowledges funding by the Deutsche Forschungsgemeinschaft (DFG, German Research Foundation) under Germany's Excellence Strategy -- EXC 2126/1 -- 390838866.
Preusser acknowledges funding by the Deutsche Forschungsgemeinschaft (DFG, German Research Foundation) through CRC TR 224 (Project B04). 
Preusser acknowledges financial support by the European Research Council (HEUROPE 2022 ADG, GA No. 101055295 – InfoEcoScience).
Preusser is grateful for the hospitality of Yale University.}}%

\author{
Axel Niemeyer\thanks{\protect Division of the Humanities and Social Sciences, California Institute of Technology, \textit{\href{mailto:niemeyer@caltech.edu}{niemeyer@caltech.edu}}.}
\and
Justus Preusser\thanks{\protect Department of Economics and IGIER, Bocconi University, \textit{\href{mailto:justus.preusser@unibocconi.it}{justus.preusser@unibocconi.it}}.}
}

\date{
  This version: \today%
}

\maketitle

\begin{abstract}%
We study allocation problems without monetary transfers where agents have correlated types, i.e., hold private information about one another. Such peer information is relevant in various settings, including science funding, allocation of targeted aid, or intra-firm allocation. Incentive compatibility requires that agents cannot improve their own allocation by misrepresenting the merits of allocating to others. We characterize optimal incentive-compatible mechanisms using techniques from the theory of perfect graphs. Optimal mechanisms improve on review panels commonly observed in practice by eliciting information directly from eligible agents and by using allocation lotteries to alleviate incentive constraints. Computational hardness results imply that exactly optimal mechanisms are impractically complex. We propose ranking-based mechanisms as a viable alternative and show that they are approximately optimal when agents are informationally small, i.e., when no single agent has information that is crucial for evaluating a large fraction of the other agents.


\end{abstract}

\vfill
\pagenumbering{gobble}
\pagebreak
\pagenumbering{arabic}

\section{Introduction}\label{sec:introduction}
This paper studies how peer information can be used to improve efficiency in allocation problems without monetary transfers. By peer information, we mean information that agents hold about the merits of allocating to other agents. Important examples include peer review in science funding, community-based targeting of social aid, or resource allocation within organizations based on information dispersed across several divisions.\footnote{For field studies on targeting with community information, see e.g. \citet{alatas_targeting_2012}, \citet{alatas2019does} or \citet{hussam2022targeting}. 
    The oil and gas company BP has been reported to use a peer review mechanism for capital expenditure authorizations across business units (\citet{goold2005making}).
    Other examples include group members selecting a leader based on nominations (e.g. \citet{alon2011sum,holzman_impartial_2013}),
    job promotions being decided based on feedback from colleagues who are also vying for a promotion (\citet{huang2019discovery}), and hiring in the gig economy relying on peer assessments of other applicants (\citet{kotturi2020hirepeer}). 
    See the survey by \citet{olckers2022manipulation} for further examples.} 
In these settings, monetary transfers play a limited role. For example, when allocating funds such as a research grant, the allocation itself is a monetary transfer; additional transfers would simply amount to a different allocation. 

Peer information is not directly available to the social planner. 
Like any other piece of private information, agents must be incentivized to truthfully report what they know about their peers. For example, when asked for a peer assessment, an agent may be tempted to discredit their peers if doing so improves their own allocation. 
Dishonest peer assessments are documented in several empirical studies, for example, when allocating loans to entrepreneurs (\citet{hussam2022targeting}), identifying workers for a promotion (\citet{huang2019discovery}), or in lab experiments where payoffs are determined based on how subjects evaluate each other's performance on a task (\citet{carpenter2010tournaments,balietti2016peer}).

In our model, a principal allocates a good among several agents.
Each agent wants the good, but the principal wants to allocate the good to the agent who needs it most or who can put it to best use. 
The principal's values from assigning the good to different agents are initially unobserved by all parties, but each agent has a signal---their type---that is arbitrarily correlated with the entire profile of the principal's values. 
An agent's type captures all private information that the agent has about themselves and their peers. 
For example, when allocating social aid in a local community, a type may summarize an agent's noisy assessment of their own and their neighbors' values in an underlying social network. 
The fact that agents' types are correlated is the main analytical challenge in our model.

To ensure incentives for honest reports, we focus on dominant-strategy incentive-compatible (DIC) mechanisms; that is, each agent finds it optimal to report their type truthfully, regardless of what the agent believes about the other agents' types and reports.
In our setting, this robust notion of incentives means that an agent's report can never increase their \emph{own} chances of getting the good. 
However, an agent's report can help determine the optimal allocation among the \emph{other} agents.

Although agents' types are plausibly correlated 
in many allocation problems without transfers, the existing literature has not studied optimal mechanisms for such settings and does not provide tools for finding them.
On the one hand, a small but growing literature replaces transfers with non-monetary screening devices but maintains the ubiquitous assumption that types are statistically independent. This assumption is necessary for the applicability of standard tools but rules out peer information because an agent’s own type is the only source of information about the value of allocating to that agent.
On the other hand, the existing literature on correlated types has focused on how transfers can be used to screen agents via their beliefs about the types of others (\citet{cremer1985optimal,cremermclean1988}); these constructions require precise knowledge of agents' beliefs and are moot in settings without transfers. In contrast to this literature, we do not exploit beliefs as a screening device and do not rely on transfers to leverage correlation. Instead, we focus on the purely informational channel of correlation---how one agent’s information can help determine the optimal allocation among the other agents.\footnote{\citet{kattwinkel2022fullybiased} is the only other paper that studies correlated types in an allocation problem without transfers. We discuss the differences to this and other papers in more detail in \Cref{sec:related_literature}.}

The following class of \emph{jury mechanisms} is an instructive starting point.
Before consulting anyone, the principal assigns some agents the role of ``jurors,'' the others the role of ``candidates.'' The jurors' reports determine which candidate gets the good. Jurors never win, and candidates are never consulted. Jury mechanisms are DIC and resemble mechanisms commonly used in practice. For example, scientific peer review is founded on the idea that reviewers should have no stake in the decision. 
But are jury mechanisms optimal? For example, is it possible to improve the mechanism by also eliciting information from the candidates? We show that jury mechanisms are optimal only in two special cases: if there are two agents and the principal can keep the good, or if there are up to three agents and the principal must allocate the good (\Cref{prop:juries}). 

Optimal mechanisms improve on jury mechanisms by better resolving a fundamental trade-off between allocation and information elicitation. 
By allocating to an agent at a given type profile, the principal commits to allocating to the agent at all type profiles that differ from the given profile only in the agent's report (otherwise the agent could influence their own allocation by misreporting).
Thus, the principal can either benefit from allocating to the agent at these type profiles or from using the variation in the agent's information to decide the allocation among others, but not from both.
Jury mechanisms resolve this trade-off by pre-assigning who is eligible to receive the good versus who provides information. However, whether an agent fits their assigned role---whether they have a high value or good information about their peers---may vary across type profiles. Optimal mechanisms resolve the trade-off more granularly: an agent may provide information at some type profiles (when they have good information) and receive the good at others (when they have a high value).

As a key step in our analysis of optimal mechanisms, we show that DIC mechanisms can be characterized using an auxiliary graph that encodes the principal's trade-off between allocation and information elicitation (\Cref{lemma:bijection}). 
A vertex of the graph represents a commitment to allocate to an agent at a set of type profiles that differ only in that agent's type (as required by DIC). An edge is a pair of commitments that promises the good to distinct agents at the same type profile. A deterministic mechanism must select non-adjacent vertices, i.e. make non-conflicting commitments. In graph-theoretic terms, deterministic DIC mechanisms are stable sets of the graph, and stochastic DIC mechanisms are fractional stable sets (they make probabilistic commitments).

We first use this graph-theoretic characterization to show that randomization is an inherent structural property of optimal DIC mechanisms.
Intuitively, randomization allows the principal to resolve the trade-off between allocation and information elicitation more flexibly because an agent may win with some probability but still influence how the remaining probability is distributed among the other agents.
However, this intuition is very incomplete: because the principal maximizes a linear objective, they never benefit from randomizing over deterministic mechanisms. Instead, the principal must consider extreme points of the set of DIC mechanisms.\footnote{The extreme-point approach has gained traction in the mechanism design literature in recent years (e.g. \citet{manelli2007multidimensional,manelli2010bayesian} and \citet{kleiner2021extreme}). However, none of the existing results can be applied to our specific setting without transfers and with correlation.}
Formally, stochastic extreme points exist if and only if there are at least three agents and at least one agent has at least three possible types (\Cref{thm:stochastic_existence}).\footnote{In the model, the number of possible types per agent is finite and at least two.} Moreover, when each agent has many possible types, essentially all extreme points are stochastic: the set of deterministic DIC mechanisms, all of which are extreme, constitutes a vanishing fraction of the set of extreme points (\Cref{thm:stochastic_prevalence}). 
The benefit of randomization can be traced to how stochastic extreme points distribute allocation probability around certain cycles---odd holes---in the auxiliary graph (\Cref{prop:hole_characterization}). 
The proofs leverage powerful results from the theory of perfect graphs (\citet{chvatal1975}, \citet{chudnovsky2006}).

We then use the graph-theoretic characterization to show that the set of optimal mechanisms eludes simple economic descriptions. 
This complexity arises because there is no straightforward way to determine who should receive the good versus who provides information: the allocation at one type profile constrains in a non-local, combinatorial way the possible allocations at other type profiles.
To make this point precise, we first focus on the conceptually simpler subset of deterministic DIC mechanisms.
The problem of finding an optimal deterministic DIC mechanism---a stable set of maximum weight in the auxiliary graph---is NP-hard (\Cref{thm:npcomplete}).\footnote{Related hardness results are known for canonical mechanism design problems with transfers and correlation. \citet{papadimitriou2011optimal,papadimitriou2015optimal} show hardness of finding an optimal deterministic ex-post IC and IR mechanism for auctioning a single good to bidders with correlated private values. \citet{papadimitriou2016complexity,papadimitriou2022complexity} show hardness of finding an optimal deterministic IC mechanism for selling to a single buyer in a repeated interaction over two periods when valuations are correlated across periods.} 
Thus, if one could give a simple economic description of deterministic DIC mechanisms, then this description would suggest a computationally efficient procedure for finding an optimal deterministic DIC mechanism, which is impossible under the widely believed $P \neq NP$ conjecture.\footnote{\citet[see their introduction]{papadimitriou2016complexity} analogously refer to computational complexity to argue that there is no simple characterization of optimal ex-post IR auctions with correlated values. \citet[Section 1.1]{hart2017approximate} argue that even computationally tractable problems may still be complex on a ``conceptual'' level ``since, even after computing the precise solution, one may not understand its structure, what it means and represents, and how it varies with the given parameters.''\label{footnote:hart_nisan}}
The complexity of deterministic DIC mechanisms extends to optimal (possibly stochastic) DIC mechanisms for two reasons.
First, every deterministic DIC mechanisms is an extreme point and hence an indispensable candidate for optimality; thus, a complete description of optimal DIC mechanisms a fortiori requires a description of deterministic mechanisms.
Second, stochastic extreme points may allocate deterministically on parts of the type space, and on these deterministic parts, they inherit the structure of deterministic mechanisms.

The complexity and stochasticity of optimal mechanisms are reminiscent of multi-dimensional screening problems (e.g., \citet{manelli2007multidimensional}, \citet{daskalakis2017strong}, \citet{lahr2024extreme}) but emerge here for an entirely different reason: the interaction of incentive constraints across agents with simple, state-independent preferences rather than a screening motive for a single agent with complex, multi-dimensional preferences.

To address the complexity of optimal mechanisms, we identify simple mechanisms that we call \emph{ranking-based mechanisms} and show that they are approximately optimal in natural environments. 
Informed by our findings about optimal mechanisms, ranking-based mechanisms use all agents' information, consider all agents as potential recipients of the good, and allocate randomly to alleviate incentive constraints.
To describe these mechanisms, we define an agent's \emph{peer value} as the principal's expected value of allocating to that agent based only on the reports of the other agents. A ranking-based mechanism ranks agents according to their peer values and randomly selects an agent above a given rank threshold. Before allocating the good, the mechanism checks for a conflict of interest: does the agent pass the rank threshold robustly, regardless of what the agent reports about their peers?
Intuitively, the agent must pass the bar in a hypothetical scenario where the agent appraises their competitors (thereby diminishing their own rank). If the agent passes the check, they receive the good. Otherwise, if a conflict of interest is detected, the principal keeps the good.

Ranking-based mechanisms are approximately optimal when agents are informationally small (\Cref{thm:ranking_based}). For a given rank threshold, the mechanism performs badly only at type profiles where relatively few agents are robustly above the threshold. At such a profile, the principal inefficiently withholds the good with high probability. The performance of the mechanism thus hinges on the impact that any individual agent's report has on the rank of their own peer value. We refer to this impact as the environment’s \emph{informational size}.\footnote{Similar notions of informational size appear in earlier work on mechanism design and general equilibrium with information asymmetries, e.g. \citet{gul1992asymptotic,mclean2002informational,mclean2015implementation,gerardi2009aggregation,andreyanov2021robust}.}
The performance of a ranking-based mechanism admits a lower bound that depends on informational size and that holds type profile-by-type profile. For vanishing informational size, the principal inefficiently withholds the good with vanishing probability. By suitably increasing the rank threshold as informational size vanishes, ranking-based mechanisms approximate the performance of optimal DIC mechanisms. 

Informational size is naturally small in many applications of interest. For example, suppose there is an underlying social network and that each agent has information only about their neighbors in this network. In the applications given earlier, this could be the researchers working on nearby subfields, the literal neighbors in a community, or the firm's divisions situated at the same geographic location. In this setup, informational size is small if the network’s neighborhoods are small relative to the overall size of the network; that is, if subfields are highly specialized, one’s neighbors constitute a small fraction of the overall community, and the divisions of the firm are spread across many locations. As another example, even if each agent holds information about all other agents and can therefore influence everyone else’s peer value, informational size is still small if no agent has information that is crucial for evaluating a large fraction of the other agents.
In these applications, ranking-based mechanisms thus demonstrate how to provide incentives for honest evaluations while sacrificing little economic efficiency. This insight extends to multi-unit allocation problems.

Next, we describe the model. In \Cref{sec:model:jury_mechanisms}, we show that jury mechanisms are optimal if there are few agents. In \Cref{sec:feasibility_graph}, we introduce the auxiliary graph. 
In \Cref{sec:DIC_characterization}, we discuss stochastic DIC mechanisms and extreme points.
In \Cref{sec:complexity}, we discuss the complexity of optimal DIC mechanisms. 
In \Cref{sec:approximate_optimality}, we turn to approximate optimality and ranking-based mechanisms.
In \Cref{sec:related_literature}, we discuss the related literature in detail.
\Cref{sec:conclusion} discusses extensions and shortcomings of our analysis. 
All proofs are in \Cref{appendix:proofs}. 
\Cref{OA} (online appendix) presents further examples and results.

\section{Model}\label{sec:model}

\paragraph*{Environment.}
A principal allocates a good among a number $n$ of agents, where $n\geq 2$.\footnote{Our main insights extend to the problem of allocating multiple units; see \Cref{sec:conclusion}.}
Let $[n] = \lbrace 1, \ldots, n\rbrace$ denote the set of agents. 
Each agent $i$ enjoys a payoff of $1$ if allocated the good and a payoff of $0$ otherwise.\footnote{Nothing in our analysis would change if an agent's payoff from getting the good would depend on the agents' private information, provided that this payoff never changes sign. Indeed, the good could be a bad (e.g. an undesirable task), in which case the payoff is negative.}
The principal's payoff from allocating to agent $i$ is given by $u_{i} \in [-1, 1]$.
The principal's payoff from keeping the good is normalized to $0$.

The payoff profile $(u_{1}, \ldots, u_{n})$ is initially unobserved by all parties.
However, each agent $i$ has a private type $\theta_{i}$ from a finite type space $\Theta_{i}$ that is informative about the payoff profile (and the types of the other agents).
Each agent $i$ has at least two possible types, $\vert\Theta_{i}\vert\geq 2$.
The joint distribution of payoffs and types is given by a Borel probability measure $\mu$ on $\Theta\times [-1, 1]^{n}$, where $\Theta$ denotes the set of type profiles with typical element $\theta = (\theta_{1}, \ldots, \theta_{n})$. 
For our analysis, the distribution $\mu$ need not be a common prior; it may simply be interpreted as the principal's belief about the structure of peer information.

\paragraph*{Notation.}
As usual, the set of type profiles of agents other than $i$ is denoted $\Theta_{-i}$ with typical element $\theta_{-i}$.
For a profile $\theta\in\Theta$, we write $\mu(\theta_{-i})$ and $\mu(\theta)$ to mean the marginal probabilities of $\theta_{-i}$ and $\theta$, respectively.
We also write $\mathbb{E}[u_{i}\vert\theta]$ for the conditional expectation of $u_{i}$ given $\theta$ (provided $ \mu(\theta)>0$, else we arbitrarily define $\mathbb{E}[u_{i}\vert\theta] = 0$).

\paragraph*{Mechanisms.}
A \emph{(direct) mechanism} is a function $q\colon\Theta\to [0, 1]^{n}$ such that $\sum_{i=1}^{n}q_{i}(\theta) \leq 1$ for all $\theta\in\Theta$. 
Here, $q_{i}$ denotes agent $i$'s winning probability.\footnote{One can alternatively interpret the good as a divisible resource. Agent $i$'s share of the resource is denoted $q_{i}$, and the principal's payoff from allocating to agent $i$ is linear in $q_{i}$.}
Since there is only one good to allocate, the sum of winning probabilities is less than $1$. 
Whenever the sum is strictly less than $1$ at some type profile, the principal keeps the good with the remaining probability.

A mechanism is \emph{dominant-strategy incentive-compatible (DIC)} if $q_{i}(\theta_{i}, \theta_{-i}) \geq q_{i}(\theta_i^{\prime}, \theta_{-i})$
for all $i\in[n]$, $\theta_{i}, \theta_{i}^{\prime}\in\Theta_{i}$ and $\theta_{-i}\in\Theta_{-i}$.
By swapping the roles of $\theta_{i}$ and $\theta_{i}^{\prime}$ in this inequality, a mechanism $q$ is DIC if and only if each agent $i$'s allocation is constant in their own report; that is, $q_{i}(\theta_{i}, \theta_{-i}) = q_{i}(\theta_i^{\prime}, \theta_{-i})$.
We henceforth drop $i$'s type from $i$'s winning probability, i.e., $q_{i}(\theta_{-i}) = q_{i}(\theta)$, since we focus on DIC mechanisms throughout the paper.
The set of DIC mechanisms is denoted by $Q$. The Revelation Principle applies.

The principal's utility from a DIC mechanism $q$ is denoted $U(q)$ and is given by
\begin{equation*}
    U(q) = \sum_{\theta\in\Theta}\sum_{i\in[n]} \mu(\theta) q_{i}(\theta_{-i}) \mathbb{E}[u_{i}\vert\theta].
\end{equation*}
A DIC mechanism is \emph{optimal} if it maximizes the principal's utility across all DIC mechanisms.
A mechanism $q$ is \emph{deterministic} if it maps to $\lbrace 0, 1\rbrace^{n}$. 
A mechanism is \emph{stochastic} if it is not deterministic.

\paragraph*{Examples of peer information.}
Our model nests various forms of peer information:
\begin{itemize}
    \item Correlated information in mechanism design is often modeled by assuming that $u_{i} = \theta_{i}$. That is, each agent privately knows their value, and information about the other agents' values is captured entirely by correlation between the values of different agents.
    \item In the peer selection literature (e.g. \citet{alon2011sum,holzman_impartial_2013}), the principal selects an agent based on nominations from their peers. We can nest the peer selection problem by assuming that each agent $i$'s type $\theta_{i}$ equals the identity $j$ of another agent. To model a principal who aims to select an agent with many nominations, we could let the payoff $u_{j}$ from allocating to an agent $j$ be given by $u_{j} = \sum_{i\in [n]\colon i\neq j}\bm{1}(\theta_{i} = j)$. Similarly, we could capture agents' submitting multiple nominations, scores, rankings, etc. See \Cref{sec:related_literature} for a detailed discussion of the relation of our paper to the peer selection literature.
    \item Suppose there is an underlying social network.
    For each $i$, let $N(i)$ denote agent $i$'s neighbors in the network. 
    Agent $i$ privately observes $u_{i}$ and a noisy signal $u_{j} + \varepsilon_{ij}$ for each of the neighbors $j\in N(i)$, where $\varepsilon_{ij}$ is some random variable.
    Thus, agent $i$'s type $\theta_i$ is $(u_{i}, (u_{j} + \varepsilon_{ij})_{j\in N(i)})$. 
\end{itemize}

\paragraph*{Peer values.}
For all $i\in [n]$ and $\theta_{-i}\in\Theta_{-i}$, we define the \emph{peer value of agent $i$ at $\theta_{-i}$} as $\bar u_{i}(\theta_{-i}) = \mathbb{E}[u_{i}\vert\theta_{-i}]$.
The peer value captures the collective prediction of others about the value $u_i$ of agent $i$.
(If $\mu(\theta_{-i}) = 0$, we arbitrarily put $\bar u_{i}(\theta_{-i}) = 0$.)

The principal's utility from a DIC mechanism can be written in terms of the peer values. For a DIC mechanism $q$, straightforward manipulations show
\begin{equation}\label{eq:principal_utility}
    U(q) = \sum_{\theta\in\Theta} \sum_{i\in[n]}\mu(\theta) q_{i}(\theta_{-i}) \bar u_{i}(\theta_{-i}).
\end{equation}
Thus, the principal can only elicit the agents' peer information.
Indeed, if agents have no peer information in the sense that for all $i$ the peer value $\bar u_{i}(\theta_{-i})$ is constant in $\theta_{-i}$, then \eqref{eq:principal_utility} implies that a constant mechanism that allocates to an agent $i$ with the highest unconditional expected value $\mathbb{E}[u_{i}]$ is optimal.

\section{Jury mechanisms}\label{sec:model:jury_mechanisms}
We begin our analysis with the class of \emph{jury mechanisms}, which resemble review panels commonly observed in practice. These mechanisms ensure DIC in the most straightforward way: those who influence the allocation never receive the good.

\begin{definition}[Jury mechanisms]\label{def:jury_mechanism}
    A mechanism $q$ is a \emph{jury mechanism} if there is a partition of the set of agents into \emph{jurors} $J$ and \emph{candidates} $C$ such that:
    \begin{enumerate}
    \item jurors are never allocated the good ($q_{i} = 0$ for all $i\in J$);
    \item candidates' reports never influence the allocation ($q$ is constant in $\theta_{i}$ for all $i\in C$).
    \end{enumerate}
\end{definition}

\begin{theorem}
\label{prop:juries}
    If $n=2$, then every DIC mechanism is a randomization over jury mechanisms; in particular, a jury mechanism is optimal.

    If $n\leq 3$, then every DIC mechanism $q$ that always allocates the good (i.e., $\sum_{i\in[n]}q_{i} = 1$) is a randomization over jury mechanisms; in particular, a jury mechanism is optimal among mechanisms that always allocate.
\end{theorem}

With two agents, a jury mechanism involves one juror who recommends whether the other agent (the candidate) should be allocated the good or whether the principal should keep the good.
With three agents and mandatory allocation, there is one juror who decides between the two remaining agents (the candidates).

Beyond these special cases, we shall see that jury mechanisms are \emph{not} generally optimal.
Intuitively,
jury mechanisms pre-assign who is eligible to receive the good versus who provides information, but whether an agent fits their assigned role---whether they have a high value or good information about their peers---may vary across type profiles. 
Thus, the principal may do better by considering all agents as sources of information and as potential recipients of the good.
In the next section, we provide a graph-theoretic characterization of DIC mechanisms to understand the trade-off between allocation and information elicitation.
The proof of \Cref{prop:juries} is also based on this graph-theoretic perspective.

Henceforth, we focus on mechanisms in which the principal may keep the good. The one-agent difference in \Cref{prop:juries} holds more generally since, intuitively, mechanisms with $n$ agents can be thought of as mechanisms with $n+1$ agents that always allocate and where one agent has no private information.
We discuss mandatory allocation in \Cref{sec:conclusion}. 

\section{A Graph-Theoretic Characterization of DIC Mechanisms}\label{sec:feasibility_graph}

In this section, we characterize DIC mechanisms via an auxiliary graph, which we will use to gain insights into the structure of optimal DIC mechanisms. As explained below, the graph captures a fundamental trade-off between allocation and information elicitation. 

Recall that when allocating to an agent at a type profile, DIC requires that the principal commits to also allocate to the agent at all type profiles that differ only in the agent's type.
Using the notation $q_{i}(\theta_{-i})$, where we omit $i$'s type from $i$'s allocation in a DIC mechanism $q$, the feasibility constraints read
\begin{equation}\label{eq:feasibility}
	\tag{Feasibility}
	\forall\theta\in\Theta,\quad
	\sum_{i=1}^{n} q_{i}(\theta_{-i}) \leq 1.
\end{equation}
The feasibility graph tracks commitments to allocate to different agents that are mutually incompatible with the feasibility constraints.
For example, the principal cannot simultaneously allocate to agent $i$ at type profile $(\theta_i,\theta_j,\theta_{-ij})$ and to another agent $j$ at type profile $(\theta_i',\theta_j',\theta_{-ij})$ since the principal would have to commit to allocate to both $i$ and $j$ at $(\theta_i',\theta_j,\theta_{-ij})$ to satisfy DIC. In other words, the commitments to allocate to agent $i$ when others report $\theta_{-i}$ and to agent $j$ when others report $\theta_{-j}^{\prime}$ are incompatible if $q_{i}(\theta_{-i})$ and $q_{j}(\theta_{-j}^{\prime})$ simultaneously appear in the same \eqref{eq:feasibility} constraint for some type profile. Define $V = \cup_{i=1}^{n}(\lbrace i\rbrace \times\Theta_{-i})$ and think of each $(i, \theta_{-i})\in V$ as indexing the commitment to agent $i$ at $\theta_{-i}$. 

\begin{definition}[Feasibility graph]
	The \emph{feasibility graph} $G$ is the graph whose vertex set is $V = \cup_{i=1}^{n}(\lbrace i\rbrace \times\Theta_{-i})$ and such that two vertices $(i, \theta_{-i})$ and $(j, \theta_{-j}^{\prime})$ are adjacent if and only if $i \neq j$ and there exists a type profile $\hat{\theta}$ such that $\hat{\theta}_{-i} = \theta_{-i}$ and $\hat{\theta}_{-j} = \theta_{-j}^{\prime}$.
\end{definition}
In \Cref{appendix:examples:feasibility_graph}, we depict the feasibility graph in an example. 

The feasibility graph captures a trade-off between allocating to an agent and using the agent's information.  
First, suppose the principal selects a vertex $(i, \theta_{-i})$ (i.e. commits to allocate to agent $i$ when others report $\theta_{-i}$). Then, the principal cannot select any vertex $(j, (\theta_{i}^{\prime}, \theta_{-ij}))$ for any $j\neq i$ or $\theta_{i}^{\prime}$ since these vertices are all adjacent to $(i, \theta_{-i})$.
Second, suppose the principal does not select $(i, \theta_{-i})$. Then, the principal can decide for each $\theta_{i}^{\prime}$ which vertex $(j, (\theta_{i}^{\prime}, \theta_{-ij}))$ to select (subject to the selected vertices' being non-adjacent to other selected vertices). In the first case, the principal benefits from allocating to $i$; in the second case, from $i$'s information.

In graph-theoretic terms, deterministic DIC mechanisms correspond to stable (or independent) sets of the feasibility graph $G$, and (possibly stochastic) DIC mechanisms correspond to fractional stable sets of the feasibility graph $G$. A \emph{stable set} of $G$ is a subset of pairwise non-adjacent vertices. A stable set $S$ of $G$ can be mapped to a deterministic DIC mechanism $q$ by setting $q_{i}(\theta_{-i})=\bm{1}((i, \theta_{-i}) \in S)$ for all $(i, \theta_{-i})\in V$. Conversely, if $q$ is deterministic and DIC, then $\lbrace v\in V\colon q(v) = 1\rbrace$ is a stable set (where $q(v) = q_{i}(\theta_{-i})$ for $v = (i, \theta_{-i})$). A \emph{fractional stable set} of $G$ is a function $q:V\to [0,1]$ such that $\sum_{v\in X} q(v)\leq 1$ for all maximal cliques $X$ in $G$. A \emph{clique} of $G$ is a subset of pairwise adjacent vertices, and a clique is \emph{maximal} if it is not contained in another clique.
Every type profile corresponds to a maximal clique since the variables appearing in one \eqref{eq:feasibility} constraint are all adjacent and since every two \eqref{eq:feasibility} constraints have at most one variable in common.
We summarize:
\begin{lemma}
\label{lemma:bijection}
	There is a bijection between deterministic DIC mechanisms and stable sets of the feasibility graph.
	There is a bijection between DIC mechanisms and fractional stable sets of the feasibility graph.
\end{lemma}
Identifying an optimal DIC mechanism is tantamount to solving a combinatorial optimization problem.
Specifically, the peer values and the distribution of types imply weights for the vertices of the feasibility graph $G$; the weight on vertex $(i, \theta_{-i})$ is $\mu(\theta_{-i})\bar u_{i}(\theta_{-i})$. The problem of finding an optimal deterministic DIC mechanism is the same as finding a stable set in $G$ with the highest cumulative weight; this is an instance of the \emph{maximum weight stable set (MWSS) problem}. The problem of finding an optimal (possibly stochastic) DIC mechanism is the \emph{fractional relaxation} of this problem.

Next, we derive properties of optimal DIC mechanisms as consequences of this graph-theoretic characterization.

\section{Stochastic Mechanisms and Extreme Points}\label{sec:DIC_characterization}

In this section, we show that randomization is an inherent structural property of optimal DIC mechanisms.
Intuitively, randomization allows the principal to resolve the trade-off between allocation and information elicitation more flexibly because an agent may win with some probability but still influence how the remaining probability is distributed among the other agents.
However, this intuition is very incomplete: because the principal maximizes a linear objective, they can never benefit from simply randomizing over deterministic mechanisms. Instead, the principal must implement a stochastic allocation \emph{conditional} on the agents' reports. Formally, these stochastic mechanisms are extreme points of the set $Q$ of DIC mechanisms.\footnote{An extreme point of $Q$ is a DIC mechanism that cannot be represented as a convex combination of other DIC mechanisms. The extreme points are the natural objects of interest because every (optimal) DIC mechanism can be represented as a convex combination of (optimal) extreme points. Further, in our setting every extreme point is exposed: there exists a joint distribution of types and values that renders this extreme points uniquely optimal among all DIC mechanisms.} We characterize when stochastic extreme points exist, show that they are prevalent when type spaces are rich, and identify why they improve on deterministic DIC mechanisms.

For conciseness, we say that a DIC mechanism is \emph{extreme} if it is an extreme point of $Q$.
All deterministic DIC mechanisms are extreme.
The next theorem addresses the converse: are all extreme DIC mechanisms deterministic?

\begin{theorem}\label{thm:stochastic_existence}
	All extreme DIC mechanisms are deterministic if and only if at least one of the following is true:
	\begin{enumerate}
		\item there are two agents ($n= 2$);
		\item all types spaces are binary ($\vert\Theta_{i}\vert = 2$ for all $i\in [n]$).
	\end{enumerate}
\end{theorem}
(Recall that the model assumes $n\geq 2$ and $\vert\Theta_{i}\vert \geq 2$ for all $i\in [n]$.)
Thus,  except in two special cases, deterministic mechanisms do not suffice for implementation and optimality. But just how prevalent are stochastic mechanisms among the extreme points?

When type spaces are large, then all but a vanishing fraction of extreme DIC mechanisms are stochastic.
To state the result, let $\deterministic Q$ denote the set of deterministic DIC mechanisms, and let $\extremepoints Q$ denote the set of extreme DIC mechanisms.
\begin{theorem}\label{thm:stochastic_prevalence}
    Fix $n\geq 3$.
	For all $\varepsilon > 0$ there exists $m\in\mathbb{N}$ such that if $\vert\Theta_{i}\vert \geq m$ for all $i\in [n]$, then 
    $
		\left\vert \deterministic Q \right\vert < \varepsilon \left\vert \extremepoints Q\right\vert.
	$
\end{theorem}

Stochastic extreme DIC mechanisms are characterized by certain cycles---odd holes---in the feasibility graph $G$.
For $k\in\mathbb{N}$, a set $\lbrace v_{1}, \ldots, v_{k}\rbrace$ of $k$ vertices of $G$ is an \emph{odd (k-)hole} if $k\geq 5$ is odd and, for all $j\in\lbrace 1, \ldots, k\rbrace$, the vertex $v_{j}$ is adjacent to $v_{j-1}$ and $v_{j+1}$ (where $v_{1} = v_{k+1}$) and to no other vertex from $\lbrace v_{1}, \ldots, v_{k}\rbrace$.
Given a DIC mechanism $q$, a \emph{stochastic component of $q$} is an inclusion-wise maximal connected set of vertices $v$ of $G$ such that $q(v)\in (0, 1)$.\footnote{Two vertices are \emph{connected} if there is a path joining them.} Say that a stochastic mechanism is \emph{simple} if it takes values in $\lbrace 0, \frac{1}{2}, 1\rbrace$, i.e. either allocates deterministically or splits the good evenly among two agents.

\begin{theorem}\label{prop:hole_characterization}
    Let $q$ be a stochastic DIC mechanism. If $q$ is extreme, then every stochastic component of $q$ contains an odd hole.
    The converse holds if $q$ is simple.
\end{theorem}

\begin{figure}
	\centering
	\begin{tikzpicture}
		\def\scalefactor{0.5}
		
		\foreach \i in {0,...,6} {
			\node[fill, circle, inner sep=1.5pt] (n\i) at ({90-\i*360/7}:\scalefactor*3cm) {};
		}
		
		\foreach \i/\label in {0/1, 1/0, 2/1, 3/0, 4/1, 5/0, 6/0} {
			\node[text=black] at ({90-\i*360/7}:\scalefactor*2.4cm) {\label};
		}
		
		\foreach \i in {0,...,6} {
			\node[text=black] at ({90-\i*360/7}:\scalefactor*3.8cm) {$\frac{1}{2}$};
		}
		
		\foreach \i in {0,...,5} {
			\pgfmathtruncatemacro{\nexti}{\i+1}
			\draw (n\i) -- (n\nexti);
		}
		\draw (n6) -- (n0);
		
	\end{tikzpicture}
	\caption{An odd hole in the feasibility graph. 
        The outer labels show how a stochastic mechanism might allocate around the hole; the inner labels show how a deterministic mechanism might allocate around the hole.} 
	\label{fig:C7_graph}
\end{figure}
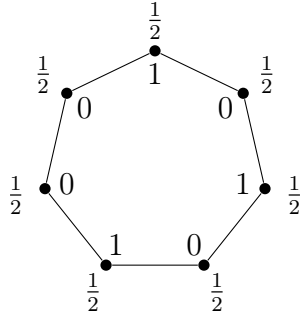

\Cref{prop:hole_characterization} shows why stochastic mechanisms can outperform deterministic mechanisms: they can distribute allocation mass more flexibly around odd holes in the feasibility graph. 
For an illustration, consider \Cref{fig:C7_graph}, which depicts an odd hole in the feasibility graph, together with the allocation probabilities of a deterministic DIC mechanism and a simple stochastic DIC mechanism. 
Recall that adjacent vertices of the feasibility graph are commitments to allocate to two distinct agents that are incompatible with the \eqref{eq:feasibility} constraints at some type profile. 
Thus, the deterministic DIC mechanism can select at most three out of seven vertices from the hole. 
However, the stochastic mechanism can select all seven vertices, each with probability $1/2$, by splitting the good between two agents at type profiles where commitments are incompatible. 
In particular, the stochastic mechanism distributes more total allocation mass around the hole than any deterministic DIC mechanism and, therefore, cannot be written as a convex combination of deterministic DIC mechanisms.
\Cref{prop:hole_characterization} confirms that the stochastic mechanism is extreme (provided that each of its stochastic components contains an odd hole).
In \Cref{appendix:examples:stochastic_extreme_point}, we elaborate on this example, showing how odd holes are embedded in the feasibility graph and how to interpret the hole in terms of the principal's trade-off between allocation and information elicitation.

The proofs of \Cref{thm:stochastic_existence,thm:stochastic_prevalence,prop:hole_characterization} are based on a theorem of \citet{chvatal1975} and the Strong Perfect Graph Theorem (\citet{chudnovsky2006}). Recall that DIC mechanisms correspond to fractional stable sets of the feasibility graph, while deterministic DIC mechanisms correspond to stable sets.
\citet{chvatal1975} shows that, for a general graph, every fractional stable set is a convex combination of (incidence vectors of) stable sets if and only if the given graph is perfect.
The Strong Perfect Graph Theorem asserts that a graph $G$ is perfect if and only if neither $G$ nor the complement of $G$ admit an odd hole.
(We define perfection and the complement of a graph in \Cref{appendix:proof_preliminaries}.)
Equipped with these results and leveraging the special structure of the feasibility graph, we prove  \Cref{thm:stochastic_existence} (existence) and \Cref{prop:hole_characterization} (characterization) by checking for the existence of odd holds and their complements. The proof of \Cref{thm:stochastic_prevalence} (prevalence) is more involved since the number of stable sets diverges as the number of possible types increases. 
However, we can show that the number of certain stable sets increases more slowly than the number of certain odd holes, which is sufficient to establish the result.

\section{The Complexity of Eliciting Peer Information}
\label{sec:complexity}

In this section, we show that optimal DIC mechanisms have a complex structure despite resolving an intuitive economic trade-off between allocation and information elicitation.
To make this point, we first focus on the conceptually simpler subset of deterministic DIC mechanisms and make precise why these mechanisms admit no tractable economic characterization (beyond the graph-theoretic one given in \Cref{sec:feasibility_graph}).
We then explain why this complexity extends to stochastic mechanisms. 

As benchmark, recall that with two agents all deterministic DIC mechanisms are jury mechanisms (which in this case are also the extreme points of the set of DIC mechanisms by \Cref{prop:juries}).
Is there an equally simple description of deterministic DIC mechanisms with three or more agents?

If $n\geq 3$, then the problem of finding an optimal deterministic DIC mechanism is NP-hard, implying significant hurdles to any simple description of deterministic DIC mechanisms. Before discussing the intuition and implications in more detail, we state the formal result.

\begin{definition}[\textsc{OptDet}-$n$]
	For $n\in\mathbb{N}$, let \textsc{OptDet}-$n$ be the following problem.
	The input consists of finite sets $\Theta_{1}, \ldots, \Theta_{n}$ and weights $w_i: \Theta_{-i}\to \mathbb{Z}$ for all $i\in [n]$.
	The problem is to find a deterministic DIC mechanism $q$ (for $n$ agents with type spaces $\Theta_{1}, \ldots, \Theta_{n}$) that maximizes $\sum_{i, \theta} w_{i}(\theta_{-i}) q_{i}(\theta_{-i})$ across all deterministic DIC mechanisms.\footnote{In a given environment, the weights of interest are given by $w_{i}(\theta_{-i}) = \mu(\theta_{-i}) u_{i}(\theta_{-i})$ for all $(i, \theta_{-i})$. Conversely, all weight vectors can result from some environment (up to rescaling). Computation requires that weights are given as integers or rational numbers.
    By fixing $n$, we ensure that complexity does not arise merely from increasing the number of agents.}
\end{definition}

\begin{theorem}\label{thm:npcomplete}
	For fixed $n\geq 3$, the problem \textsc{OptDet}-$n$ is NP-hard.
\end{theorem}

For an intuition, recall the principal's trade-off between allocation and information elicitation.
The key difficulty is that the allocation at one type profile constrains in a non-local, combinatorial way the allocation at other type profiles.
For instance, by allocating to an agent $i$ at a type profile $\theta = (\theta_i,\theta_j,\theta_{-ij})$, the principal can no longer allocate to any agent $j$ at type profile $(\theta_i',\theta_j',\theta_{-ij})$ for all $\theta_i'$ and $\theta_j'$. Consequently, even if at $\theta$ agent $i$ has the highest peer value $\bar{u}_{i}(\theta_{-i}) = \mathbb{E}[u_i|\theta_{-i}]$, we cannot conclude that the optimal mechanism allocates to $i$ because doing so constrains the allocation that can be made at other type profiles. 
The allocations at these type profiles, in turn, constrain the allocations on various other sets of type profiles, and so on.
Moreover, if $i$ has relevant peer information---i.e., if the peer values $\bar{u}_{j}(\theta_{i}^{\prime}, \theta_{-ij}) = \mathbb{E}[u_j|\theta_i',\theta_{-ij}]$ differ significantly across $j$ and $\theta_i'$---it is not obvious how the principal should capitalize on this information precisely because it constrains the decision at further type profiles.

For the proof, recall that deterministic DIC mechanisms correspond to stable sets of the feasibility graph; the problem of finding an optimal deterministic DIC mechanisms is the maximum-weight stable set (MWSS) problem on the feasibility graph. For general graphs, it is known that the MWSS problem is NP-hard. The content of \Cref{thm:npcomplete} is that the same conclusion holds even for feasibility graphs. 
The rough idea behind the reduction argument is that edges in any graph can be simulated via induced paths in a feasibility graph in a way that preserves the structure of stable sets for appropriately chosen weights.

Beyond the direct implications for computation, \Cref{thm:npcomplete} shows that there is no conceptually simple description of deterministic DIC mechanisms.
This interpretation rests on the premise that a conceptually simple description would imply an efficient algorithm for finding an optimal deterministic DIC mechanism. 
Unless $P=NP$, however, such an efficient algorithm does not exist. 
Although there is no formal notion of conceptual complexity in the mechanism design literature, computational complexity is arguably the best available proxy. 
The complexity of allocating with peer information is in contrast to many allocation problems with independent types. There, optimal mechanisms often admit conceptually simple economic descriptions that naturally suggest efficient computational solutions.\footnote{
For example, consider the allocation problem without transfers in \citet{benporath2014}, where the principal can verify agents' types at a cost. \citet{benporath2014} show that optimal mechanisms are characterized by two parameters: a type cutoff and a ``favored'' agent. This characterization immediately yields a computationally efficient solution by exhaustive search over the two parameters. More generally, \citet{vohra2012optimization} shows how a number of problems with and without transfers correspond to polymatroid optimization problems with respect to reduced-form allocations, which can be solved efficiently via a greedy algorithm. See e.g. \citet{cai2012algorithmic} and \citet{alaei2019efficient} for details regarding computation.}

The lack of a conceptually simple description of deterministic DIC mechanisms extends to optimal (possibly stochastic) DIC mechanisms for two reasons.
First, as alluded to earlier, every deterministic DIC mechanisms is extreme and hence an indispensable candidate for optimality; thus, a complete description of optimal DIC mechanisms a fortiori requires a description of deterministic mechanisms.
Second, stochastic extreme points inherit the structure of deterministic mechanisms on the parts of the type space where they allocate deterministically.
To give but one example, take the class of extreme DIC mechanisms where the report of a dummy agent decides whether the principal uses a stochastic extreme DIC mechanism or a deterministic DIC mechanism for the remaining agents.
A description of this class of $n$-agent extreme DIC mechanisms entails a description of all deterministic $(n-1)$-agent DIC mechanisms.\footnote{More generally, fix any stochastic extreme DIC mechanism. Consider the subgraph of the feasibility graph induced by the vertices that are not in the neighborhood of any stochastic component of the feasibility graph. The restriction of the mechanism to this subgraph can be identified with a stable set of the subgraph. This stable set can be replaced arbitrarily with other stable sets of the subgraph to produce new stochastic extreme DIC mechanisms. These stable sets have the same combinatorial structure as the stable sets of the entire feasibility graph, i.e. of deterministic DIC mechanisms. (In fact, the stable sets of the subgraph are themselves stable sets of the entire feasibility graph.)}

\begin{remark}
    One may wonder if more can be said about deterministic mechanisms by restricting to simple type spaces, e.g., binary type spaces. 
    This is the special case identified in \Cref{thm:stochastic_existence} where deterministic DIC mechanisms are optimal among all DIC mechanisms.
    With $n$ agents and binary types, the feasibility graph is the line graph of the $n$-dimensional hypercube. In particular, the problem of determining an optimal deterministic DIC mechanism corresponds to finding an optimal weighted hypercube matching. 
    Hypercube matchings are investigated in the mathematical literature; simple descriptions are not to be expected.    
    For example, the number of these matchings is unknown except for small $n$ (\citet{ostergaard2013enumerating}). There is a description of all (perfect) matchings for $n\leq4$ in \citet{fink2009connectivity}.
\end{remark}

\section{Approximate Optimality and Ranking-based Mechanisms}\label{sec:approximate_optimality}

In this section, we shift our focus from optimal DIC mechanisms to approximately optimal ones. Our earlier findings show that optimal DIC mechanisms lack closed-form descriptions. As a result, optimal mechanisms may be difficult to use and institutionalize in practice. In contrast, the \emph{ranking-based mechanisms} that we propose here are simple, yet retain key properties of optimal mechanisms: they use all agents' information, consider all agents as potential recipients of the good, and randomize the allocation to alleviate incentive constraints.
Moreover, it is easy to see when and why ranking-based mechanisms perform well. Specifically, their performance depends on the extent to which any individual agent’s report influences the peer values of others. As we will argue, this influence is naturally small when there are many agents and, in this case, ranking-based mechanisms are approximately optimal among all DIC mechanisms.

Ranking-based mechanisms aim to allocate to agents who are ranked highly in terms of their peer value $\bar{u}_i(\theta_{-i})=\mathbb{E}[u_i|\theta_{-i}]$. 
Given a type profile $\theta$, the agent with the highest peer value receives rank $1$, the second highest receives rank $2$, and so on. Ties are broken lexicographically, favoring agent $i$ over agent $j$ if $i < j$. It is convenient to normalize all ranks to be in $[0, 1]$ by dividing by the number $n$ of agents. Let $r_i(\theta)$ denote $i$'s rank at $\theta$.\footnote{ 
	Formally, 
	$r_{i}(\theta) = \frac{1}{n}\left\vert\left\lbrace j\in [n]\colon \bar{u}_{j}(\theta_{-j}) > \bar{u}_{i}(\theta_{-i})\right\rbrace\right\vert + \frac{1}{n} \left\vert\left\lbrace j\in [n]\colon (\bar{u}_{j}(\theta_{-j}) = \bar{u}_{i}(\theta_{-i}))\land (i \geq j)\right\rbrace\right\vert.$
	}

It is not incentive compatible to naively allocate the good to an agent with a high peer value: agents can misreport their type to diminish the peer values of other agents, thereby improving their own rank.
Define 
\begin{equation*}
	r_{i}^{\ast}(\theta_{-i}) = \max_{\theta_{i}\in\Theta_{i}} r_{i}(\theta_{i}, \theta_{-i})
\end{equation*}
to be $i$'s \emph{robust rank at $\theta_{-i}$}.
Intuitively, the robust rank is obtained in a hypothetical scenario where agent $i$ appraises their competitors, thereby diminishing their own rank.

A ranking-based mechanisms can be interpreted to work in two steps.
First, the principal randomly selects an agent ranked, say, in the top $10\%$ of peer values. Second, the principal allocates to the selected agent if and only if the agent's \emph{robust} rank is also within the top $10\%$.
Intuitively, the principal checks for a conflict of interest: does the agent pass the bar regardless of what the agent claims about their peers?
\begin{definition}[Ranking-based mechanism]
	Let $p \in (0, 1)$.
	The \emph{ranking-based mechanism $q^{p}$ with threshold $p$} is defined for all $i\in [n]$ and $\theta_{-i}\in\Theta_{-i}$ by:
	\begin{equation*}
		q^{p}_{i}(\theta_{-i}) = 
		\begin{cases}
			\frac{1}{pn}\quad&\text{if $r_{i}^{\ast}(\theta_{-i}) \leq p$ \text{and} $u_{i}(\theta_{-i}) \geq 0$;}
			\\
			0\quad&\text{else.}
		\end{cases}
	\end{equation*}
\end{definition}
Ranking-based mechanisms are DIC: $q^{p}_{i}(\theta_{-i})$ does not depend on agent $i$'s type report.
Moreover, $q^{p}$ is a feasible mechanism, never allocating more than is available: the number of agents with a robust rank better than $p$ is at most the number of agents with a rank better than $p$ which, in turn, is at most $pn$ by definition of the rank.

A crucial factor for the performance of ranking-based mechanisms is the impact that any individual agent's report has on their own rank.
We refer to this impact as the informational size.
Informational size is crucial since the mechanism may inefficiently withhold the good with high probability at type profiles where many agents have a unilateral misreport that raises their rank above the rank threshold.

\begin{definition}
For $\theta\in\Theta$, the \emph{informational size at $\theta$} is denoted $\delta(\theta)$ and is given by
\begin{equation*}
    \delta(\theta) =
    \max\limits_{i\in [n], \theta_{i}^{\prime}\in\Theta_{i}} \left\vert r_{i}(\theta_{i}, \theta_{-i}) - r_{i}(\theta_{i}^{\prime}, \theta_{-i})\right\vert.
\end{equation*}
\end{definition}

Informational size is small if no agent has a large influence on the peer values of a large fraction of other agents. Small informational size is a natural assumption in environments with many agents, as illustrated by the following two examples.

\begin{example}
\label{example:infosize1}
Informational size is small if each agent only influences the peer values of few others (but possibly has a large influence on those).
Suppose there is an underlying social network. For each $i$, let $N(i)$ denote agent $i$'s neighbors in the network.
Agent $i$'s type is informative only about the values and types of their neighbors $N(i)$.
If the network is sparse in the sense that the degree $\vert N(i)\vert$ of every agent $i$ is small relative to the size $n$ of the network, then informational size is also small since each agent's report only influences the peer values of their neighbors. 
(Specifically, agent $i$ can impact their own rank by at most $\vert N(i)\vert / n$.) 
Besides the examples given in the introduction, a sparse network may also arise from information acquisition constraints: for example, in peer review, each researcher cannot feasibly assess all other researchers. 
\end{example}

\begin{example}
Informational size may also be small if each agent can influence the peer values of many others, but influences each given peer value only to a small degree.
Suppose each agent $i$ privately observes a noisy signal $u_{j} + \varepsilon_{ij}$ for \emph{every} other agent $j$, where $\varepsilon_{ij}$ is some random variable. If the variables $\varepsilon_{ij}$ are conditionally i.i.d. across $i$ for every $j$, then informational size is small when there are many agents, except at a vanishing number of type profiles where peer values are highly concentrated.
Alternatively, assume each agent's type specifies for every other agent a numerical rating between, say, 1 and 5. Types could be correlated across agents. If the principal's values are simply the average peer ratings, then the informational size is small whenever there are many agents, except at a vanishing number of type profiles where peer values are highly concentrated.
\end{example}
	
\begin{remark}\label{remark:info_smallness_notions}
    Similar notions of informational size appear in earlier work on mechanism design and general equilibrium with informational asymmetries, e.g. \citet{gul1992asymptotic,mclean2002informational,mclean2015implementation,andreyanov2021robust,gerardi2009aggregation}. In those models, the agents have private signals about an unobserved state. Informational size measures the impact of individual signals on the posterior about the state conditional on the full profile of private signals. For our allocation problem, it is instead natural to define informational size via peer values and ranks.
\end{remark}

The next result shows that ranking-based mechanisms become approximately optimal as informational size vanishes probabilistically. The fact that informational size is only required to vanish probabilistically allows for a small number of type profiles where informational size is large, such as type profiles where peer values are highly concentrated.

The result uses a mild technical condition.
\begin{definition}[Regularity]\label{def:regularity}
A sequence $(n, \Theta^{n}, \mu^{n})_{n\in\mathbb{N}}$ of environments with associated peer values $(\bar{u}^{n})_{n\in\mathbb{N}}$ is \emph{regular} if for all $\varepsilon > 0$ there exists $p \in (0, 1)$ such that
\begin{equation*}
\lim_{n\to\infty}\mu^{n}\left\lbrace\theta\in\Theta^{n}\colon 
    \left\vert\left\lbrace j\in [n]\colon
    \bar{u}_{j}^{n}(\theta_{-j}) + \varepsilon\geq \max_{i\in [n]}\bar{u}_{i}^{n}(\theta_{-i})\right\rbrace\right\vert
    \geq pn 
    \right\rbrace = 1.
\end{equation*}
\end{definition}
Regularity says that the number of agents who are competitive with the top-ranked agent is proportional to the total number of agents.  
Regularity is consistent with the narrative that it is difficult to distinguish the very best agents by peer evaluations alone (see e.g. the discussion of \citet{fang2016casadevall} in the context of science funding).
Nevertheless, even if regularity is violated but informational size is small, ranking-based mechanism with a threshold $p$ close to $0$ still allocate to the top agents with overwhelming probability---they may just fail to pick up on the extraordinary value of the very best agent.

\begin{theorem}\label{thm:ranking_based}
    Let $(n, \Theta^{n}, \mu^{n})_{n\in\mathbb{N}}$ be a regular sequence of environments. Suppose the associated informational size $(\delta^{n})_{n\in\mathbb{N}}$ converges to $0$ in probability; that is, for all $d > 0$,
    \begin{equation*}
        \lim_{n\to\infty}\mu^{n}\left\lbrace\theta\in\Theta^{n}\colon \delta^{n}(\theta) > d\right\rbrace = 0.        
    \end{equation*}
    Then, the difference between the principal's expected utility from an optimal ranking-based mechanism and an optimal DIC mechanism vanishes as $n\to\infty$.
\end{theorem}

The proof goes as follows. Fixing a threshold $p \in (0, 1)$ and letting the informational size vanish in probability, we show that the ex-ante probability that the ranking-based mechanism $q^{p}$ withholds the good vanishes. Conditional on allocating, $q^{p}$ allocates to an agent with rank below $p$. Under regularity, if $p$ is close to $0$, the peer value of such an agent is close to the highest peer value (with high probability). 
Recall from \eqref{eq:principal_utility} that the principal's expected utility in a DIC mechanism is determined by the peer values.
In particular, no DIC mechanism does better than always allocating to the highest peer value. This upper bound is approximately attained by the optimal ranking-based mechanism, which completes the proof.
For every fixed number  $n$ of agents, the proof also implies lower bounds on the performance of ranking-based mechanisms in terms of $p$ and informational size.

A broader insight is that incentives matter little for eliciting peer information when agents are informationally small.
For fixed informational size, it is typically impossible to identify and allocate to the agent with the highest peer value. 
The reason is that, by DIC, the principal cannot elicit an agent's information about how the agent is ranked relative to others and then use this information to decide whether to allocate to the agent.
Optimally resolving this tension requires complex DIC mechanisms (\Cref{sec:complexity}).
But, if informational size is small, then an individual agent's information is less important for how the agent is ranked relative to others, and hence there is little loss from ignoring this information when deciding the agent's allocation.
Thus, simple mechanisms such as ranking-based mechanisms can approximately attain the upper bound of always allocating to the agent with the highest peer value.

Ranking-based mechanisms are also approximately optimal for allocating \emph{multiple} units of a good to agents with unit demand. \Cref{thm:ranking_based} generalizes not only to sequences of environments along which the ratio of the number of units to the number of agents vanishes, but also to sequences along which this ratio is bounded away from zero (\Cref{appendix:multiunit_allocation:ranking}).

\begin{remark}
    The performance of jury mechanisms in allocation problems with many agents is less clear: 
    vanishing informational size alone does not imply that jury mechanisms are approximately optimal.
    In \Cref{appendix:jury_mechanisms}, we describe environments along the lines of \Cref{example:infosize1} in which agents are informed only about their neighbors in a network. In these environments, the performance of jury mechanisms remains bounded away from the performance of optimal mechanisms as informational size vanishes.

    Nevertheless, jury mechanisms perform well in special environments in which there is little loss from pre-assigning jurors and candidates. 
    In the \Cref{appendix:jury_mechanisms}, we show that jury mechanisms are approximately optimal with many agents for the problem of allocating a \emph{single} unit if, from an ex-ante perspective, agents are exchangeable as suppliers of information about others or exchangeable as recipients of the good. 
\end{remark}

\section{Related Literature}\label{sec:related_literature}

This paper contributes to the literature on correlation in mechanism design, on allocation without transfers, on peer selection, and on the role of randomization in mechanism design.

In settings with transfers, \citet{cremer1985optimal,cremermclean1988} and \citet{mcafee1992correlated} show how the principal can exploit correlation to extract all private information without leaving information rents to the agents.\footnote{See also \citet{lopomo2022detectability} for a recent treatment.}
The principal offers each agent a menu of lotteries regarding the types of the other agents.
Since different types of an agent hold different beliefs about the types of others, different types self-select into different lotteries (under a condition on the correlation structure).
These constructions are infeasible in our setting since there are no monetary transfers and since we demand robustness to beliefs.\footnote{\citet{cremer1985optimal,cremermclean1988} also consider full surplus extraction in \emph{dominant-strategy} IC mechanisms, but there they still allow agents' participation decisions to depend on their beliefs.}

Without transfers, \citet{kattwinkel2022fullybiased} characterize Bayesian IC (BIC) mechanisms with two agents and correlated types when the principal must allocate the good.
By contrast, we focus on dominant-strategy IC mechanisms with many agents.
Since all DIC mechanisms are BIC, our complexity results suggest that a simple characterization of BIC mechanisms may be difficult to find for a general number of agents.
We are not aware of other papers on allocation problems without transfers where information is correlated across agents. Other papers instead consider correlation in the form of an external signal that the principal can use to cross-check an agent's type report (\citet{kattwinkel2019,bloch2023selecting,kattwinkel2023costless,pereyra2023optimal}).
  
A recent literature on allocation problems without transfers studies non-monetary screening devices in settings with independent types (i.e. there is no peer information).
These devices include verification,\footnote{See \citet{benporath2014,epitropou2019optimal,erlanson2019note,erlanson2024optimal,halac2020commitment,patel2022costly}.}
ex-post punishments,\footnote{See \citet{mylovanov2017optimal,li2020mechanism}.}
promises of future allocations,\footnote{See \citet{kovac2013optimal,guo2018dynamic,lipnowski2020repeated,li2023dynamics}.}
evidence,\footnote{See \citet{benporath2019,ben2023sequential}.} 
costly signaling,\footnote{See \citet{condorelli2012,chakravartykaplan2013,akbarpour2023comparison}.} 
heterogeneous risk attitudes,\footnote{See \citet{ortoleva2021diverse}.}
investment and falsification,\footnote{See \citet{augias2023non,perez2022test,perez2023fraud,li2024screening}.} and
allocative externalities.\footnote{See \citet{bhaskar2019resource,goldluecke2023multiple}.}  
An important direction for future research is to combine peer information (i.e. correlated types) with such non-monetary devices. Towards this goal, we contribute by analyzing the natural theoretical benchmark with correlation and without other devices and by developing novel tools for this analysis.

We also contribute to the literature on peer selection at the intersection of economics and computer science (for a survey, see \citet{olckers2022manipulation}).
In this literature, the principal aims to select an agent based on nominations (or grades, rankings, etc.) from their peers.
One difference to this literature is that we do not start with nominations as primitives. Instead, the principal's payoffs from the allocation depend on arbitrary private information that agents hold about one another.
Nominations are a special case in which an agent's nomination equals their type.
The other key difference is that we take an optimal mechanism design approach.
\begin{itemize}
    \item One substrand, following \citet{de2008impartial,holzman_impartial_2013}, takes an axiomatic approach.
    The central axiom---\emph{impartiality}---is equivalent to DIC: one's own submitted nomination should have no impact on one's own chances of being selected.
    Thus, our results yield novel insights about the set of impartial nomination rules, including a graph-theoretic characterization of the set and its extreme points, as well as a complexity-theoretic result about deterministic rules.
    Moreover, our \Cref{prop:juries} generalizes results by \citet[Proposition 2.i]{holzman_impartial_2013} and \citet[Theorem 5, obtained in the context of exchange economies]{kato2004non}. Their results can be interpreted in our context as showing that all three-agent \emph{deterministic} DIC mechanisms that always allocate the good are jury mechanisms. \Cref{prop:juries} extends this characterization to stochastic mechanisms.
    \item Another substrand, following \citet{alon2011sum}, provides approximation guarantees relative to various benchmarks.
    We instead focus on expected values, leading to different results and calling for different techniques.\footnote{For example, jury mechanisms can be optimal in our setup, but the \emph{$2$-partition mechanism} of \citet{alon2011sum}, an analogue of jury mechanisms, is not optimal by their criterion (\citet{fischer2015optimal}). Our focus on expected values lets us focus on extreme DIC mechanisms, which have so far gone unstudied.}
    The paper by \citet{kurokawa2015impartial} deserves special mention; their \emph{credible subset mechanism} shares similarities with our ranking-based mechanisms. In their model, motivated by conference peer review, $n$ agents apply to $k$ slots and each agents provides a numerical grade for $m$ other agents. The credible subset mechanism selects $k$ agents from those that would make it into the top $k$ based on average peer grades if agents gave the worst possible grades for others. Similar to ranking-based mechanisms, incentive compatibility requires that sometimes no agent be selected. The credible subset mechanism provides a good approximation guarantee relative to naively selecting the $k$ best agents if the number of slots $k$ is large relative to the number of grades $m$ per agent. In contrast, our ranking-based mechanisms perform well for general models of peer information and without restrictions on the number of slots or ratings per agent, provided that agents are informationally small.
\end{itemize}

\citet{bloch2021practice, bloch2022friend} and \citet{baumann2023robust} study models in which agents hold information about their neighbors' allocation values in a social network. \citet{bloch2021practice,bloch2022friend} assume that agents observe ordinal comparisons between their neighbors' values. They ask for which networks the principal can reconstruct the ordinal ranking of agents from their reports if each agent desires a high rank.
\citet{baumann2023robust} assumes that each agent perfectly observes their neighbors' values and that information is partially verifiable.
Baumann asks for which networks the principal can fully implement the first-best of selecting the highest value agent (under various equilibrium notions).

Finally, we contribute to the literature on the gap between stochastic and deterministic mechanisms (e.g. \citet{chen2019equivalence,jarman2017deterministic,mora2022deterministic,pycia2015decomposing,budish2013designing,kovavc2009stochastic}) by showing that, in our setting with correlation and no transfers, deterministic mechanisms typically do not suffice for implementation and optimality.
Our techniques based on the theory of perfect graphs are novel to the mechanism design literature and may be useful for understanding stochastic mechanisms in other settings with combinatorial constraints. 

\section{Concluding Remarks}
\label{sec:conclusion}

We conclude by discussing extensions and shortcomings of our analysis, including the case of mandatory allocation, the allocation of multiple goods, restricted environments, costly information acquisition, nepotism, and coalition-proofness.

\paragraph*{Mandatory allocation.}
We have considered mechanisms where the principal may withhold the good from the agents. The principal may wish to do so for two reasons. First, the (opportunity) cost of providing the good may exceed the benefit from allocating. Second, withholding the good can help address incentive constraints; for example, when there are three agents, jury mechanisms are optimal among DIC mechanisms that always allocate (\Cref{prop:juries}) while optimal DIC mechanisms that withhold the good can be markedly more complex and involve randomization (\Cref{thm:stochastic_existence,thm:stochastic_prevalence,prop:hole_characterization,thm:npcomplete}).

Nevertheless, in some applications, the principal may be constrained to always allocate the good to the agents.
For example, it may untenable to withhold social aid or the assignment of an important task.
Another example is a ratchet effect where the principal allocates all available resources because the principal expects that leftover resources are interpreted as a signal of reduced needs
(see e.g. \citet{liebman2017expiring}).

We provide analogues for our results when the principal must allocate (\Cref{appendix:mandatory_allocation}).
For the analogues of the characterization and complexity results (\Cref{sec:DIC_characterization,sec:complexity}), the only change is to raise the threshold for the number of agents by one. For example, if there are at least \emph{four} agents and each agent has many possible types, then stochastic extreme DIC mechanisms that always allocate are prevalent among the extreme points of the set DIC mechanisms that always allocate.
For an intuition, think of the principal keeping the object as the principal allocating to a default agent $n+1$ whose report never influences the allocation.
(This intuition is incomplete because, in the mandatory allocation case, there exist DIC mechanisms without any default agent.)
Similarly, ranking-based mechanisms can be modified by designating one or more default agents. 

\paragraph*{Multiple Goods.}
Our qualitative insights extend to a setting in which the principal has $k$ units available (e.g. several research grants) and allocates to each agent at most one unit. 
Ranking-based mechanisms remain approximately optimal for allocating multiple units (\Cref{appendix:multiunit_allocation:ranking}). 
In the graph-theoretic characterization of DIC mechanisms, stable sets are now replaced with subgraphs that contain no clique of size greater than $k$. Finding an optimal deterministic mechanism remains NP-hard if $k\leq n-2$ (\Cref{appendix:multiunit_allocation:hardness}). In particular, this result implies the existence of stochastic extreme points.

\paragraph*{Restricted environments.}
Without restrictions on the environment---the joint distribution of the agents' types and the principal's payoffs---all extreme points of the set of DIC mechanisms are candidates for optimality.
We have argued that these mechanisms admit no simple description.
Hence it may be worth investigating whether there are meaningful restricted environments with a more tractable subset of candidate mechanisms (i.e. a restricted set of objective functionals that only exposes a subset of the extreme points).
We note that peer information is uninteresting if the environment is ``too simple.'' For example, consider a standard symmetric environment in which each agent knows their own allocation value and these values are exchangeable random variables: if the good must be allocated, then it is optimal to ignore all reports and pick a winner at random (\Cref{appendix:symmetry_nothing_goes}).

\paragraph*{Costly information acquisition.}
Agents may incur significant costs when acquiring information to evaluate their peers (e.g., when evaluating a grant proposal).
These costs may strengthen the case for ranking-based mechanisms: if each agent can feasibly acquire information only about a few other agents, then each agent is informationally small if there are many agents. It would be interesting to study information acquisition costs, perhaps even participation costs (e.g., the costs of preparing an application), more explicitly in the context of peer mechanisms.

Recent discussions in science funding suggest to randomize the allocation among agents receiving the best peer reviews since it can be too costly for reviewers to further distinguish among the best proposals (\citet{fang2016casadevall}).
Although our model does not feature information acquisition costs, ranking-based mechanisms still randomize among the top agents to ensure DIC. Thus, providing incentives for honest evaluations can be seen as a supporting argument for randomization in science funding.

\paragraph*{Nepotism.}
We have assumed that each agent only cares about their own allocation.
In practice, however, agents may also care about the allocations of others, such as their friends and family when it comes to the allocation of targeted aid (for empirical evidence, see e.g. the survey by \citet{olckers2022manipulation}). In this case, an analogous notion of DIC might require that each agent must influence neither their own allocation nor the allocation of their friends.
One could then analyze optimal mechanisms based on a modified feasibility graph.\footnote{In this graph, each vertex specifies an agent $i$ and types of all agents except $i$ and $i$'s friends. As in the feasibility graph, such a vertex should be viewed as a commitment to allocate to agent $i$. Vertices are adjacent if the commitments potentially violate feasibility.} Ranking-based mechanisms would continue to work well if each agent has few friends relative to the total number of agents and is also informationally small for the ranks of their friends.

\paragraph*{Coalition-proofness.}
In our model, only constant mechanisms are coalition-proof, where coalition-proofness means that no subset of agents can improve its total allocation probability via a joint misreport.\footnote{Consider a coalition of two agents. Coalition-proofness implies DIC, meaning no agent can change their own allocation. The total allocation probability in a two-person coalition must be fixed, and therefore no agent can change the allocation of their coalition partner either. 
The impossibility obtains analogously for coalition-proofness based on Pareto improvements.}
This result is not too surprising since we have abstracted away from any screening devices beyond the allocation itself that may be available to the principal.
To understand collusion more generally, it would be interesting to consider a dynamic allocation problem where collusion is sustained via repeated interactions (see \citet{shah2022challenges} for a discussion of collusion rings in peer review).

\appendix

\section{Proofs}\label{appendix:proofs}

\subsection{Preliminaries}\label{appendix:proof_preliminaries}

We recall some definitions and present some basic facts about the feasibility graph.
\paragraph*{Definitions.}
For a moment, let $G$ with vertices $V$ and edges $E$ be a simple undirected graph.

Given a subset $V^{\prime}$ of vertices, the \emph{subgraph of $G$ induced by $V^{\prime}$}, denoted $G[V^{\prime}]$, is the graph on vertices $V^{\prime}$ where every two vertices in $V^{\prime}$ are adjacent if and only if they are adjacent in $G$.
A graph is an \emph{induced subgraph of $G$} if it is induced by a subset of vertices.

We next define perfection.
The \emph{clique number} of $G$ is the maximum cardinality $\vert C\vert$ across cliques $C$ of $G$.
For an integer $k$, a partition of $V$ into $k$ non-empty stable sets is a \emph{$k$-node-colouring}.
The \emph{chromatic number} of $G$ is the smallest integer $k$ such that $G$ admits a $k$-node-colouring.
The graph $G$ is \emph{perfect} if for all induced subgraphs $(V^{\prime}, E^{\prime})$ of $G$ the clique number of $(V^{\prime}, E^{\prime})$ equals the chromatic number of $(V^{\prime}, E^{\prime})$.

The complement $\bar{G}$ of $G$ is the graph on the same set of vertices as $G$ but where every pair of vertices is adjacent in $\bar{G}$ if and only if the pair is non-adjacent in $G$.

The \emph{fractional stable set polytope of $G$}, denoted $QSTAB(G)$, is the set of functions $q\colon V\to [0, 1]$ such that all maximal cliques $X$ satisfy $\sum_{v\in X}q(v) \leq 1$. 
The \emph{stable set polytope of $G$}, denoted $STAB(G)$, is the convex hull of the set of vectors that are incidence vectors of stable sets of $G$.

\paragraph*{Notation and basic facts.}

In what follows, $G$ refers to the feasibility graph.

For some proofs with three or more agents it will be convenient to denote vertices as follows. Given a type profile $\theta$, we write $\theta_{-123}$ to mean the agents other than $1$ to $3$ (if such other agents exist). We also write $(\cdot, \theta_{2}, \theta_{3}, \theta_{-123})$ to mean the vertex $(1, \theta_{-1})$, we write $(\theta_{1}, \cdot, \theta_{3}, \theta_{-123})$ to mean $(2, \theta_{-2})$, and so on.

We say a type profile $\theta$ \emph{contains} the vertices $\lbrace (i, \theta_{-i})\colon i\in[n]\rbrace$.
If $(i, \theta_{-i})$ and $(j, \theta_{-j}^{\prime})$ are adjacent vertices, then $\hat{\theta} = (\theta_{i}^{\prime}, \theta_{j}, \theta_{-ij})$ is the unique type profile containing $(i, \theta_{-i})$ and $(j, \theta_{-j}^{\prime})$.
The map $\theta\mapsto \lbrace (i, \theta_{-i})\colon i\in[n]\rbrace$ is a bijection from types profiles to maximal cliques.
It follows that for all two adjacent vertices $v$ and $v^{\prime}$, there is a unique maximal clique containing both $v$ and $v^{\prime}$.

\subsection{Proof of \headercref{Theorem}{{prop:juries}}}

If $n=2$ and $q$ is a DIC mechanism that always allocates, then $q_{1}(\theta_{2}) + q_{1}(\theta_{1}) = 1$ for all $(\theta_{1}, \theta_{2})\in\Theta$.
In particular, $q$ is constant.
A constant mechanism is a convex combination of constant deterministic mechanisms. All constant mechanisms are jury mechanisms.

For $n=2$, a DIC mechanism (that need not always allocate) is a special DIC mechanism with three agents that always allocates and where one agent is a default agent whose report does not influence the allocation but absorbs all residual allocation probability.

Thus, we prove the claim if $n=3$ for DIC mechanisms that always allocate.
In what follows, let $n=3$, and let $\bar{Q}$ denote set of DIC mechanisms that always allocate.

In our terminology, \citet[Proposition 2.ii]{holzman_impartial_2013} show that all deterministic mechanisms in $\bar{Q}$ are jury mechanisms.
Hence, to prove that all mechanisms in $\bar{Q}$ are convex combinations of jury mechanisms, it suffices to show that all extreme points of $\bar{Q}$ are deterministic.
Thus, let $q\in\bar{Q}$ be stochastic.
We show $q$ is not an extreme point of $\bar{Q}$.

Let $\tilde{V}$ denote the subset of vertices $v$ such that $q(v)\in (0, 1)$.
We construct non-empty disjoint subsets $R$ (red) and $B$ (blue) of $\tilde{V}$ such that, for all maximal cliques $X$, either $X\cap (R\cup B)=\emptyset$ or $\vert X\cap R\vert = \vert X\cap B\vert = 1$.
Recall that we may identify maximal cliques with type profiles.
If $X\cap (R\cup B)=\emptyset$, we say that the type profile (associated with $X$) is \emph{uncolored}; if $\vert X\cap R\vert = \vert X\cap B\vert = 1$, we say that the type profile is \emph{two-colored}.

The existence of such sets $R$ and $B$ imply that $q$ is not an extreme point.
Indeed, for a sufficiently small number $\varepsilon > 0$, both $q + \varepsilon(\bm{1}_{R} - \bm{1}_{B})$ and $q - \varepsilon(\bm{1}_{R} - \bm{1}_{B})$ are in $\bar{Q}$, and $q$ is a convex combination of $q + \varepsilon(\bm{1}_{R} - \bm{1}_{B})$ and $q - \varepsilon(\bm{1}_{R} - \bm{1}_{B})$.
Here, $\bm{1}_{R}$ and $\bm{1}_{B}$, respectively, denote the indicator functions for $R$ and $B$, respectively.

We assume without loss that there exists a type profile $(\theta_{1}^{0}, \theta_{2}^{0}, \theta_{3}^0)$ such that the vertices $(\cdot, \theta_{2}^{0}, \theta_{3}^{0})$ and $(\theta_{1}^{0}, \cdot, \theta_{3}^{0})$ are in $\tilde{V}$.
Define $T_{1} = \lbrace \theta_{1}\in\Theta_1\colon (\theta_{1}, \cdot, \theta_{3}^{0}) \notin \tilde{V}\rbrace$ and $T_{2} = \lbrace \theta_{2}\in\Theta_2\colon (\cdot, \theta_{2}, \theta_{3}^{0}) \notin \tilde{V}\rbrace$.
Let $T_{1}^{c} = \Theta_{1}\setminus T_{1}$ and $T_{2}^{c} = \Theta_{2}\setminus T_{2}^{c}$.
Both $T_{1}^{c}$ and $T_{2}^{c}$ are non-empty since $(\cdot, \theta_{2}^{0}, \theta_{3}^{0})$ and $(\theta_{1}^{0}, \cdot, \theta_{3}^{0})$ are in $\tilde{V}$.

First, if $T_{1}$ and $T_{2}$ are both empty, then the sets $R = \lbrace (\theta_{1}, \cdot, \theta_{3}^0)\colon \theta_{1}\in\Theta_{1}\rbrace $ and $B = \lbrace (\cdot, \theta_{2}, \theta_{3}^{0})\colon \theta_{2}\in\Theta_{2}\rbrace $ have the desired properties. 

Henceforth, let $T_{1}\neq \emptyset$ (the case $T_{2}\neq \emptyset$ being analogous).
We distinguish two cases. 

\textbf{Case 1.} \emph{Let $T_{2} \neq \emptyset$.}
Define $\tilde{\Theta}_{3}$ as the set of $\theta_{3}\in\Theta_{3}$ such that for all $\theta_{1}\in T_{1}^{c}$ and $\theta_{2}\in T_{2}^{c}$ we have $(\theta_{1}, \cdot, \theta_{3}) \in \tilde{V}$ and $(\cdot, \theta_{2}, \theta_{3})\in\tilde{V}$.
Note $\tilde{\Theta}_{3}$ is non-empty since $\theta_{3}^{0}\in \tilde{\Theta}_{3}$.
We make the following observations.
\begin{itemize}
    \item \textbf{Observation 1.1.} \emph{If $(\theta_{1}, \theta_{2})\in T_{1}\times T_{2}$, then $(\theta_{1}, \theta_{2}, \cdot)\notin \tilde{V}$.} 
    
    Indeed, if $(\theta_{1}, \theta_{2}, \cdot)\in \tilde{V}$, then only agent $3$ enjoys an interior winning probability at the profile $(\theta_{1}, \theta_{2}, \theta_{3}^{0})$, which is impossible since the object is always allocated.
    \item \textbf{Observation 1.2.} \emph{If $(\theta_{1}, \theta_{2})\in (T_{1}\times T_{2}^{c})\cup (T_{1}^{c}\times T_{2})$, then $(\theta_{1}, \theta_{2}, \cdot)\in \tilde{V}$.} 
    
    Indeed, the assumption $(\theta_{1}, \theta_{2})\in (T_{1}\times T_{2}^{c})\cup (T_{1}^{c}\times T_{2})$ implies that exactly one of agent $1$ and agent $2$ enjoys an interior winning probability at the profile $(\theta_{1}, \theta_{2}, \theta_{3}^{0})$; hence agent $3$ must also enjoy an interior winning probability at this profile.
    \item \textbf{Observation 1.3.} \emph{If $\theta_{3}\notin \tilde{\Theta}_{3}$, then all $\theta_{1}\in T_{1}$ and $\theta_{2}\in T_{2}$ satisfy $(\theta_{1}, \cdot, \theta_{3})\in\tilde{V}$ and $(\cdot, \theta_{2}, \theta_{3})\in\tilde{V}$.}

    Indeed, let $\theta_{3}\notin \tilde{\Theta}_{3}$.
    By definition of $\tilde{\Theta}_{3}$, there exists $\theta_{1}^{\prime}\in T_{1}^{c}$ or $\theta_{2}^{\prime}\in T_{2}^{c}$ such that $(\theta_{1}^{\prime}, \cdot, \theta_{3}) \notin \tilde{V}$ or $(\cdot, \theta_{2}^{\prime}, \theta_{3})\notin\tilde{V}$.
    We first prove the claim in the case in which there is $\theta_{2}^{\prime}\in T_{2}^{c}$ such that $(\cdot, \theta_{2}^{\prime}, \theta_{3})\notin\tilde{V}$.

    Let $\theta_{1}\in T_{1}$. We show $(\theta_{1}, \cdot, \theta_{3})\in\tilde{V}$.
    Since $(\theta_{1}, \theta_{2}^{\prime}) \in T_{1}\times T_{2}^{c}$, we have $(\theta_{1}, \theta_{2}^{\prime}, \cdot) \in \tilde{V}$ (Observation 1.2).
    Since also $(\cdot, \theta_{2}^{\prime}, \theta_{3})\notin\tilde{V}$ (by assumption), feasibility at $(\theta_{1}, \theta_{2}^{\prime}, \theta_{3})$ requires $(\theta_{1}, \cdot, \theta_{3})\in\tilde{V}$, as desired.

    Now let $\theta_{2}\in T_{2}$. We show $(\cdot, \theta_{2}, \theta_{3})\in\tilde{V}$.
    By assumption, $T_{1}$ is non-empty.
    Find $\theta_{1}\in T_{1}$.
    By the previous paragraph, $(\theta_{1}, \cdot, \theta_{3})\in\tilde{V}$.
    Since $(\theta_{1}, \theta_{2}) \in T_{1}\times T_{2}$, we know $(\theta_{1}, \theta_{2}, \cdot)\notin \tilde{V}$ (Observation 1.1).
    Thus, since $(\theta_{1}, \cdot, \theta_{3})\in\tilde{V}$, feasibility at $(\theta_{1}, \theta_{2}, \theta_{3})$ requires $(\cdot, \theta_{2}, \theta_{3})\in\tilde{V}$, as desired.

    We have proven the claim in the case in which there is $\theta_{2}^{\prime}\in T_{2}^{c}$ such that $(\cdot, \theta_{2}^{\prime}, \theta_{3})\notin\tilde{V}$.
    If there exists $\theta_{1}^{\prime}\in T_{1}^{c}$ such that $(\theta_{1}^{\prime}, \cdot, \theta_{3}) \notin \tilde{V}$, the argument is analogous with the roles of agents $1$ and $2$ switched. Switching the roles is valid since the claims proven so far hold symmetrically for both agents $1$ and $2$ and since both $T_{1}$ and $T_{2}$ are non-empty.
\end{itemize}
We next define our candidates for $R$ and $B$.
We first define the following:
\begin{align*}
    \forall\theta_{3}\in \tilde{\Theta}_{3},\quad
    R^{\prime}(\theta_{3}) &= \lbrace (\cdot, \theta_{2}, \theta_{3})\colon \theta_{2} \in T_{2}^{c}\rbrace
    \quad\text{and}\quad
    B^{\prime}(\theta_{3}) = \lbrace (\theta_{1}, \cdot, \theta_{3})\colon \theta_{1}\in T_{1}^{c}\rbrace;
    \\
    \forall\theta_{3}\notin \tilde{\Theta}_{3},\quad
    R^{\prime}(\theta_{3}) &= \lbrace (\theta_{1}, \cdot, \theta_{3})\colon \theta_{1}\in T_{1}\rbrace
    \quad\text{and}\quad
    B^{\prime}(\theta_{3}) = \lbrace (\cdot, \theta_{2}, \theta_{3})\colon \theta_{2}\in T_{2}\rbrace.
\end{align*}
Let $R_{0} =  \lbrace (\theta_{1}, \theta_{2}, \cdot)\colon \theta_{1}\in T_{1}^{c}, \theta_{2}\in T_{2}\rbrace$ and $B_{0} = \lbrace (\theta_{1}, \theta_{2}, \cdot)\colon \theta_{1}\in T_{1}, \theta_{2}\in T_{2}^{c}\rbrace$.
Finally, let $R = R_{0} \cup (\cup_{\theta_{3}\in\Theta_{3}} R^{\prime}(\theta_{3}))$ and $B = B_{0} \cup (\cup_{\theta_{3}\in\Theta_{3}} B^{\prime}(\theta_{3}))$.

We next verify $R\cup B\subseteq \tilde{V}$.
Observation 1.2 asserts $R_{0}\cup B_{0}\subseteq \tilde{V}$.
Next, for all $\theta_{3}\in \tilde{\Theta}_{3}$, the definition of $\tilde{\Theta}_{3}$ immediately implies $R^{\prime}(\theta_{3})\cup B^{\prime}(\theta_{3})\subseteq \tilde{V}$, and for all $\theta_{3}\notin \tilde{\Theta}_{3}$ Observation 1.3 implies $R^{\prime}(\theta_{3})\cup B^{\prime}(\theta_{3})\subseteq \tilde{V}$. The sets $R$ and $B$ are disjoint and non-empty (since $T_{1}^{c}$, $T_{2}^{c}$ and $\tilde{\Theta}_{3}$ are all non-empty).

To complete the proof in Case 1, it remains to verify all type profiles are two-colored or uncolored.
By inspection, one may verify the following:
If $\theta_{3}\in \tilde{\Theta}_{3}$, then all type profiles in $T_{1}\times T_{2}\times\lbrace\theta_{3}\rbrace$ are uncolored, whereas all other type profiles in $\Theta_{1}\times\Theta_{2}\times\lbrace\theta_{3}\rbrace$ are two-colored.
Conversely, if $\theta_{3}\notin \tilde{\Theta}_{3}$, then all type profiles in $T^{c}_{1}\times T^{c}_{2}\times\lbrace\theta_{3}\rbrace$ are uncolored, whereas all other type profiles in $\Theta_{1}\times\Theta_{2}\times\lbrace\theta_{3}\rbrace$ are two-colored.

\textbf{Case 2.} \emph{Let $T_{2} = \emptyset$.}
Thus, all $\theta_{2}\in\Theta_{2}$ satisfy $(\cdot, \theta_{2}, \theta_{3}^{0})\in\tilde{V}$.
Define $\hat{\Theta}_{3}$ as the set of $\theta_{3}\in\Theta_{3}$ such that all $\theta_{2}\in\Theta_{2}$ satisfy $(\cdot, \theta_{2}, \theta_{3})\in\tilde{V}$.
Define $\hat{T}_{1}$ as the set of $\theta_{1}\in\Theta_{1}$ for which there exists $\theta_{3}\in \hat{\Theta}_{3}$ such that $(\theta_{1}, \cdot, \theta_{3})\notin\tilde{V}$.
Notice that $\hat{\Theta}_{3}$ is non-empty since $\theta_3^0\in \hat{\Theta}_{3}$; the set $\hat{T}_{1}$ may or may not be empty.
We observe the following:
\begin{itemize}
    \item \textbf{Observation 2.1.} \emph{If $(\theta_{1}, \theta_{2})\in \hat{T}_{1}\times\Theta_{2}$, then $(\theta_{1}, \theta_{2}, \cdot)\in\tilde{V}$.}
    
    Indeed, by the definition of $\hat{T}_{1}$, there is $\theta_{3}\in \hat{\Theta}_{3}$ such that  $(\theta_{1}, \cdot, \theta_{3})\notin \tilde{V}$. By definition of $\hat{\Theta}_{3}$, we have $(\cdot, \theta_{2}, \theta_{3})\in \tilde{V}$. Hence feasibility at $(\theta_{1}, \theta_{2}, \theta_{3})$ requires $(\theta_{1}, \theta_{2}, \cdot)\in\tilde{V}$.
    
    \item \textbf{Observation 2.2.} \emph{If $\theta_{3}\notin \hat{\Theta}_{3}$, then all $\theta_{1}\in\hat{T}_{1}$ satisfy $(\theta_{1}, \cdot, \theta_{3})\in\tilde{V}$.}
    
    Indeed, by definition of $\hat{\Theta}_{3}$, there exists $\theta_{2}\in\Theta_{2}$ such that $(\cdot, \theta_{2}, \theta_{3})\notin\tilde{V}$. Observation 2.1 implies $(\theta_{1}, \theta_{2}, \cdot)\in\tilde{V}$. Hence feasibility at $(\theta_{1}, \theta_{2}, \theta_{3})$ requires $(\theta_{1}, \cdot, \theta_{3})\in\tilde{V}$.
\end{itemize}
We next define our candidates for $R$ and $B$.
We first define the following:
\begin{align*}
    \forall \theta_{3}\in \hat{\Theta}_{3},\quad R^{\prime}(\theta_{3}) &= \lbrace (\cdot, \theta_{2}, \theta_{3})\colon \theta_{2}\in\Theta_{2}\rbrace;
    \\
    \forall \theta_{3}\notin \hat{\Theta}_{3},\quad R^{\prime}(\theta_{3}) &= \lbrace (\theta_{1}, \cdot, \theta_{3})\colon \theta_{1}\in\hat{T}_{1}\rbrace;
    \\
    R &= \cup_{\theta_{3}\in\Theta_{3}}R^{\prime}(\theta_{3});
    \\
    B &= \lbrace (\theta_{1}, \cdot, \theta_{3}) \colon \theta_{1}\notin \hat{T}_{1}, \theta_{3}\in \hat{\Theta}_{3}\rbrace \cup \lbrace (\theta_{1}, \theta_{2}, \cdot) \colon \theta_{1} \in\hat{T}_{1}, \theta_{2}\in \Theta_{2}\rbrace.
\end{align*}
Observations 2.1 and 2.2 and the definition of $\hat{\Theta}_{3}$ imply $R\cup \lbrace (\theta_{1}, \theta_{2}, \cdot) \colon \theta_{1}\in\hat{T}_{1}, \theta_{2}\in\Theta_{2}\rbrace\subseteq\tilde{V}$.
The definition of $\hat{T}_{1}$ implies $\lbrace (\theta_{1}, \cdot, \theta_{3}) \colon \theta_{1}\notin \hat{T}_{1}, \theta_{3}\in \hat{\Theta}_{3}\rbrace\subseteq\tilde{V}$.
The sets $R$ and $B$ are clearly disjoint, and they are non-empty since $\hat{\Theta}_{3}$ is non-empty.

To complete the proof in Case 2, it remains to show all type profiles are two-colored or uncolored.
By inspection of $R$ and $B$, one may verify the following:
If $\theta_{3}\in \hat{\Theta}_{3}$, then all type profiles in $\Theta_{1}\times\Theta_{2}\times\lbrace \theta_{3}\rbrace$ are two-colored; if $\theta_{3}\notin \hat{\Theta}_{3}$, then all type profiles in $\hat{T}_{1}\times\Theta_{2}\times\lbrace \theta_{3}\rbrace$ are two-colored whereas all type profiles in $\hat{T}_{1}^{c}\times\Theta_{2}\times\lbrace \theta_{3}\rbrace$ are uncolored. \qed

\subsection{Proofs for \headercref{Section}{{sec:DIC_characterization}}}
In this part of the appendix, we prove \Cref{thm:stochastic_existence,thm:stochastic_prevalence,prop:hole_characterization}. We shall use \Cref{prop:hole_characterization} in the proof of \Cref{thm:stochastic_prevalence}, and hence we present the proof of \Cref{thm:stochastic_prevalence} last.
\subsubsection{Auxiliary results}

As indicated in the main text, the existence of stochastic extreme DIC mechanisms is connected to odd holes in the feasibility graph.
The goal of the next few auxiliary results is to establish that all extreme DIC mechanisms are deterministic if and only if the feasibility graph does not have an odd hole of length $7$ or greater (\Cref{lemma:stochastic_hole_link}).

For $i\in [n]$, define $V_{i} = \lbrace i\rbrace\times\Theta_{-i}$ as the set of \emph{$i$-vertices}.
Given a vertex $v = (i, \theta_{-i})$ and $j$ distinct from $i$, we say $\theta_{j}$ is the \emph{type of $j$ at $v$}.
\begin{lemma}\label{lemma:type_flips_along_path}
    Let $v, v^{\prime}, v^{\prime\prime}$ be distinct vertices, where $v^{\prime}$ is an $i$-vertex.
    Let $v^{\prime}$ be adjacent to $v$ and $v^{\prime\prime}$.
    If $v$ and $v^{\prime\prime}$ are non-adjacent, then $i$'s type at $v$ differs from $i$'s type at $v^{\prime\prime}$.
\end{lemma}

\begin{proof}[Proof of \Cref{lemma:type_flips_along_path}]
    The type profile containing $v$ and $v^{\prime}$ coincides with the type profile containing $v^{\prime}$ and $v^{\prime\prime}$ in the types of all agents other than $i$. If these two profiles were to also agree in $i$'s type, then $v$ and $v^{\prime\prime}$ would either be adjacent or coincide.
\end{proof}

\begin{lemma}\label{lemma:hole_nonexistence}
    The feasibility graph does not admit an odd $5$-hole.
    The complement of the feasibility graph does not admit an odd hole.
\end{lemma}

\begin{proof}[Proof of \Cref{lemma:hole_nonexistence}]
    Towards a contradiction, let $G$ admit an odd $5$-hole $(v_{1}, \ldots, v_{5})$.
    Recall that two vertices are adjacent only if they belong to different agents.
    Since the hole contains five vertices, there is an agent such that the hole contains exactly one vertex of this agent.
    Without loss, let the vertex belonging to this agent be the vertex $v_{2}$.
    Since $(v_{3}, v_{4}, v_{5}, v_{1})$ is a path that contains no $i$-vertex, the type of agent $i$ is constant across $(v_{3}, v_{4}, v_{5}, v_{1})$.
    However, since $H$ is a hole, vertex $v_{2}$ is adjacent to $v_{1}$ and $v_{3}$ while $v_{1}$ and $v_{3}$ are non-adjacent.
    Thus \Cref{lemma:type_flips_along_path} implies that $i$'s type at $v_{1}$ differs from $i$'s type at $v_{3}$.
    Contradiction.    
    Thus $G$ does not admit an odd $5$-hole.

    Towards a contradiction, let the complement of $G$ admit an odd hole $(v_{1}, \ldots, v_{k})$ for some $k\geq 5$.
    If $k = 5$, then $(v_{1}, v_{3}, v_{5}, v_{2}, v_{4})$ is an odd $5$-hole in $G$; contradiction.
    Thus let $k \geq 7$.
    By definition of the complement, $\lbrace v_{1}, v_{4}, v_{6}\rbrace$ and $\lbrace v_{2}, v_{4}, v_{6}\rbrace$ are cliques in $G$.
    We know that there is a unique maximal clique of $G$ containing $v_{4}$ and $v_{6}$.
    Thus this clique also contains both $v_{1}$ and $v_{2}$.
    In particular, either $v_{1} = v_{2}$, or $v_{1}$ and $v_{2}$ are adjacent in $G$; in either case, we have a contradiction to the assumption that $(v_{1}, \ldots, v_{k})$ is an odd hole in the complement of $G$.
    Thus the complement of $G$ does not admit an odd hole.
\end{proof}

\begin{lemma}\label{lemma:stochastic_hole_link}
    All extreme DIC mechanisms are deterministic if and only if the feasibility graph does not admit an odd hole of length seven or more.
\end{lemma}

\begin{proof}[Proof of \Cref{lemma:stochastic_hole_link}] 
    Recall that the set of DIC mechanisms equals $QSTAB(G)$, while the convex hull of the set of deterministic DIC mechanisms equals $STAB(G)$.
    Theorem 3.1 of \citet{chvatal1975} implies that $STAB(G) = QSTAB(G)$ holds if and only if $G$ is perfect.
    According to the Strong Perfect Graph Theorem (\citet{chudnovsky2006}), the feasibility graph is perfect if and only if neither $G$ nor its complement admit an odd hole.
    The claim follows from \Cref{lemma:hole_nonexistence}.   
\end{proof}

\subsubsection{Proof of \headercref{Theorem}{{thm:stochastic_existence}}}
    Let $n = 2$. Recall that two vertices are adjacent only if they belong to distinct agents. It follows that $G$ is bipartite, and hence does not admit an odd hole. Thus \Cref{lemma:stochastic_hole_link} implies that all extreme DIC mechanisms are deterministic.

    Let all type spaces be binary. 
    We show that $G$ does not admit an odd hole, so that \Cref{lemma:stochastic_hole_link} implies that all extreme DIC mechanisms are deterministic.
    Without loss of generality, relabel the type spaces such that $\Theta_{i} = \lbrace 0 , 1\rbrace$ for all $i\in N$.
    Let $(v_{1}, \ldots, v_{k})$ be an induced cycle; that is, for all $\ell\in\lbrace 1, \ldots, k\rbrace$, vertex $v_{\ell}$ is adjacent to $v_{\ell-1}$ and $v_{\ell+1}$, and non-adjacent to all other vertices of the cycle (where $v_{k+1} = v_{1}$ is understood).
    We show $k$ is even.
    It suffices to show that for all agents $i$ the cycle contains an even number of $i$-vertices.
    We use two observations, valid for all $\ell\in\lbrace 1, \ldots, k\rbrace$.
    First, if none of $v_{\ell-1}$, $v_{\ell}$ and $v_{\ell+1}$ are $i$-vertices, then $i$'s type is constant across these vertices.
    Second, if $v_{\ell}$ is an $i$-vertex, then implies \Cref{lemma:type_flips_along_path} that the type of $i$ at $v_{\ell-1}$ differs from type of $i$ at $v_{\ell+1}$.
    Since agent $i$ has two possible types, the two observations together imply that the cycle contains an even number of $i$-vertices.

    Now let $n\geq 3$ and suppose at least one type space is non-binary.
    We show there exists an odd hole.
    By possibly relabeling the agents and the type spaces, suppose $\Theta_{1}$ and $\Theta_{2}$ contains $\lbrace 0, 1\rbrace$, and that $\Theta_{3}$ contains $\lbrace 0, 1, 2\rbrace$.
    Fix an arbitrary profile $\theta_{-123}$ of agents other than $1$, $2$ and $3$ (if such agents exists).
    The following seven vertices form an odd hole:
    \begin{align*}
        & v_{1} = (\cdot, 0, 0, \theta_{-123}), \hfill
        & v_{5} = (0, 1, \cdot, \theta_{-123}), \\
        & v_{2} = (1, \cdot, 0, \theta_{-123}), \hfill
        & v_{6} = (0, \cdot, 2, \theta_{-123}), \\
        & v_{3} = (1, 1, \cdot, \theta_{-123}), \hfill
        & v_{7} = (0, 0, \cdot, \theta_{-123}). \\
        & v_{4} = (\cdot, 1, 1, \theta_{-123}), \hfill &
        \tag*{\qed}
    \end{align*}

\subsubsection{Proof of \headercref{Theorem}{{prop:hole_characterization}}}
    First, let $q$ be extreme and let $K$ be a non-empty stochastic component of $q$.
    Let $G[K]$ denote the subgraph induced by $K$.
    
    Note that, if $v\in K$, $v^{\prime}\notin K$ and $q(v^{\prime}) > 0$, then $v$ and $v^{\prime}$ are non-adjacent.
    Indeed, if $v$ and $v^{\prime}$ are adjacent and $q(v^{\prime}) \in (0, 1)$, then $v^{\prime}\notin K$ contradicts the fact that $K$ is a component; if $v$ and $v^{\prime}$ are adjacent and $q(v^{\prime}) = 1$, then we have a contradiction to feasibility.
    
    It follows that the restriction of $q$ to $K$ is an extreme point of the fractional stable set polytope of $G[K]$.
    By the Strong Perfect Graph Theorem and Theorem 3.1 of \citet{chvatal1975}, the graph $G[K]$ or the complement of $G[K]$ admit an odd hole.
    \Cref{lemma:hole_nonexistence} implies the complement of $G[K]$ does not admit an odd hole.
    Thus $G[K]$ admits an odd hole.

    It remains to show that $q$ is extreme if $q$ only takes values in $\lbrace 0, \frac{1}{2}, 1\rbrace$ and every stochastic component of $q$ contains an odd hole.
    To that end, let $q^{\prime}$ be a DIC mechanism receiving non-zero weight in a convex combination that equals $q$.
    Clearly, $q$ and $q^{\prime}$ agree on the set of vertices to which $q$ assigns $0$ or $1$.
    Now consider a stochastic component $K$ of $q$.
    Thus $q(v) = \frac{1}{2}$ for all $v\in K$.
    We show $q^{\prime}(v) = \frac{1}{2}$ for all $v\in K$.
    We proceed in two steps.
    
    First, by assumption, the component $K$ contains an odd hole $(v_{1}, \ldots, v_{k})$ for some integer $k$.
    By construction, we have $q(v_{\ell}) + q(v_{\ell+1}) = 1$ for all $\ell \in \lbrace 1, \ldots, k\rbrace$ (where $v_{k+1} = v_{1}$).
    Since $v_{\ell}$ and $v_{\ell+1}$ are adjacent, all DIC mechanisms $q^{\prime\prime}$ from the convex combination satisfy $q^{\prime\prime}(v_{\ell}) + q^{\prime\prime}(v_{\ell+1}) \leq 1$.
    Thus $q^{\prime}(v_{\ell}) + q^{\prime}(v_{\ell+1}) = 1$ for all $\ell \in \lbrace 1, \ldots, k\rbrace$.
    Using that $k$ is odd, we infer $q^{\prime}(v_{1}) = 1 - q^{\prime}(v_{2}) = \ldots = q^{\prime}(v_{k}) = 1 - q^{\prime}(v_{1})$.
    Thus $q^{\prime}$ equals $\frac{1}{2}$ on the hole $(v_{1}, \ldots, v_{k})$.
    
    Second, since $K$ is a component and since there is an odd hole in $K$ where $q$ and $q^{\prime}$ agree, it suffices to show the following: if $v$ and $v^{\prime}$ in $K$ are adjacent and $q^{\prime}(v) = \frac{1}{2}$, then $q^{\prime}(v^{\prime}) = \frac{1}{2}$.
    To that end, note $q(v) = q(v^{\prime}) = \frac{1}{2}$.
    Clearly, all DIC mechanisms $q^{\prime\prime}$ from the convex combination satisfy $q^{\prime\prime}(v) + q^{\prime\prime}(v^{\prime}) \leq 1$.
    Since $q(v) + q(v^{\prime}) = 1$, we infer $q^{\prime}(v) + q^{\prime}(v^{\prime}) = 1$.
    Since $q^{\prime}(v) = \frac{1}{2}$ by assumption, we conclude $q^{\prime}(v^{\prime}) = \frac{1}{2}$. 
    \qed

\subsubsection{Proof of \headercref{Theorem}{{thm:stochastic_prevalence}}}
Let $\mathcal{S}$ denote the set of stable sets of the feasibility graph.
Recall that there is a bijection between stable sets and deterministic DIC mechanisms.
Thus $\vert\mathcal{S}\vert = \vert \deterministic Q\vert$.

Fix $n\geq 3$.
Let $m = \min_{i\in [n]} \vert\Theta_{i}\vert$.
By possibly relabelling the type spaces, let each type space contain $\lbrace 1, \ldots, m\rbrace$.
To prove \Cref{thm:stochastic_prevalence}, it suffices to show that if $m\geq 6$, then
\begin{equation*}
    \vert\extremepoints Q\vert \geq \vert \mathcal{S}\vert \left(1 + \frac{m-2}{3(n+1)^{9}}\right).
\end{equation*}
Let $m^{\prime}$ denote the largest integer multiple of $3$ that is less than $m$.
Note $m^{\prime} \geq m - 2$.

We begin by constructing a family of odd holes.
Fix an arbitrary type profile $\theta_{-123} = (\theta_{4}, \ldots, \theta_{n})$ of agents other than $1$, $2$ and $3$ (if such agents exist).
For $k\in\lbrace 3, 6, \ldots, m^{\prime}\rbrace$, let $H(k)$ denote the set consisting of the following nine vertices:
\begin{align*}
\begin{alignedat}{4}
v_{1} &= (\cdot && ,\ k-2 && ,\ k-2 && ,\ \theta_{-123}), \\
v_{2} &= (k-1 && ,\ \cdot && ,\ k-2 && ,\ \theta_{-123}), \\
v_{3} &= (k-1 && ,\ k-1 && ,\ \cdot && ,\ \theta_{-123}), \\
v_{4} &= (\cdot && ,\ k-1 && ,\ k-1 && ,\ \theta_{-123}), \\
v_{5} &= (k && ,\ \cdot && ,\ k-1 && ,\ \theta_{-123}), \\
\end{alignedat}
\qquad
\begin{alignedat}{4}
v_{6} &= (k && ,\ k && ,\ \cdot && ,\ \theta_{-123}), \\
v_{7} &= (\cdot && ,\ k && ,\ k && ,\ \theta_{-123}), \\
v_{8} &= (k-2 && ,\ \cdot && ,\ k && ,\ \theta_{-123}), \\
v_{9} &= (k-2 && ,\ k-2 && ,\ \cdot && ,\ \theta_{-123}). \\
\end{alignedat}
\end{align*}
By inspection, $H(k)$ is an odd hole.
Let $\mathcal{H} = \lbrace H(k)\colon k\in\lbrace 3, 6, \ldots, m^{\prime}\rbrace\rbrace$.
Note $\vert\mathcal{H}\vert = \frac{m^{\prime}}{3}$.

We use some further auxiliary definitions.
Recall that, for the feasibility graph, for each pair of adjacent vertices there is a unique maximal clique that contains both of them.
In particular, for all $H\in\mathcal{H}$, for each of the nine pairs of adjacent vertices in $H$, there is a unique maximal clique containing the pair.
Let $\mathcal{X}_{H}$ denote the nine maximal cliques obtained in this way, and let $V_{H} = \cup_{X\in\mathcal{X}_{H}} X$ denote the vertices contained in these cliques (including vertices that are not themselves in $H$).
Let $N(H)$ denote the neighborhood of $H$; that is, the vertices adjacent to at least one vertex in $H$.
Note $V_{H} \subseteq N(H)$.
Further, let $L_{H}$ denote the number of stable sets in the subgraph induced by $V_{H}$. 

We next construct a family of stochastic extreme points.
Given $H\in\mathcal{H}$ and $S\in\mathcal{S}$, define $q_{H, S}\colon V\to [0, 1]$ as follows:
for all $v\in  H \cup (N(H)\cap (S\setminus V_{H}))$, let $q_{H, S}(v) = 1/2$; for all $v\in  S\setminus N(H)$, let $q_{H, S}(v) = 1$; else, let $q_{H, S}(v) = 0$. 
We later verify that $q_{H, S}$ is a stochastic extreme DIC mechanism for all $H\in\mathcal{H}$ and $S\in\mathcal{S}$.

We next provide a number of estimates for the number $\vert\lbrace q_{H, S}\colon H\in\mathcal{H}, S\in\mathcal{S}\rbrace\vert$ of stochastic extreme points just constructed.

Fixing an arbitrary $H\in \mathcal{H}$, let $Q_{H} = \lbrace q_{H, S}\colon S\in\mathcal{S}\rbrace$.
Given $q \in Q_{H}$, let
$q^{-1}_{H}(q) = \lbrace S\in\mathcal{S}\colon q_{H, S} = q\rbrace$.
We claim that $\vert \mathcal{S}\vert \leq \vert Q_{H}\vert L_{H}$ holds, where we recall $L_{H}$ that denotes the number of stable sets in the subgraph induced by $V_{H}$.
To that end, we make two observations.
First, inspecting the definition of $q_{H, S}$, if $S$ and $S^{\prime}$ are two stable sets such that $S \setminus V_{H} = S^{\prime} \setminus V_{H}$, then $q_{H, S} = q_{H, S^{\prime}}$.
Second, since $S$ is stable, the set $S \cap V_{H}$ is itself a stable set on the subgraph induced by $V_{H}$.
The two observations imply that $\vert q^{-1}_{H}(q)\vert \leq L_{H}$ holds for all $q\in Q_{H}$.
Finally, we note that $\lbrace q^{-1}_{H}(q)\colon q \in Q_{H}\rbrace$ partitions $\mathcal{S}$. 
Hence $\vert \mathcal{S}\vert \leq \vert Q_{H}\vert L_{H}$.

We next claim that all $H\in\mathcal{H}$ satisfy $L_{H}\leq (n+1)^{9}$.
Recall the definition $V_{H} = \cup_{X\in \mathcal{X}_{H}} X$, where $\mathcal{X}_{H}$ is a set of nine maximal cliques.
Each maximal clique of the feasibility graph contains $n$ vertices, and a stable set selects from each maximal clique at most one vertex (or selects nothing).
Hence $L_{H}\leq (n+1)^{9}$. 
    
Lastly, we claim that if $H$ and $H^{\prime}$ in $\mathcal{H}$ are distinct, then $Q_{H}\cap Q_{H^{\prime}} = \emptyset$.
Equivalently, for arbitrary $S, S^{\prime}\in\mathcal{S}$ and $k, k^{\prime}\in\lbrace 3, \ldots, m^{\prime}\rbrace$, we have $q_{H(k), S} = q_{H(k^{\prime}), S^{\prime}}$ only if $k = k^{\prime}$.
To that end, let $S, S^{\prime}\in\mathcal{S}$ and $k, k^{\prime}\in\lbrace 3, \ldots, m^{\prime}\rbrace$ and $q_{H(k), S} = q_{H(k^{\prime}), S^{\prime}}$. 
Fix two adjacent vertices $v$ and $\hat{v}$ in $H$.
Thus $q_{H(k), S}(v) = q_{H(k), S}(\hat{v}) = q_{H(k^{\prime}), S^{\prime}}(v) = q_{H(k^{\prime}), S^{\prime}}(\hat{v}) = \frac{1}{2}$. 
Note at least one of $v$ and $\hat{v}$ must be in $H(k^{\prime})$; indeed, else $q_{H(k^{\prime}), S^{\prime}}(v) = q_{H(k^{\prime}), S^{\prime}}(\hat{v}) = \frac{1}{2}$ requires that both be in $S^{\prime}$, contradicting stability of $S^{\prime}$.
Without loss, let $v\in H(k^{\prime})$.
In particular, $v\in H(k) \cap H(k^{\prime})$.
Suppose $v$ is an $i$-vertex, take $j \in \lbrace 1, 2, 3\rbrace\setminus \lbrace i\rbrace$ and consider $j$'s type at $v$.\footnote{Recall that for $i\in [n]$ we call $V_{i} = \lbrace i\rbrace\times\Theta_{-i}$ the set of $i$-vertices, and $j$'s type at $(i, \theta_{-i})$ means $\theta_{j}$. For example, if $v = (\cdot, k-2, k-2, \theta_{-123})$, then $v$ is a $1$-vertex and the types of agents $2$ and $3$ are both $k-2$ at $v$.}
Since $v\in H(k) \cap H(k^{\prime})$, agent $j$'s type is in both $\lbrace k, k+1, k+2\rbrace$ and $\lbrace k^{\prime}, k^{\prime}+1, k^{\prime}+2\rbrace$.
Since $k$ and $k^{\prime}$ are integer multiples of $3$, we conclude $k = k^{\prime}$.

Collecting our work, we get
\begin{align*}
\vert\lbrace q_{H, S}\colon H\in\mathcal{H}, S\in\mathcal{S}\rbrace\vert
= \sum_{H\in\mathcal{H}} \vert Q_{H}\vert 
\geq \sum_{H\in\mathcal{H}} \frac{\vert \mathcal{S}\vert}{L_{H}} 
\geq \sum_{H\in\mathcal{H}} \frac{\vert \mathcal{S}\vert}{(n+1)^{9}} 
&= \frac{m^{\prime}}{3} \frac{\vert \mathcal{S}\vert}{(n+1)^{9}} 
\\
&\geq \frac{m-2}{3} \frac{\vert \mathcal{S}\vert}{(n+1)^{9}} 
.
\end{align*}
Finally, using that for all $H\in\mathcal{H}$ and $S\in\mathcal{S}$ the DIC mechanism $q_{H, S}$ is stochastic and extreme, we infer the promised inequality
\begin{equation*}
    \vert\extremepoints Q\vert \geq \vert\mathcal{S}\vert + \vert\lbrace q_{H, S}\colon H\in\mathcal{H}, S\in\mathcal{S}\rbrace\vert \geq \vert \mathcal{S}\vert \left(1 + \frac{m-2}{3(n+1)^{9}}\right)
    .
\end{equation*}

Finally, we show that $q_{H, S}$ is an extreme DIC mechanism for all $H\in\mathcal{H}$ and $S\in\mathcal{S}$.

We first show $q_{H, S}$ is a well-defined DIC mechanism.
To show $q_{H, S}$ is well-defined, we have to argue that all maximal cliques $X$ satisfy $q(X) \leq 1$.
First, suppose $X$ contains a vertex $v$ in $S\setminus N(H)$.
It follows that $X$ contains no vertex in $H$ (else $v$ would be in $N(H)$) nor a vertex in $(N(H)\cap (S\setminus V_{H}))$ (by stability of $S$). 
Hence, in this case, $q_{H, S}$ assigns $1$ to $v$ and $0$ to all other vertices in $X$.
Next, suppose $X$ only contains vertices in $H \cup (N(H)\cap (S\setminus V_{H}))$. 
Now $X$ can contain at most two vertices from $H$ (since $H$ is an odd hole) and at most one vertex in $N(H)\cap (S\setminus V_{H})$ (by stability of $S$).
Further, if $X$ contains a vertex in $N(H)\cap (S\setminus V_{H})$, then it contains at most one vertex in $H$ (by definition of $V_{H}$).
Since all vertices in $H \cup (N(H)\cap (S\setminus V_{H}))$ are assigned $\frac{1}{2}$, we infer $q(X) \leq 1$.
Lastly, if $X$ contains no vertices in $(S\setminus N(H)) \cup (H \cup (N(H)\cap (S\setminus V_{H})))$, then clearly $q(X) = 0$ holds.

\Cref{prop:hole_characterization} implies $q_{H, S}$ is extreme since $q_{H, S}$ maps to $\lbrace 0, \frac{1}{2}, 1\rbrace$, and since $H \cup (N(H)\cap (S\setminus V_{H}))$ is the unique stochastic component of $q_{H, S}$ and contains the odd hole $H$. \qed

\subsection{Proof of \headercref{Theorem}{{thm:npcomplete}}}
To show that \textsc{OptDet}-$n$ is NP-hard if $n\geq 3$, it suffices to show that the following decision problem, \textsc{Det}-$3$, is NP-complete.
\begin{definition}[\textsc{Det}-$n$]
    The input consists of an integer $k$, finite sets $\Theta_{1}, \ldots, \Theta_{n}$ and weights $w_i: \Theta_{-i}\to \lbrace 0, 1\rbrace$ for all $i\in [n]$.
    The problem is to decide whether there is a deterministic DIC mechanism $q$ (for $n$ agents with respective type spaces $\Theta_{1}, \ldots, \Theta_{n}$) such that $\sum_{i, \theta} w_{i}(\theta_{-i}) q_{i}(\theta_{-i}) \geq k$.
\end{definition}
Recall that \textsc{StableSet} refers to the following decision problem.
The input is an integer $\hat{k}$ and a (simple undirected) graph $\hat{G}$.
The problem is to determine whether $\hat{G}$ admits a stable set of cardinality $\hat{k}$ or greater.
\textsc{StableSet} is NP-complete (\citet[Theorem 15.23]{korte2018combinatorial}).
To show that \textsc{Det}-3 is NP-complete, we show that \textsc{StableSet} polynomially reduces to \textsc{Det}-3. 

Let $(\hat{G}, \hat{k})$ be an instance of \textsc{StableSet}.
We next construct our candidate instance of \textsc{Det}-3.
Let $\hat{V}$ and $\hat{E}$ denote the vertex and edge sets of $\hat{G}$.
Fix an arbitrary linear order $<$ on $\hat{V}$ (say, by enumeration).
Define $\Theta_{1} = \hat{E}$ and $\Theta_{2} = \Theta_{3} = \hat{V}$.
The resulting feasibility graph is denoted $G$, with vertices $V$ and edges $E$.
Next, for all edges $e$ of $\hat{G}$, define a set $P(e)$ of vertices in $G$ as follows: denote the edge $e$ as $e = v v^{\prime}$ such that $v < v^{\prime}$, and define
\begin{equation*}
    P(e)=
    \left(
    (\cdot,v,v),
    (e,v,\cdot),
    (e,\cdot,v^{\prime}),
    (\cdot,v^{\prime},v^{\prime})
    \right).
\end{equation*}
Notice that $P(e)$ is an induced path in $G$ from $(\cdot,v,v)$ to $(\cdot,v^{\prime},v^{\prime})$.
We refer to $(\cdot,v,v)$ and $(\cdot,v^{\prime},v^{\prime})$ as the \emph{endpoints} of $P(e)$; the vertices $(e,v,\cdot)$ and $(e,\cdot,v^{\prime})$ are the \emph{interior} vertices of $P(e)$.
(Likewise, a vertex is an \emph{endpoint} if for some $e$ it is an endpoint of $P(e)$; a vertex is \emph{interior} if for some $e$ it is an interior vertex of $P(e)$.)
Next, define $w\colon V\to\lbrace 0, 1\rbrace$ as follows.
For all $v\in V$, let $w(v) = 1$ if there is an edge $e$ in $\hat{E}$ such that $v\in P(e)$; else, let $w(v) = 0$.
Finally, let $k = \hat{k} + \vert\hat{E}\vert$.

To complete the proof, we argue that $\hat{G}$ admits a stable set of cardinality $\hat{k}$ or greater if and only if $G$ admits a weighted stable set of weight $k$ or greater.

Let $\hat{S}$ be a stable set in $\hat{G}$ of cardinality $\hat{k}$ or greater.
Construct a stable set $S$ in $G$ as follows.
Denote $T = \lbrace (\cdot, v, v)\colon v\in \hat{S}\rbrace$, and include $T$ in $S$.
Next, for all edges $e\in\hat{E}$, find an interior vertex $\omega$ of $P(e)$ that is non-adjacent to all vertices in $T$, and include $\omega$ in $\hat{S}$; such a vertex $\omega$ exists since $\hat{S}$ is stable, meaning that the two endpoints of $P(e)$ are not both included in $T$.
Clearly, $S$ has weight of at least $k=\hat{k} + \vert \hat{E}\vert$.
Moreover, $S$ is stable since $T$ is stable, since the interior vertices are chosen to be non-adjacent from $T$, and since interior vertices of distinct paths are non-adjacent.

Now suppose $G$ admits a stable set $S$ of weight $\hat{k} + \vert \hat{E}\vert$ or greater.

In an auxiliary step, we argue that $G$ admits a stable set having weight $\hat{k} + \vert \hat{E}\vert$ or greater and such that for all $e\in\hat{E}$ the stable set contains at most one endpoint of $P(e)$.
Indeed, suppose for some $e\in\hat{E}$ both endpoints of $P(e)$ are in a stable set of $G$.
Note that none of the interior vertices of $P(e)$ are in the stable set, by stability.
Now consider the following adjustment to the stable set: choose one of the endpoints of $P(e)$, find the adjacent interior vertex in $P(e)$, and in the stable set replace the chosen endpoint by the interior vertex.
The resulting set is stable since in $G$ the chosen interior vertex is only adjacent to the other interior vertex of $P(e)$ (which we already noted is not in the stable set) and to the removed endpoint of $P(e)$.
Further, the adjustment leaves the total weight unchanged.
By repeating this adjustment a finite number of times, we obtain a stable set with the claimed properties.

Thus suppose $G$ admits a stable set with weight $\hat{k} + \vert \hat{E}\vert$ or greater and such that for all $e\in\hat{E}$ the stable set $S$ contains at most one endpoint of $P(e)$. 
Note that $S$ contains at most $\vert\hat{E}\vert$ interior vertices.
Thus $S$ contains at least $\hat{k}$ endpoint vertices.
Now define $\hat{S} = \lbrace v\in\hat{V}\colon (\cdot, v, v)\in S\rbrace$.
The set $\hat{G}$ is stable in $\hat{G}$ since, for all $e\in E$, at most one endpoint of $P(e)$ is in $S$.
Moreover, since $S$ contains at least $\hat{k}$ endpoint vertices, we conclude $\hat{S}$ has cardinality of at least $\hat{k}$. \qed

\subsection{Proof of \headercref{Theorem}{{thm:ranking_based}}}

    Fix $n\in\mathbb{N}$ and consider the environment $(n, \Theta^{n}, \mu^{n})$.
    For all $i\in [n]$, let $\bar{u}_{i}^{n}$ denote $i$'s peer value in this environment. Let $r_{i}^{n}$ and $r_{i}^{n, \ast}$, respectively, denote $i$'s rank and robust rank, respectively.
    For $p\in (0, 1)$, let $q^{n,p}$ denote the ranking-based mechanism with threshold $p$. Let $U^{n}$ denote the principal's expected utility (as a function on the set of DIC mechanisms).
    
    We show that for all $\varepsilon > 0$ there exists $p\in (0, 1)$ such that 
    \begin{equation*}
        \left\vert U^{n}(q^{n,p}) - \sum_{\theta\in\Theta^{n}} \mu^{n}(\theta) \max\left(0, \max_{i\in [n]}\bar{u}_{i}^{n}(\theta_{-i})\right) \right\vert < \varepsilon    
    \end{equation*}
    for all but finitely many $n$.
    This claim proves the theorem since, for every environment, no DIC mechanism yields a higher expected utility than the utility from always allocating to an agent with the highest positive peer value; indeed, recall from \eqref{eq:principal_utility} that every $n$ and every DIC mechanism $q$ in the $n$'th environment, we have 
    \begin{equation*}
    U^{n}(q) = \sum_{\theta\in\Theta^{n}} \sum_{i\in[n]}\mu^{n}(\theta) q_{i}(\theta_{-i}) \bar{u}^{n}_{i}(\theta_{-i})\leq \sum_{\theta\in\Theta^{n}} \sum_{i\in[n]}\mu^{n}(\theta) \max\left(0, \max_{i\in [n]} \bar{u}^{n}_{i}(\theta_{-i})\right).
    \end{equation*}

    In the $n$'th environment, given a type profile $\theta\in\Theta^{n}$ and $p\in (0, 1)$, let $\bar{u}^{n}(p, \theta)$ denote the smallest peer value among the agents $i$ whose rank $r_{i}(\theta)$ is below $p$; that is, $\bar{u}^{n}(p, \theta) = \min\lbrace \bar{u}_{i}^{n}(\theta_{-i}) \colon i\in [n] \land r_{i}(\theta) \leq p\rbrace$. 
    At every type profile $\theta$, every agent who is allocated the good by $q^{n, p}$ has a peer value of at least $\bar{u}^{n}(p, \theta)$.
    
    For all $n\in\mathbb{N}$ and $p\in (0, 1)$ and $\theta\in\Theta^{n}$, we have the following lower bound on the principal's payoff at $\theta$ from $q^{p, n}$:
    \begin{align*}
        &U^{n}(q^{p, n}) \\
        = 
        &
        \sum_{\theta\in\Theta^{n}}\sum_{i\in [n]} \mu^{n}(\theta) \max(0, \bar{u}_{i}^{n}(\theta_{-i})) \frac{\bm{1}_{(r_{i}^{n, \ast}(\theta_{-i}) \leq p)}}{n p}
        \\
        \geq 
        &
        \sum_{\theta\in\Theta^{n}}\mu^{n}(\theta) \max\left(0, \bar{u}^{n}(p, \theta)\right) \sum_{i\in [n]}  \frac{1}{n p} \bm{1}_{(r_{i}^{n, \ast}(\theta_{-i}) \leq p)}
        \\
        \geq
        &
        \sum_{\theta\in\Theta^{n}}\mu^{n}(\theta)\max\left(0, \bar{u}^{n}(p, \theta)\right) \sum_{i\in [n]} \frac{1}{n p} \bm{1}_{(r_{i}^{n}(\theta) \leq p - \delta^{n}(\theta))}        
        \\
        \geq
        &
        \sum_{\theta\in\Theta^{n}}\mu^{n}(\theta) \max\left(0, \bar{u}^{n}(p, \theta)\right)
        -
        1 + \sum_{\theta\in\Theta^{n}}\mu^{n}(\theta) \sum_{i\in [n]}\frac{1}{n p} \bm{1}_{( r_{i}^{n}(\theta) \leq p - \delta^{n}(\theta))}
        ,
    \end{align*}
    where the last line uses that all peer values are in $[-1, 1]$. 

    Since no two agents have the same rank (the rank entails a tie-breaking rule), we have
    \begin{equation*}
        \sum_{\theta\in\Theta^{n}}\mu^{n}(\theta) \sum_{i\in [n]} \frac{1}{n p} \bm{1}_{( r_{i}^{n}(\theta) \leq p - \delta^{n}(\theta))}
        =
        \sum_{\theta\in\Theta^{n}}\mu^{n}(\theta) \frac{\lfloor n (p - \delta^{n}(\theta))\rfloor}{np},
    \end{equation*}
    where $\lfloor n (p - \delta^{n}(\theta))\rfloor$ means the largest integer weakly less than $n (p - \delta^{n}(\theta))$.
    Since $(\delta^{n})_{n\in\mathbb{N}}$ converges to $0$ in probability, for fixed $p\in (0, 1)$ and as $n\to\infty$ we have
    \begin{equation*}
        \sum_{\theta\in\Theta^{n}}\mu^{n}(\theta) \frac{\lfloor n (p - \delta^{n}(\theta))\rfloor}{np} \to 1.
    \end{equation*}

    Returning to the lower bound for $U^{n}(q^{p, n})$, we may complete the proof by showing that for all $\varepsilon > 0$ there exists $p \in (0, 1)$ such that
    \begin{equation}\label{eq:thm:ranking_based:close_quantiles}
        \lim_{n\to\infty}
        \sum_{\theta\in\Theta^{n}}\mu^{n}(\theta) \left(\max\left(0, \max_{i\in [n]} \bar{u}_{i}^{n}(\theta_{-i})\right) - \max\left(0, \bar{u}^{n}(p, \theta)\right)\right)
        < \varepsilon
        .
    \end{equation}
    Let $\varepsilon > 0$.
    Regularity implies that there exists $p \in (0, 1)$ such that
    \begin{equation*}
        \lim_{n\to\infty}\mu^{n}\left\lbrace\theta\in\Theta^{n}\colon 
		\left\vert\left\lbrace j\in [n]\colon
		\bar{u}_{j}^{n}(\theta_{-j}) + \varepsilon\geq \max_{i\in [n]}\bar{u}_{i}^{n}(\theta_{-i})\right\rbrace\right\vert
		\geq pn 
		\right\rbrace = 1.
    \end{equation*}
    Given a profile $\theta$, recall that $\bar{u}^{n}(p, \theta)$ is the smallest peer value among the agents $i$ whose rank is below $p$.
    If $\theta$ is a type profile such that at least $n p$ agents have a peer value within $\varepsilon$ of the highest peer value $\max_{i\in [n]}\bar{u}_{i}^{n}(\theta_{-i})$, then $\bar{u}^{n}(p, \theta) \geq \max_{i\in [n]}\bar{u}_{i}^{n}(\theta_{-i}) - \varepsilon$.
    Hence
    \begin{equation*}
        \lim_{n\to\infty} \mu^{n}\left\lbrace \theta\in\Theta^{n} \colon \bar{u}^{n}(p, \theta) \geq \max_{i\in [n]}\bar{u}_{i}^{n}(\theta_{-i}) - \varepsilon\right\rbrace = 1. 
    \end{equation*}
        Since all peer values are in $[-1, 1]$, for all $\varepsilon > 0$ there is $p \in (0, 1)$ satisfying \eqref{eq:thm:ranking_based:close_quantiles}. \qed

\addcontentsline{toc}{section}{References}

\newrefcontext[sorting=nyt]
\printbibliography
\pagebreak

\section{Online Appendix}\label{OA}

This appendix is organized as follows.
\Cref{appendix:examples} presents examples of the feasibility graph and stochastic extreme DIC mechanisms.
\Cref{appendix:jury_mechanisms} discusses the (failure of) approximate optimality of jury mechanisms.
\Cref{appendix:mandatory_allocation} considers the principal's problem with mandatory allocation.
\Cref{appendix:multiunit_allocation} considers the principal's problem with multiple units.
\Cref{appendix:symmetry_nothing_goes} shows that the principal's problem is uninteresting under classical symmetry assumptions.

\subsection{Examples of the Feasibility Graph and Stochastic Extreme DIC Mechanisms}\label{appendix:examples}
\subsubsection{The Feasibility Graph}\label{appendix:examples:feasibility_graph}
\begin{figure}[ht]
	\centering
	\includegraphics[width=\textwidth,page=3]{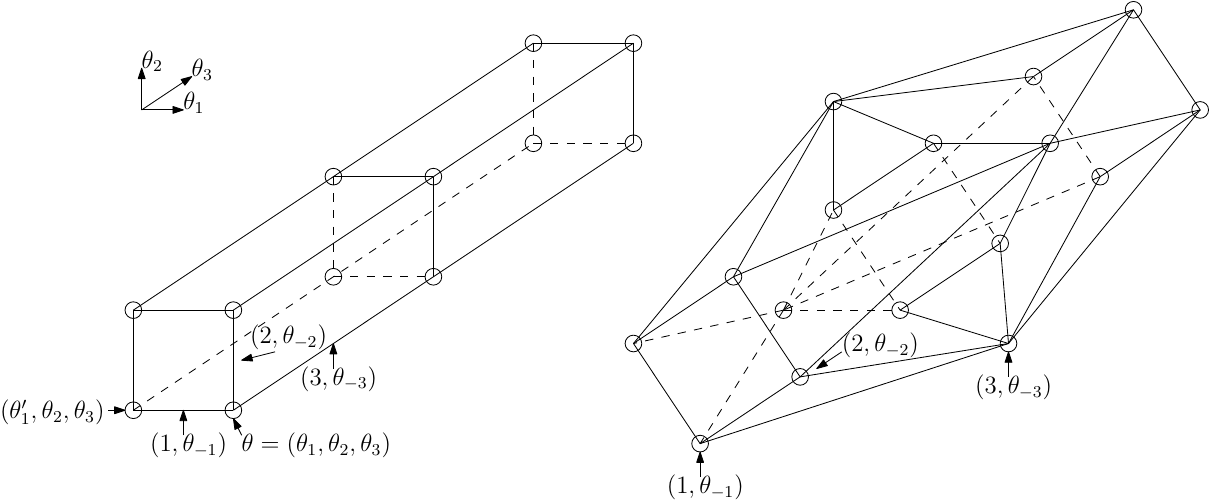}
	\caption{The set of type profiles (left) and the feasibility graph (right)  in an example with $n=3$, $\vert\Theta_{1}\vert = \vert\Theta_{2}\vert= 2$ and $\vert\Theta_{3}\vert = 3$.}
	\label{fig:feasibility_graph_example}
\end{figure}

\Cref{fig:feasibility_graph_example} shows the feasibility graph in an example with three agents, where agents $1$ and $2$ each have two possible types, and agent $3$ has three possible types.
The left panel shows the set of type profiles, here depicted as a $2\times 2\times 3$ grid.
Agent $1$'s type varies horizontally, agent $2$'s type vertically, and agent $3$'s type diagonally.
The right panel shows the feasibility graph.
To understand the connection between the panels, think of a vertex $(i, \theta_{-i})$ of the feasibility graph as a set of type profiles along which the types of agents other than $i$ are fixed at $\theta_{-i}$ while $i$'s type varies across all of $\Theta_{i}$.
In the left panel, for example, the vertex $(1, \theta_{-1})$ is thus depicted as the line connecting the profiles $\theta = (\theta_{1}, \theta_{2}, \theta_{3})$ and $(\theta_{1}^{\prime}, \theta_{2}, \theta_{3})$.
Two vertices are adjacent if and only if the corresponding sets of type profiles intersect; for example, the vertices $(1, \theta_{-1})$, $(2, \theta_{-2})$ and $(3, \theta_{-3})$ are all adjacent since the corresponding sets of type profiles intersect at $\theta$.
In the right panel, vertices are depicted as circles (instead of lines) and lines (both solid and dashed ones) indicate adjacencies.

\subsubsection{A Stochastic Extreme DIC Mechanism}\label{appendix:examples:stochastic_extreme_point}

\begin{figure}[ht]
    \centering
    \includegraphics[width=\textwidth,page=5]{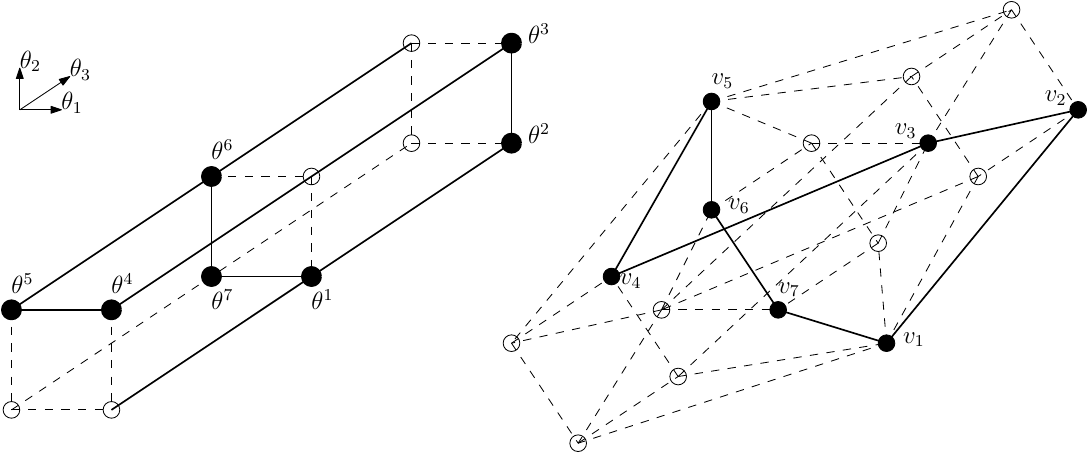}
    \caption{The right panel shows an odd hole $v_{1}, \ldots, v_{7}$ in the feasibility graph. The left panel shows the seven type profiles $\theta^{1}, \ldots, \theta^{7}$ that each contain two vertices of the hole.}
    \label{fig:stochastic_example}
\end{figure}

Continuing with the example from \Cref{appendix:examples:feasibility_graph}, suppose there are three agents, where agents $1$ and $2$ each have two possible types, and agent $3$ has three possible types.
The left panel of \Cref{fig:stochastic_example} shows the space of type profiles.
Agent $1$'s type varies horizontally, agent $2$'s type vertically, and agent $3$'s type diagonally.

The right panel of \Cref{fig:stochastic_example} shows an odd hole $v_{1}, \ldots, v_{7}$ in the feasibility graph.
Let us say that a profile $\theta$ \emph{contains} the vertices $\lbrace (i, \theta_{-i})\colon i\in [n]\rbrace$.
For each $\ell \in \lbrace 1, \ldots, 7\rbrace$, the type profile $\theta^{\ell}$ indicated in the left panel of the figure contains the vertices $v_{\ell-1}$ and $v_{\ell}$ (where $v_{0} = v_{7}$ is understood).
For example, type profile $\theta^{1}$ contains $v_{7}$ and $v_{1}$, and $\theta^{2}$ contains $v_{1}$ and $v_{2}$, and so on.

Our candidate stochastic extreme DIC mechanism $q$ assigns $\frac{1}{2}$ to all vertices $v_{1}, \ldots, v_{7}$ of the odd hole, and $0$ to all other vertices.
Translated to the left panel, this means that, for each $\ell\in \lbrace 1, \ldots, 7\rbrace$, at $\theta^{\ell}$ the mechanism flips a coin between the two agents who have vertices from the hole which are contained in $\theta^{\ell}$. For example, at $\theta^{1}$ the mechanism flips a coin between agents $1$ and $3$, at $\theta^{2}$ between agents $3$ and $2$, and so on.

In the main text, we intuited that randomization helps resolve the trade-off between allocating to an agent and using the agent's information since the agent can simultaneously win with some probability but nevertheless influence how the remaining probability is distributed among the other agents.
We see this in the stochastic mechanism $q$.
For example, agent $3$ wins with probability $1/2$ at both $\theta^{1}$ and $\theta^{2}$. These two type profiles differ only in agent $3$'s type. Depending on agent $3$'s report, the remaining probability $1/2$ is assigned to agent $1$ (at $\theta^{1}$) or to agent $2$ (at $\theta^{2}$). 
In this sense, the principal enjoys both the benefit from allocating to agent $3$ and agent $3$'s information.

To sharpen the intuition, suppose the weights $w_{i}(\theta_{-i}) = \mu(\theta_{-i})u_{i}(\theta_{-i})$ are such that $w_{i}(\theta_{-i})$ equals $1$ if the vertex $(i, \theta_{-i})$ is in the odd hole $\lbrace v_{1}, \ldots, v_{7}\rbrace$, and equals $-1$ otherwise.
In particular, at $\theta^{1}$ the principal wishes to select agent $1$ or $3$, while at $\theta^{2}$ the principal wishes to select agent $2$ or $3$.
Thus, here it is valuable to allocate to agent $3$, and agent $3$'s type also contains valuable information.

Let us show that for these weights the stochastic DIC mechanism $q$ described earlier is in fact uniquely optimal (and therefore extreme).
The principal's expected utility from $q$ is given by $\sum_{v} w(v) q(v) = 7/2$.
Now let $q^{\prime}$ be an optimal DIC mechanism.
We show $q = q^{\prime}$.
Since $q^{\prime}$ is optimal, it holds $\sum_{v} w(v) q^{\prime}(v) \geq \sum_{v} w(v) q(v) = 7/2$. 
For vertices outside $\lbrace v_{1}, \ldots, v_{7}\rbrace$, the weight $w$ is strictly negative.
Since decreasing the allocation probability at a vertex only slackens feasibility, we must have $q^{\prime}(v) = 0$ for all $v$ outside $\lbrace v_{1}, \ldots, v_{7}\rbrace$.
Thus $\sum_{v} w(v) q^{\prime}(v) = \sum_{\ell=1}^{7} q^{\prime}(v_{\ell})$ and  it suffices to show $q^{\prime}(v_{\ell}) = 1/2$ for all $\ell\in [7]$.
For all $\ell\in [7]$, feasibility requires $q^{\prime}(v_{\ell-1}) + q^{\prime}(v_{\ell}) \leq 1$ since $v_{\ell}$ and $v_{\ell-1}$ are adjacent (where again $v_{0} = v_{7}$).
Therefore, for all $\ell\in [7]$, the inequality $7/2 \leq q^{\prime}(v_{1}) + \ldots + q^{\prime}(v_{7})$ requires $7/2 \leq q^{\prime}(v_{\ell}) + 3$, and thus $1/2 \leq q^{\prime}(v_{\ell})$.
Using again $q^{\prime}(v_{\ell-1}) + q^{\prime}(v_{\ell}) \leq 1$, we conclude $q(v_{\ell}) = 1/2$ for all $\ell\in [7]$, as desired.

\subsection{Approximate Optimality of Jury Mechanisms}\label{appendix:jury_mechanisms}

Here, we discuss when jury mechanisms are approximately optimal with many agents.
We recall the definition:
\begin{definition}[Jury mechanisms]
    A mechanism $q$ is a \emph{jury mechanism} if there is a partition of the set of agents into \emph{jurors} $J$ and \emph{candidates} $C$ such that:
    \begin{enumerate}
    \item jurors are never allocated the good ($q_{i} = 0$ for all $i\in J$);
    \item candidates' reports never influence the allocation ($q$ is constant in $\theta_{i}$ for all $i\in C$).
    \end{enumerate}
\end{definition}
Jury mechanisms resemble peer review procedures used in practice. Therefore, it is interesting to understand when jury mechanisms perform well.

In \Cref{appendix:jury_mechanisms:example}, we show that vanishing informational size alone does not imply that jury mechanisms are approximately optimal with many agents.
Then, in \Cref{appendix:jury_mechanisms:sufficient_conditions}, we formalize two notions of exchangeability under which jury mechanisms are approximately optimal with many agents.

\subsubsection{Vanishing Informational Size is Insufficient for Approximate Optimality of Jury Mechanisms}\label{appendix:jury_mechanisms:example}

In this section, we show that jury mechanisms need not be approximately optimal along regular sequences of environments with vanishing informational size.
The basic building block is an environment with three agents that we later expand to have many agents.
The key properties of the environment are as follows:
\begin{enumerate}
    \item At each type profile in the support of the type distribution, exactly two agents have a peer value of $1$, while the other agent has a peer value of $0$. Further, an agent's peer value is $0$ if the reports of the others is outside the support. 
    \item The information of an individual agent correctly identifies only with probability $2/3$ which of the other two agents has a peer value of $1$ (possibly both).
    \item Each individual agent has a peer value of $1$ at only $2/3$ of all type profiles in the support of the type distribution.
\end{enumerate}
In such an environment with three agents, the argument for why ranking-based mechanisms outperform jury mechanisms is as follows. 
Informational size is at most $2/3$ at every profile since agents influence the ranks only through the tie-breaking rule.
Property $1$ implies that the ranking-based mechanism with threshold $p =2/3$ always selects an agent with peer value $1$; indeed, an agent with peer value $1$ has rank $2/3$ or less regardless of what this agent reports, while an agent with peer value $0$ has rank $1$; thus at each profile the two agents with peer value $1$ both enjoy a winning probability $1/(np) = 1/2$.
However, properties 2 and 3 imply that each jury mechanism obtains at most $2/3$.
Indeed, if there is a single juror, then property (2) implies the juror finds the best candidate only with probability $2/3$.
For fewer or more jurors, the jury mechanism is constant, and hence property (3) implies that the constantly chosen candidate is the best only with probability $2/3$.

An explicit environment with the above three properties is as follows.
For each of the three agents, the type space is $\lbrace 1, 2, 3\rbrace$.
Consider the following odd $9$-hole $H$ in the feasibility graph:
\begin{equation*}
        (\cdot, 1, 1), (2, \cdot, 1), (2, 2, \cdot), (\cdot, 2, 2), (3, \cdot, 2), (3, 3, \cdot), (\cdot, 3, 3), (1, \cdot, 3), (1, 1, \cdot).
\end{equation*}
Let $T$ denote the set of type profiles traversed by this odd hole; that is, 
\begin{equation*}
    T = \lbrace (1, 1, 1), (2, 1, 1), (2, 2, 1), (2, 2, 2), (3, 2, 2), (3, 3, 2), (3, 3, 3), (1, 3, 3), (1, 1, 3)\rbrace.
\end{equation*}
The distribution of type profiles is uniform over $T$ (i.e., $\mu(\theta) = \frac{1}{9}\bm{1}(\theta\in T)$ for all $\theta\in\Theta$) and peer values are given by $\bar u_{i}(\theta_{-i}) = \bm{1}((i, \theta_{-i}) \in H)$.
Properties (1) and (3) are immediate from the construction.
For property (2), it suffices to note that for every distinct agents $i$ and $j$ and type $\theta_{i}$, agent $j$ has a peer value of $1$ at no more than two of the three type profiles where $i$'s type is $\theta_{i}$; for example, for $i=3$ and $\theta_{3} = 1$, agent $1$ has a peer value of $1$ only at $(1, 1, 1)$ and $(2, 1, 1)$ (but not at $(2, 2, 1)$), while agent $2$ has a peer value of $1$ only at $(2, 1, 1)$ and $(2, 2, 1)$ (but not at $(1, 1, 1)$).

We now expand the above example to have many agents and so that informational size vanishes as the number of agents diverges and regularity holds.
Let the number of agents $n$ be an integer multiple $\ell$ of $3$.
Partition the agents into $\ell$ groups of $3$ so that within each group the agents are labeled with three consecutive integers (the labeling only matters for the tie-breaking rule).
For each group, fix an auxiliary environment for three agents as above.
Define the environment (for $n$ agents) by letting all variables from the auxiliary environments be independent across groups. (Thus, for example, Nature takes $\ell$ independent uniform draws from the set $T$ defined above.)
Informational size is at most $2/n$ at every type profile in the support since each agent cannot influence the peer values of the agents belonging to other groups, and since within each group each agent influences the rank only through the tie-breaking rule.\footnote{For this argument, the partition of the agents into groups should be chosen so that the agents within a group are labeled with three consecutive integers. Since ties are broken in favor of agent $i$ versus agent $j$ if and only if $i < j$, it follows that each agent can influence their own rank by at most $2/n$.} 
Further, as $n\to\infty$ the sequence of environments is regular since two thirds of the agents all have the highest peer value of $1$.
As before, a ranking-based mechanism (with, say, threshold $2/3$) always selects an agent with the highest peer value.
Now consider a jury mechanism.
Fix a realization of the types of the jurors.
For every group $\ell$, the jurors find a candidate from group $\ell$ whose peer value is $1$ only with probability $2/3$; this is because the types of jurors outside of group $\ell$ are uninformative about the candidates in group $\ell$, and because the jurors from group $\ell$ find a candidate whose peer value is $1$ only with probability $2/3$.
There is only one good to allocate.
Thus, regardless from which group the principal selects a candidate using the types of the jurors, the principal finds a candidate whose peer value is $1$ only with probability $2/3$.
Thus, the expected utility of each jury mechanism is at most $2/3$.

\subsubsection{Sufficient Conditions for Approximate Optimality of Jury Mechanisms}\label{appendix:jury_mechanisms:sufficient_conditions}

In this part of the appendix, we formalize two notions of exchangeability under which jury mechanisms are approximately optimal with many agents.
We focus on environments in which agents receive conditionally independent signals about the values of each other.

An \emph{information structure with conditionally independent signals} specifies the following for every $i\in\mathbb{N}$:
\begin{enumerate}
    \item a distribution $f_{i}$ whose support $\supp f_{i}$ is finite and contained in $[-1, 1]$;
    \item for every $j\in\mathbb{N}$ distinct from $i$ a finite set $S_{i}^{j}$ and a function $g_{i}^{j}$ that assigns to each $u_{i}\in \supp f_{i}$ a distribution over $S_{i}^{j}$.
\end{enumerate}
Given an information structure with conditionally independent signals and $n\in\mathbb{N}$, the environment with $n$ agents works as follows.
The values $u_{1}, \ldots, u_{n}$ are independent across agents $1$ to $n$ with respective marginals $f_{1}, \ldots, f_{n}$. 
Conditional on $u_{i}$, each agent $j$ other than $i$ observes a signal $\sigma_{i}^{j}$ about $u_{i}$ drawn from $g_{i}^{j}(\cdot\vert u_{i})$.
Thus, each agent $j$'s type is $(\sigma_{i}^{j})_{i \in [n]\colon i\neq j}$. 
Conditional on $(u_{1}, \ldots, u_{n})$, the signals $(\sigma_{i}^{j})_{i\neq j}$ are independent.
Thus, the probability of a profile $((u_{i})_{i\in [n]}, (\sigma_{i}^{j})_{i,j\in[n]\colon i\neq j})$ is $\prod_{i, j\in [n]\colon i\neq j} f_{i}(u_{i}) g_{i}^{j}(\sigma_{i}^{j}\vert u_{i})$.

Nothing in the analysis would change if an agent $i$ also observed their own value since this information cannot be used to determine agent $i$'s allocation (by DIC) and since agent $i$'s value is uninformative about the other agents' values (by independence of the values).

Fix an information structure with conditionally independent signals.
\begin{itemize}
    \item \emph{Agents are exchangeable as suppliers of information} if for all $i\in \mathbb{N}$ the set of possible signals $S_{i}^{j}$ and the conditional distributions $(g_{i}^{j}(\cdot\vert u_{i}))_{u_{i}\in\supp f_{i}}$ do not depend on $j$.
    
    Intuitively, for all $i$ and $j$, the information that $j$ can potentially provide about $i$ does not depend systematically on $j$'s identity.

    \item \emph{Agents are exchangeable as recipients of the good} if the marginal distributions $f_{i}$ do not depend on $i$ and for all $j\in \mathbb{N}$ the set of possible signals $S_{i}^{j}$ and all conditional distributions $(g_{i}^{j}(\cdot\vert u))_{u\in \supp f_{i}}$ do not depend on $i$.
    
    Intuitively, for all $i$ and $j$, neither the value of allocating to $i$ nor the information that $j$ can potentially provide about $i$ depend systematically on $i$'s identity.
\end{itemize}

\begin{theorem}\label{thm:jury_approx_optimality}
    Consider an information structure with conditionally independent signals in which agents are exchangeable suppliers of information or in which agents are exchangeable as recipients of the good.
    As the number of agent diverges, the difference between the principal's expected utility from an optimal jury mechanism and an optimal DIC mechanism vanishes.
\end{theorem}
We provide a proof sketch that is straightforward but notationally tedious to verify. For this proof, it is crucial that the information structure is fixed. (By contrast, our result for ranking-based mechanisms, \Cref{thm:ranking_based}, does not require a fixed information structure.)

Suppose agents are exchangeable as suppliers of information.
Let $n\in\mathbb{N}$.
In the environment with $n$ agents, let $\bar{U}^{n}$ denote the expected utility from always allocating to an agent with the highest peer value.
Each DIC mechanism in the environment with $n$ agents obtains at most $\bar{U}^{n}$.
We show that $\bar{U}^{n}$ is attainable in the environment with $2n$ agents using a jury mechanism.
The sequence $(\bar{U}^{n})_{n\in\mathbb{N}}$ converges,\footnote{The sequence converges since it is bounded and increasing. Boundedness is immediate since values are in $[-1, 1]$. Monotonicity follows since the information structure is held fixed as the number of agents increases. Thus, the principal has the option of ignoring agents who are added to the environment.} and hence the theorem follows.
For all $n$, the value $\bar{U}^{n}$ is generated by the following mechanism $q^{n}$ (which need not be DIC): the principal collects the entire profile of signals $(\sigma_{i}^{j})_{i, j\in [n]\colon i\neq j}$, and then allocates to an agent $i$ for which the expected value of $u_{i}$ conditional on $(\sigma_{i}^{j})_{j\in [n]\colon i\neq j}$ is maximal across $i\in [n]$.
We shall replicate $q^{n}$ using the following jury mechanism with $2n$ agents: agents $1$ to $n$ are the candidates, agents $n+1$ to $2n$ are the jurors.
For each $j\in [n]$, the principal collects from juror $n+j$ only the signals $(\sigma_{i}^{n+j})_{i\in [n]\colon i\neq j}$.
The principal treats the signals thus obtained in the same way that $q^{n}$ would treat the signals of agent $j$.
That is, the principal selects a candidate $i$ with the highest expected value $u_{i}$ conditional on $(\sigma_{i}^{n+j})_{i\in [n]\colon i\neq j}$.
The only difference between this jury mechanism and $q^{n}$ is that for each $i\in [n]$ the conditional expected value of agent $i$ is computed using $n-1$ signals originating from agents $n+1$ to $2n$ instead of $n-1$ signals originating from $[n]\setminus\lbrace i\rbrace$.
But agents are exchangeable as suppliers of information, meaning that the signals about $u_{i}$ are conditionally iid. across \emph{all} agents other than $i$.
Thus, the origin of the signals is irrelevant for determining the conditional expected value of $u_{i}$.
Thus, the described jury mechanism yields the same expected utility as $q^{n}$.

If agents are exchangeable as recipients of the good, a similar argument applies.
We now construct a jury mechanism in which agents $n+1$ to $2n$ substitute for agents $1$ to $n$ whenever one of the latter is selected to receive the good by $q^{n}$.
That is, agents $1$ to $n$ are the jurors, agents $n+1$ to $2n$ are the candidates.
For all $j\in [n]$, the principal collects from juror $j$ only the signals $(\sigma_{i}^{j})_{i\in \lbrace n+1, \ldots, 2n\rbrace\colon i\neq n+j}$.\footnote{We exclude the signal if $i=n+j$ since $q^{n}$ computes each agent's conditional expected value using $n-1$ signals. 
Thus, to replicate $q^{n}$, we deliberately choose the jury mechanism to also use only $n-1$ signals for each conditional expected value.}
The principal treats the signal of $j \in \lbrace 1, \ldots, n\rbrace$ about $i\in\lbrace n+1, \ldots, n\rbrace$ in the same way that $q^{n}$ would treat the signal of $j$ about agent $n-i$.
Since agents are exchangeable as recipients, the conditional value of allocating to agent $i\in\lbrace n+1, \ldots, n\rbrace$ given the elicited signal profile equals the conditional value of allocating to agent $n-i$ if the same profile of signals had been reported about about $n-i$.
Thus, the described jury mechanism achieves the same expected utility as $q^{n}$.

\subsection{Mandatory Allocation}\label{appendix:mandatory_allocation}

\subsubsection{Preliminaries}

Let $\bar{Q}$ denote the set of DIC mechanisms $q$ that always allocate, i.e. $\sum_{i\in [n]} q_{i}(\theta_{-i}) = 1$ holds for all $\theta\in\Theta$.
Let $\extremepoints \bar{Q}$ denote the set of extreme points of $\bar{Q}$.

It holds $\extremepoints \bar{Q}\subseteq\extremepoints Q$ since a DIC mechanism that always allocates can only be represented as a convex combination of other DIC mechanisms that always allocate.

As before, let $G$ denote the feasibility graph, and let $V = \cup_{i\in [n]} (\lbrace i\rbrace)\times\Theta_{-i}$ denote its vertices.
Let $\bar{\mathcal{S}}$ denote the set of stable sets $S$ of $G$ such that $\vert S\cap X\vert = 1$ holds for all maximal cliques $X$.
The map $q\mapsto \lbrace v\in V \colon q(v) = 1\rbrace$ is a bijection from $\bar{Q}$ to $\bar{\mathcal{S}}$.

\subsubsection{Results}

The next theorem characterizes when all extreme points of $\bar{Q}$ are deterministic. The only difference to the characterization for $Q$ (\Cref{thm:stochastic_existence}) is in the threshold value for the number of agents.
\begin{theorem}\label{thm:mandatory_allocation:stochastic_existence}
    All extreme points of $\bar{Q}$ are deterministic if and only if at least one of the following is true:
    \begin{enumerate}
        \item there are at most three agents ($n\leq 3$);
        \item all type spaces are binary ($\vert\Theta_{i}\vert \leq 2$ for all $i\in [n]$).
    \end{enumerate}
\end{theorem}
(Recall that the model assumes $n\geq 2$ and $\vert\Theta_{i}\vert \geq 2$ for all $i\in [n]$.
\begin{proof}[Proof of \Cref{thm:mandatory_allocation:stochastic_existence}]
    The proof of \Cref{prop:juries} shows that all extreme points are deterministic if $n\leq 3$.
    If all type spaces are binary, the claim follows from \Cref{thm:stochastic_existence} and the inclusion $\extremepoints \bar{Q}\subseteq \extremepoints Q$.
    Finally, suppose there are at least four agents and at least one type space is non-binary. Without loss, let agent $1$ have a non-binary type space. \Cref{thm:stochastic_existence} implies there exists a stochastic extreme point of the set of DIC mechanisms (that need not always allocate) for agents $1$, $2$, and $3$. View this mechanism as a DIC mechanism that always allocates for agents $1$ to $n$ where all reports of agents $4$ to $n$ are ignored. This mechanisms is an extreme point of $\bar{Q}$.
\end{proof}

\begin{remark}
    In \Cref{lemma:stochastic_hole_link}, we showed that $Q$ admits a stochastic extreme point if and only if $G$ admits an odd hole of length $7$ or greater. Only one direction of this equivalence carries over to $\bar{Q}$.
    Namely, if $\bar{Q}$ admits a stochastic extreme point, then $G$ admits an odd hole of length $7$ or greater. However, if $n=3$, then all extreme points of $\bar{Q}$ are deterministic even though $G$ admits an odd hole of length $7$ or greater if at least one type space is non-binary.
\end{remark}

The next theorem provides an analogue of \Cref{thm:stochastic_prevalence}:
essentially all extreme points of $\bar{Q}$ are stochastic if $n\geq 4$ and type spaces are large.
For the sake of simplicity, we focus on the case where the agents' type spaces all have the same cardinality.
\begin{theorem}\label{thm:mandatory_allocation:stochastic_prevalence}
    Fix $n\geq 4$.
    Suppose the agents have a common type space ($\Theta_{1} = \ldots = \Theta_{n}$) which has cardinality $m\in\mathbb{N}$.
    For all $\varepsilon > 0$ there exists $m_{\varepsilon}\in\mathbb{N}$ such that if $m \geq m_{\varepsilon}$, then 
    $
		\left\vert \deterministic \bar{Q} \right\vert < \varepsilon \left\vert \extremepoints \bar{Q}\right\vert.
	$
\end{theorem}
The proof, presented further below, is more challenging than the one for \Cref{thm:stochastic_prevalence}.

Next, we provide a characterization of stochastic extreme points via odd holes, analogously to \Cref{prop:hole_characterization}.
As in the main text, for a mechanism $q\in\bar{Q}$ a \emph{stochastic component of $q$} is an inclusion-wise maximal connected set of vertices $v$ of $G$ such that $q(v)\in (0, 1)$.
\begin{theorem}\label{prop:mandatory_allocation:hole_characterization}
    Let $q$ be a stochastic DIC mechanism that always allocates. If $q$ is an extreme point of $\bar{Q}$, then every stochastic component of $q$ contains an odd hole.
    The converse holds if $q(v) \in \lbrace 0, \frac{1}{2}, 1\rbrace$ for all $v\in V$.
\end{theorem}
\begin{proof}[Proof of \Cref{prop:mandatory_allocation:hole_characterization}]
    First, let $q\in \extremepoints \bar{Q}$.
    Since $\extremepoints \bar{Q}\subseteq \extremepoints Q$, \Cref{prop:hole_characterization} implies that each stochastic component of $q$ contains an odd hole.
    Second, suppose $q$ maps to $\lbrace 0, \frac{1}{2}, 1\rbrace$ and that each of its components admits an odd hole.
    Since $q \in \bar{Q}$ and $\bar{Q}\subseteq Q$, \Cref{prop:hole_characterization} implies $q$ is an extreme point of $Q$.
    Using again $q \in \bar{Q}$ and $\bar{Q}\subseteq Q$, it follows that $q$ is also an extreme point of $\bar{Q}$.
\end{proof}

Next, we show that the problem of determining an optimal deterministic DIC mechanisms that always allocates is NP-hard if $n\geq 4$.
This result follows immediately from \Cref{thm:npcomplete}, but we include the definitions for the sake of completeness.
\begin{definition}[\textsc{OptDetMA}-$n$]
	For $n\in\mathbb{N}$, let \textsc{OptDetMA}-$n$ be the following optimization problem (``MA'' stands for mandatory allocation).
	The input consists of finite sets $\Theta_{1}, \ldots, \Theta_{n}$ and weights $w_i: \Theta_{-i}\to \mathbb{Z}$ for all $i\in [n]$.
	The problem is to find a deterministic DIC mechanism $q$ (for $n$ agents with respective type spaces $\Theta_{1}, \ldots, \Theta_{n}$) that always allocates and that maximizes $\sum_{i, \theta} w_{i}(\theta_{-i}) q_{i}(\theta_{-i})$ across all deterministic DIC mechanisms that always allocate.
\end{definition}

\begin{theorem}\label{thm:mandatory_allocation:npcomplete}
	If $n\geq 4$, then \textsc{OptDetMA}-$n$ is NP-hard.
\end{theorem}

\begin{proof}[Proof of \Cref{thm:mandatory_allocation:npcomplete}]
    Consider the instances of \textsc{OptDetMA}-$n$ where at least one agent has a singleton type space and a weight constantly equal to $0$. Each such instance corresponds to an instance of \textsc{OptDet}-$(n-1)$. Since $n\geq 4$, \Cref{thm:npcomplete} implies that \textsc{OptDet}-$(n-1)$ is NP-hard. Thus \textsc{OptDetMA}-$n$ is also NP-hard.
\end{proof}

\begin{remark}
    If $n\leq 3$, then optimal deterministic must-allocate DIC mechanisms are computable in polynomial time.
    Indeed, according to \Cref{prop:juries} it suffices to restrict attention to jury mechanisms that always allocate. 
    It is easy to see that the optimal juror can be found in polynomial time. 
\end{remark}

\subsubsection{Proof of \headercref{Theorem}{{thm:mandatory_allocation:stochastic_prevalence}}}

We introduce some useful terminology.
For all $i\in [n]$, the set $V_{i} = \lbrace i\rbrace\times\Theta_{-i}$ is the set of \emph{$i$-vertices}.
Given a vertex $v = (i, \theta_{-i})$ and $j$ distinct from $i$, we say $\theta_{j}$ is the \emph{type of $j$ at $v$}.
Given distinct $i$ and $j$, two $i$-vertices and are \emph{$j$-translates} if there is a $j$-vertex that is adjacent to both of them; equivalently, the $i$-vertices coincide exactly except for possibly in the type of agent $j$.
To be sure, an $i$-vertex is a $j$-translate of itself.
A $j$-vertex is adjacent to an $i$-vertex $v$ only if it is adjacent to all $j$-translates of $v$.

Recall $\vert\bar{\mathcal{S}}\vert = \vert \deterministic \bar{Q}\vert$.
Recall also that the agents have a common type space with cardinality $m$.
By possibly relabelling types, thus $\Theta_{i} = \lbrace 1, \ldots, m\rbrace$ for all $i\in [n]$.

We show that for all integers $L \geq 4$ there exists $m_{L}\in\mathbb{N}$ such that if $m \geq m_{L}$, then $\vert \bar{\mathcal{S}}\vert \left(1 + \frac{L}{n}\right) \leq\vert \extremepoints \bar{Q}\vert$, which proves \Cref{thm:mandatory_allocation:stochastic_prevalence}.
In what follows, fix an integer $L\geq 4$.

We now define the central notion of the proof---\emph{regularity}---which identifies odd holes and stable sets that intersect in a particular way.
(This regularity notion has no connection to the regularity notion which we discussed in the context of ranking-based mechanisms.)
\begin{definition}[Regularity]
Let $S\in\bar{\mathcal{S}}$.
Let $H$ be a set of vertices.
Let $(v_{1}, \ldots, v_{9})$ be an odd hole.
Let $i, j, k, \ell\in[n]$ be distinct. 
The tuple $(S, H, v_{1}, \ldots, v_{9}, i, j, k, \ell)$ is \emph{regular} if
\begin{enumerate}
    \item $H$ is the set of $\ell$-translates of $\lbrace v_{1}, \ldots, v_{9}\rbrace$, and $H\subseteq V_{i}\cup V_{j}\cup V_{k}$ holds.
    \item For all vertices $\omega$, if $\omega$ is in $S$ and adjacent to two distinct vertices in $H$, then $\omega$ is an $\ell$-vertex.
    \item For all vertices $\omega$, if $\omega$ is an $\ell$-vertex that is contained in a maximal clique that also contains two vertices in $H$, then $\omega\in S$.
\end{enumerate}
The triple $\bm{r} = (S, H, \ell)$ is \emph{regular} if there exist $\lbrace v_{1}, \ldots, v_{9}\rbrace$ and $i, j, k\in[n]$ such that $(S, H, v_{1}, \ldots, v_{9}, i, \ldots, \ell)$ is regular.
\end{definition}
If $(S, H, \ell)$ is regular, the choice $(v_{1}, \ldots, v_{9}, i, j, k)$ is unique up to the type of agent $\ell$ (which is held constant across $v_{1}, \ldots, v_{9})$, and the labelling of the vertices $v_{1}, \ldots, v_{9}$.

We next address the existence of regular tuples.
\begin{lemma}\label{lemma:regular_triple_existence}
    There exists $m_{L}\in\mathbb{N}$ such that if $m\geq m_{L}$, then for all $S\in\bar{\mathcal{S}}$ there exists $\ell\in [n]$ and at least $L$ distinct sets $H$ such that $(S, H, \ell)$ is regular.
\end{lemma}
\begin{proof}[Proof of \Cref{lemma:regular_triple_existence}]
    We use the following special case of Theorem 9.2 of \citet{promel2013ramsey}.
    \begin{lemma}\label{lemma:pidgeonhole}
        There exists $m_{L}\in\mathbb{N}$ such that, if $m\geq m_{L}$, then for all $q\colon\Theta\to [n]$ there exist $\Theta_{1}^{\ast}\subseteq\Theta_{1}, \ldots\Theta_{n}^{\ast}\subseteq\Theta_{n}$ such that $q$ is constant on $\times_{i=1}^{n}\Theta_{i}^{\ast}$, and such that $\min \lbrace\vert\Theta_{1}^{\ast}\vert, \ldots, \vert\Theta_{n}^{\ast}\vert\rbrace \geq L$ holds.   
    \end{lemma}
    
    Recall that there is bijection between $\bar{Q}$ and $\bar{\mathcal{S}}$.
    Hence:
    \begin{corollary}\label{cor:stable_pidgeonhole}
        There exists $m_{L}\in\mathbb{N}$ such that, if $m\geq m_{L}$, then for all $S\in\bar{\mathcal{S}}$ there exists $\ell\in [n]$ and $\Theta_{1}^{\ast}\subseteq\Theta_{1}, \ldots\Theta_{n}^{\ast}\subseteq\Theta_{n}$ such that $\min \lbrace\vert\Theta_{1}^{\ast}\vert, \ldots, \vert\Theta_{n}^{\ast}\vert\rbrace \geq L$ and such that all $\theta_{-\ell}\in \times_{i\neq\ell}\Theta_{i}^{\ast}$ satisfy $(\ell, \theta_{-\ell}) \in S$.
    \end{corollary}
    
    Let $m_{L}$ meet the conclusion of \Cref{cor:stable_pidgeonhole}, and let $m\geq m_{L}$.
    Let $S\in\bar{\mathcal{S}}$, and let $\Theta_{1}^{\ast}\times\ldots\times\Theta_{n}^{\ast}$ be as in the conclusion of \Cref{cor:stable_pidgeonhole}.
    Now let $i$, $j$, and $k$ be three agents distinct from $\ell$ (recall $n\geq 4$).
    Let $\theta_{-} \in \times_{\iota\notin\lbrace i, j, k, \ell\rbrace} \Theta_{\iota}^{\ast}$ be an arbitrary profile of types of agents other than $i$, $j$, $k$, and $\ell$ (assuming such agents exist).
    Let $\theta_{\ell}\in\Theta_{\ell}$ be arbitrary.
    
    We next describe the construction of a family of regular tuples.
    Since $L\geq 4$, for all $\iota\in\lbrace i, j, k\rbrace$, we may select three distinct types $\theta_{\iota}^{1}, \theta_{\iota}^{2}, \theta_{\iota}^{3}$ from $\Theta_{\iota}^{\ast}$.
    Now consider the following set of vertices (here, we denote, say, vertices of agent $i$ by $(\cdot, \theta_{j}^{1}, \theta_{k}^{1}, \theta_{\ell}, \theta_{-})$, etc.):
    \begin{align}\label{eq:thm:mandatory_allocation:stochastic_prevalence:hole_construction}
    \begin{aligned}
        v_{1} &= (\cdot, \theta_{j}^{1}, \theta_{k}^{1}, \theta_{\ell}, \theta_{-}) \\
        v_{2} &= (\theta_{i}^{2}, \cdot, \theta_{k}^{1}, \theta_{\ell}, \theta_{-}) \\
        v_{3} &= (\theta_{i}^{2}, \theta_{j}^{2}, \cdot, \theta_{\ell}, \theta_{-}) \\
        v_{4} &= (\cdot, \theta_{j}^{2}, \theta_{k}^{2}, \theta_{\ell}, \theta_{-}) \\    
        v_{5} &= (\theta_{i}^{3}, \cdot, \theta_{k}^{2}, \theta_{\ell}, \theta_{-}) \\
        v_{6} &= (\theta_{i}^{3}, \theta_{j}^{3}, \cdot, \theta_{\ell}, \theta_{-}) \\
        v_{7} &= (\cdot, \theta_{j}^{3}, \theta_{k}^{3}, \theta_{\ell}, \theta_{-}) \\
        v_{8} &= (\theta_{i}^{1}, \cdot, \theta_{k}^{3}, \theta_{\ell}, \theta_{-}) \\
        v_{9} &= (\theta_{i}^{1}, \theta_{j}^{1}, \cdot, \theta_{\ell}, \theta_{-}) .
        \end{aligned}
    \end{align}
    By inspection, $(v_{1}, \ldots, v_{9})$ is an odd hole.
    Let $H$ denote this odd hole and all its $\ell$-translates.
    We show that the tuple $(S, H, v_{1}, \ldots, v_{9}, i, j, k, \ell)$ is regular for each selection of 
    $\theta_{i}^{1}, \theta_{i}^{2}, \theta_{i}^{3}, \theta_{j}^{1}, \theta_{j}^{2}, \theta_{j}^{3}, \theta_{k}^{1}, \theta_{k}^{2}, \theta_{k}^{3}$.
    That is, we show
    \begin{enumerate}
        \item $H$ is the set of $\ell$-translates of $\lbrace v_{1}, \ldots, v_{9}\rbrace$, and $H\subseteq V_{i}\cup V_{j}\cup V_{k}$ holds.
        \item For all vertices $\omega$, if $\omega$ is in $S$ and adjacent to two distinct vertices in $H$, then $\omega$ is an $\ell$-vertex.
        \item For all vertices $\omega$, if $\omega$ is an $\ell$-vertex that is contained in a maximal clique that also contains two vertices in $H$, then $u\in S$.
    \end{enumerate}
    Property (1) is immediate from the construction of $H$.

    Consider property (2).
    Consider a vertex $\omega$ that is not an $\ell$-vertex but that is adjacent to two distinct vertices $v$ and $v^{\prime}$ in $H$.
    We show $u\notin S$.
    Assume $v$ is an $i$-vertex (the other cases being similar).
    We first claim $v^{\prime}$ is not an $i$-vertex.
    Towards a contradiction, suppose $v^{\prime}$ is an $i$-vertex.
    Since $v$ and $v^{\prime}$ are $\ell$-translates of vertices in $v_{1}, \ldots, v_{9}$, there are two distinct $i$-vertices $\tilde{v}$ and $\tilde{v}^{\prime}$ in $v_{1}, \ldots, v_{9}$ that are both adjacent to some $\ell$-translate of $\omega$.
    However, since $\tilde{v}$ and $\tilde{v}^{\prime}$ are both $i$-vertices, inspection of \eqref{eq:thm:mandatory_allocation:stochastic_prevalence:hole_construction} shows that $\tilde{v}$ and $\tilde{v}^{\prime}$ must differ in at least two types (of agents other than $i$), implying that there is no vertex adjacent to both of them; contradiction.
    Thus $v^{\prime}$ is not an $i$-vertex.
    Assume $v^{\prime}$ is a $j$-vertex (the other case where $v^{\prime}$ is a $k$-vertex being similar).
    Now let $\omega$ be a $\tilde{k}$-vertex.
    We know $\tilde{k}\neq i$ (since $\omega$ would fail to be adjacent to $v$) and $\tilde{k} \neq j$ (since $\omega$ would fail to be adjacent to $v^{\prime}$).
    Thus the types of all agents other than $i$ and $\tilde{k}$ agree at $\omega$ and $v$, and the types of all other agents other than $j$ and $\tilde{k}$ agree at $\omega$ and $v^{\prime}$.
    In particular, the types of agents other than $\tilde{k}$ are in the respective spaces $\Theta_{1}^{\ast}, \ldots, \Theta_{n}^{\ast}$.
    We also have $\tilde{k}\neq \ell$ by assumption.
    It follows that $\omega$ is adjacent to an $\ell$-vertex in which the agents other than $\ell$ all have types in the respective spaces $\Theta_{1}^{\ast}, \ldots, \Theta_{n}^{\ast}$; in particular, this $\ell$-vertex is in $S$.
    Since $S$ is stable, we conclude that $\omega$ is not in $S$.

    Turning to property (3), consider an $\ell$-vertex $\omega$ adjacent to two adjacent vertices in $H$.
    Since $H$ includes all its $\ell$-translates, $\omega$ is adjacent to two vertices in $\lbrace v_{1}, \ldots, v_{9}\rbrace$.
    Suppose these are the vertices $v_{1}$ and $v_{2}$; the arguments for the other cases are similar.
    The unique $\ell$-vertex adjacent to $v_{1}$ and $v_{2}$ is the vertex$(\theta_{i}^{2}, \theta_{j}^{1}, \theta_{k}^{1}, \cdot, \theta_{-i})$.
    Given the choice of the sets $\Theta_{1}^{\ast}, \ldots, \Theta_{n}^{\ast}$, this $\ell$-vertex is in $S$, as promised.
    
    Lastly, by varying the initial choices of $(\theta_{i}^{1}, \theta_{i}^{2}, \theta_{i}^{3})$, $(\theta_{j}^{1}, \theta_{j}^{2}, \theta_{j}^{3})$, and $(\theta_{k}^{1}, \theta_{k}^{2}, \theta_{k}^{3})$, respectively, from $\Theta_{i}^{\ast}$, $\Theta_{j, m}^{\ast}$, and $\Theta_{k, m}^{\ast}$, respectively, and using that each of $\Theta_{i}^{\ast}$, $\Theta_{j, m}^{\ast}$, and $\Theta_{k, m}^{\ast}$ contains at least $L$ elements, we obtain $L$ distinct sets $H$ (in fact, we obtain more than $L$, but $L$ is all we need for the argument).
\end{proof}

Let $m_{L}$ meet the conclusion of \Cref{lemma:regular_triple_existence}.
As described earlier, we complete the proof by showing that if $m\geq m_{L}$, then
\begin{align*}
    \vert \bar{\mathcal{S}}\vert \left(1 + \frac{L}{n}\right) \leq\vert \extremepoints \bar{Q}\vert. 
\end{align*}
In what follows, let $m\geq m_{L}$.

For a later part of the argument, it shall be more convenient to work with regular triples $(S, H, \ell)$ where $\ell$ is fixed to be agent $n$.
To that end, define $\bar{\mathcal{S}}^{\ast}$ as the set of $S\in \bar{\mathcal{S}}$ for which there exist at least $L$ distinct sets $H$ such that $(S, H, n)$ is regular.
If $(S, H, \ell)$ is some regular triple, then permuting the roles of agents $\ell$ and $n$ gives another regular triple; this step uses that all agents have the same type space.
Thus:
\begin{align}\label{eq:normalized_regular_triples}
    \vert\bar{\mathcal{S}}\vert \leq n \vert \bar{\mathcal{S}}^{\ast}\vert.
\end{align}

We make some more auxiliary definitions.
Let $\bm{r} = (S, H, \ell)$ be regular.
\begin{enumerate}
    \item Let $N(H)$ denote the set of vertices that are adjacent to at least one vertex in $H$.
    \item Let $\Omega_{\bm{r}}$ denote the $\ell$-vertices that are contained in a maximal clique that also contains two vertices in $H$; that is, the $\ell$-vertices considered in point (3) of the definition of regularity. Regularity requires $\Omega_{\bm{r}} \subseteq S$.
    \item Let $V_{\ell}\setminus N(H)$ be the set of $\ell$-vertices that are non-adjacent to all vertices in $H$.
    \item Let $V^{\ast}_{\bm{r}} = (S\setminus \Omega_{\bm{r}}) \cup H \cup (V_{\ell}\setminus N(H))$, and let $G[V^{\ast}_{\bm{r}}]$ be the subgraph induced by $V^{\ast}_{\bm{r}}$.
    \item Let $\mathcal{K}_{\bm{r}}$ denote the set of connected components of $G[V^{\ast}_{\bm{r}}]$. Define $K_{\bm{r}} = \cup\lbrace K^{\prime}\in\mathcal{K}_{\bm{r}}\colon K^{\prime}\cap H \neq \emptyset\rbrace$ as the set of vertices in those connected components that have a non-empty intersection with $H$.
\end{enumerate}

We next describe a family of candidate stochastic extreme points.
\begin{definition}\label{def:candidate_stochastic_extreme_points}
For all regular $\bm{r} = (S, H, \ell)$, let $q_{\bm{r}}\colon V\to [0, 1]$ be the function defined by
\begin{align*}
    \forall_{v\in V},\quad
    q_{\bm{r}} =
    \begin{cases}
        \frac{1}{2},\quad &\text{if $v\in K_{\bm{r}}$}\\
        1,\quad &\text{if $v\in S\setminus (\Omega_{\bm{r}}\cup K_{\bm{r}})$}\\
        0,\quad &\text{else}.
    \end{cases}
\end{align*}
\end{definition}

\begin{lemma}\label{lemma:feasibility}
    If $\bm{r} = (S, H, \ell)$ is regular, then $q_{\bm{r}}$ is feasible and a stochastic extreme point of $\bar{Q}$.
\end{lemma}

\begin{proof}[Proof of \Cref{lemma:feasibility}]
    To begin with, provided we can show that $q_{\bm{r}}$ is in $\bar{Q}$, it follows immediately from \Cref{prop:mandatory_allocation:hole_characterization} that $q_{\bm{r}}$ is a stochastic extreme point of $\bar{Q}$.
    Thus we show $q_{\bm{r}}\in\bar{Q}$.

    Let $X$ be a maximal clique.
    We have to show $q_{\bm{r}}(X) = 1$, where we denote $q_{\bm{r}}(X) = \sum_{v\in X} q(v)$.

    We first observe:
    \begin{enumerate}
        \item it holds $\vert X\cap H\vert \leq 2$. Indeed, suppose towards a contradiction $\vert X\cap H\vert \geq 3$. By definition of $H$, there is an odd hole $\tilde{H}$ such that $H$ is exactly the set of $\ell$-translates of $\tilde{H}$. Hence we infer from $\vert X\cap H\vert \geq 3$ that $\tilde{H}$ contains a clique of three vertices, contradicting that $\tilde{H}$ is a hole.
        \item $H\cap S = \emptyset$. Indeed, each vertex in $H$ is adjacent to another vertex in $H$. Hence $H\cap S = \emptyset$ follows since each maximal clique contains an $\ell$-vertex, the definition of $\Omega_{\bm{r}}$, and the inclusion $\Omega_{\bm{r}}\subseteq S$.
    \end{enumerate}
    Since $S\in\bar{\mathcal{S}}$, there exists $v\in X\cap S$.
    To show $q_{\bm{r}}(X) = 1$, we proceeds in several steps.

    First, suppose $v \in \Omega_{\bm{r}}$.
    By definition of $\Omega_{\bm{r}}$, there are distinct vertices $u, u^{\prime}\in H$ adjacent to $v$.
    Since $H$ contains all its $\ell$-translates, we may assume $\omega$ and $\omega^{\prime}$ are both in $X$.
    Further, we have $X\cap (S\setminus \lbrace v\rbrace )= \emptyset$ (since $v\in S$ and $S$ is stable) and $X\cap (H\setminus\lbrace u, u^{\prime}\rbrace) = \emptyset$ (since $\vert X\cap H\vert \leq 2$) and $X\cap (V_{\ell}\setminus N(H)) = \emptyset$ (since $X\cap H\neq \emptyset$).
    Hence $q_{\bm{r}}$ assigns $0$ to all vertices in $X\setminus\lbrace u, u^{\prime}\rbrace$.
    By definition of $K_{\bm{r}}$, we have $u, u^{\prime}\in K_{\bm{r}}$, and hence $q_{\bm{r}}(X) = 1$.
    
    In what follows, suppose $v\notin \Omega_{\bm{r}}$.
    
    As an intermediate claim, we show that $X\cap(V_{\ell}\setminus N(H))\neq\emptyset$ holds if and only if $X\cap H = \emptyset$ holds.
    Indeed, if $X\cap(V_{\ell}\setminus N(H))\neq\emptyset$, then we definitionally have $X\cap H = \emptyset$.
    Conversely, if $X\cap H= \emptyset$, then the $\ell$-vertex in $X$ (each maximal clique contains a unique $\ell$-vertex) is in $V_{\ell}\setminus N(H)$ since $H$ is closed with respect to $\ell$-translations.

    The intermediate claim implies there is a vertex $\omega\in X\cap (H \cup (V_{\ell}\setminus N(H)))$.    
    
    As a second intermediate claim, we show $$X\cap \left((S\setminus \Omega_{\bm{r}}) \cup H \cup (V_{\ell}\setminus N(H))\right) = \lbrace \omega, v\rbrace.$$
    Let $\omega^{\prime}$ in $X$  be a vertex distinct from $\omega$ and $v$.
    We know $\omega^{\prime}\notin S$ (since $v\in S$ and $S$ is stable).
    We know $\omega^{\prime}\notin H$; for if $\omega^{\prime}\in H$, then $\omega$ would also be in $H$ (by the intermediate claim), implying that $X$ contains two vertices in $H$; this contradicts $v\notin \Omega_{\bm{r}}$.
    Finally, $\omega^{\prime}\notin V_{\ell}\setminus N(H)$; for if $\omega^{\prime}\in V_{\ell}\setminus N(H)$, then $\omega$ would also be in $V_{\ell}\setminus N(H)$ (by the first intermediate claim), implying that $X$ contains two distinct $\ell$-vertices, which is impossible.
    Thus $X\cap \left((S\setminus \Omega_{\bm{r}}) \cup H \cup (V_{\ell}\setminus N(H))\right) = \lbrace \omega, v\rbrace$.

    We know show $q_{\bm{r}}(\omega) + q_{\bm{r}}(v) = 1$.
    This establishes $q_{\bm{r}}(X) = 1$ since $$X\cap \left((S\setminus \Omega_{\bm{r}}) \cup H \cup (V_{\ell}\setminus N(H))\right) = \lbrace \omega, v\rbrace.$$
    If $\omega\neq v$, then since $\omega$ and $v$ are adjacent and in $G[V^{\ast}_{\bm{r}}]$ it must be that $\omega$ and $v$ are in the same connected component of $G[V^{\ast}_{\bm{r}}]$.
    (If $\omega = v$, then of course $\omega$ and $v$ are also in the same connected component.)
    In particular, we have $\omega\in K_{\bm{r}}$ if and only if $v\in K_{\bm{r}}$.
    Therefore, if $\omega\neq v$ and $v\in K_{\bm{r}}$, then $q_{\bm{r}}(X) = q_{\bm{r}}(\omega) + q_{\bm{r}}(v) = \frac{1}{2}$; if $\omega\neq v$ and $v\notin K$, then $q_{\bm{r}}(X) = q_{\bm{r}}(v) = 1$.
    Lastly, consider the case $\omega = v$.
    From the choice of $\omega$ we get $v\in S \cap (H \cup (V_{\ell}\setminus N(H))$.
    We know $S\cap H = \emptyset$ holds, and hence $v\in V_{\ell}\setminus N(H)$.
    It follows that $v$ is non-adjacent to all other vertices in $S$ (by stability), non-adjacent to all other vertices in $(V_{\ell}\setminus N(H))$ (since all $\ell$-vertices are non-adjacent), and non-adjacent to all other vertices in $H$ (since $v\in V_{\ell}\setminus N(H)$).
    Hence $\lbrace v\rbrace$ is a connected component of $G[V^{\ast}_{\bm{r}}]$.
    In particular, $v$ cannot be contained in a component that intersects $H$ since such a component must contain more than one vertex (e.g., it contains an $\ell$-translate of $v_{1}, \ldots, v_{9}$).
    Hence $q_{\bm{r}}(X) = q_{\bm{r}}(v) = 1$.
\end{proof}

We next establish that the mapping $\bm{r}\mapsto q_{\bm{r}}$ is injective for regular $\bm{r}$ that use the same agent $\ell$.
\begin{lemma}\label{lemma:uniqueness}
    Let $\bm{r} = (S, H, \ell)$ and $\bm{r}^{\prime} = (S^{\prime}, H^{\prime}, \ell^{\prime})$ be regular.
    If $q_{\bm{r}} = q_{\bm{r^{\prime}}}$ and $\ell = \ell^{\prime}$, then $\bm{r} = \bm{r}^{\prime}$.
\end{lemma}

\begin{proof}[Proof of \Cref{lemma:uniqueness}]
    We first prove $H = H^{\prime}$.
    We prove $H \subseteq H^{\prime}$, the other inclusion being analogous.
    Towards a contradiction, let $v\in H\setminus H^{\prime}$.
    Recall that $H$ consists exactly of all $\ell$-translates of some odd hole, and that $H$ contains no $\ell$-vertices.
    In particular, there are vertices $\omega_{1}, \ldots, \omega_{9}$ in $H$ that form an odd hole and such that $v = \omega_{1}$ holds.
    Note $q_{\bm{r}} = q_{\bm{r^{\prime}}}$ assigns $\frac{1}{2}$ to all vertices $\omega_{1}, \ldots, \omega_{9}$.
    Hence $\omega_{1}, \ldots, \omega_{9}$ are all in $S^{\prime}\cup H^{\prime}\cup (V_{\ell}^{\prime}\setminus N(H^{\prime}))$.
    Since $\omega_{1}, \ldots, \omega_{9}$ are also all in $H$, and since $\ell = \ell^{\prime}$ and $H\cap V_{\ell} = \emptyset$ hold, we infer that $\omega_{1}, \ldots, \omega_{9}$ are all in $S^{\prime}\cup H^{\prime}$.
    Now $\omega_{1} = v \notin H^{\prime}$ requires $\omega_{1}\in S^{\prime}$, and hence $\omega_{2}, \omega_{9}\in H^{\prime}$.
    By regularity of $\bm{r}^{\prime}$, we conclude that $\omega_{1}$ is an $\ell^{\prime}$-vertex.
    In view of $\omega_{1} = v \in H$ and $\ell = \ell^{\prime}$, this gives a contradiction since $H$ contains no $\ell$-vertices.
    
    Lastly, we prove $S = S^{\prime}$.
    The assumption $q_{\bm{r}} = q_{\bm{r^{\prime}}}$ implies $S\setminus (\Omega_{\bm{r}} \cup K_{\bm{r}}) = S^{\prime}\setminus (\Omega_{\bm{r^{\prime}}} \cup K_{\bm{r^{\prime}}})$ and $K_{\bm{r}} = K_{\bm{r^{\prime}}}$.
    Using $H = H^{\prime}$, a moment's thought reveals $\Omega_{\bm{r}} = \Omega_{\bm{r}^{\prime}}$.
    Recalling also the definitions of $K_{\bm{r}}$ and $K_{\bm{r}^{\prime}}$, the equalities $H = H^{\prime}$ and $K_{\bm{r}} = K_{\bm{r}^{\prime}}$ and $\Omega_{\bm{r}} = \Omega_{\bm{r}^{\prime}}$ also imply $(S\cap K_{\bm{r}}) \setminus \Omega_{\bm{r}} = (S^{\prime} \cap K_{\bm{r}})\setminus \Omega_{\bm{r}}$.
    Since $K_{\bm{r}}$ contains no vertex in $\Omega_{\bm{r}}$, we also have $(S\cap \Omega_{\bm{r}}) \setminus K_{\bm{r}} = S\cap \Omega_{\bm{r}}$ and $(S^{\prime}\cap \Omega_{\bm{r}}) \setminus K_{\bm{r}} = S^{\prime}\cap \Omega_{\bm{r}}$ and $S\cap \Omega_{\bm{r}} \cap K_{\bm{r}} = S^{\prime}\cap \Omega_{\bm{r}} \cap K_{\bm{r}} = \emptyset$.
    Since also $S\setminus (\Omega_{\bm{r}} \cup K_{\bm{r}}) = S^{\prime}\setminus (\Omega_{\bm{r^{\prime}}} \cup K_{\bm{r^{\prime}}})$ and $(S\cap K_{\bm{r}}) \setminus \Omega_{\bm{r}} = (S^{\prime} \cap K_{\bm{r}})\setminus \Omega_{\bm{r}}$ (as argued earlier), we conclude $S = S^{\prime}$.
\end{proof}

We are now ready to prove
\begin{align*}
    \vert \bar{\mathcal{S}}\vert \left(1 + \frac{L}{n}\right) \leq\vert \extremepoints \bar{Q}\vert. 
\end{align*}
Recall that we defined $\bar{\mathcal{S}}^{\ast}$ as the set of $S\in \bar{\mathcal{S}}$ for which there exist at least $L$ distinct sets $H$ such that $(S, H, n)$ is regular.
Consider the correspondence $R$ that assigns to each $S\in\bar{\mathcal{S}}^{\ast}$ such a set of $L$ distinct tuples $H$.
\Cref{lemma:feasibility,lemma:uniqueness} imply that the mapping $\bm{r}\mapsto q_{\bm{r}}$ is an injection from the graph of $R$ to the set of stochastic extreme points.
We observed in \eqref{eq:normalized_regular_triples} that $\vert\bar{\mathcal{S}}\vert \leq n \vert \bar{\mathcal{S}}^{\ast}\vert$ holds.
Hence
\begin{align*}
    \left\vert \extremepoints\bar{Q}\right\vert \geq \vert\bar{\mathcal{S}}\vert + \left\vert \lbrace \bm{r}\colon (S, \bm{r})\in\graph R\rbrace\right\vert
    &=
    \vert\bar{\mathcal{S}}\vert + \sum_{S\in\bar{\mathcal{S}}^{\ast}} \left\vert  R(S) \right\vert
    \\
    &\geq
    \vert\bar{\mathcal{S}}\vert + \sum_{S\in\bar{\mathcal{S}}^{\ast}} L
    \\
    &\geq \vert\bar{\mathcal{S}}\vert \left(1 + \frac{L}{n}\right)
    ,
\end{align*}
as desired.

\subsection{Multi-unit allocation}\label{appendix:multiunit_allocation}

There are $k$ units, where $1 \leq k \leq n$.
Each agent desires at most one unit.
The principal does not gain from allocating more than one unit to an agent.
A mechanism is now a function $q\colon\Theta\to [0, 1]^{n}$ such that $\sum_{i\in [n]} q_{i}(\theta) \leq k$.
Here, $q_{i}$ is the probability that agent $i$ is allocated a unit, and we focus without loss on mechanisms that allocate at most one unit to each agent.
As before, DIC requires that $q_{i}$ be constant in $i$'s report.

For a type profile $\theta$, let $u_{i}(\theta_{-i}) = \mathbb{E}[u_{i}\vert\theta_{-i}]$ denote agent $i$'s peer value.
The principal's utility from a DIC mechanism $q$ is denoted $U(q)$ and equals:
\begin{equation*}
    U(q) = \frac{1}{k}\sum_{\theta\in\Theta}\sum_{i\in [n]} \mu(\theta) q_{i}(\theta_{-i}) \bar{u}_{i}(\theta_{-i}),
\end{equation*}
where we normalize by the number of units. This normalization matters since, in the analysis of ranking-based mechanisms, we shall allow the number of units to diverge.

We next discuss ranking-based mechanisms (\Cref{appendix:multiunit_allocation:ranking}) and then the results regarding complexity and randomization (\Cref{appendix:multiunit_allocation:hardness}).

\subsubsection{Ranking-based mechanisms}\label{appendix:multiunit_allocation:ranking}

We define agents' ranks and robust ranks exactly as in the single-unit case from the main text.
That is, given $\theta\in\Theta$ and $i\in [n]$, let $r_{i}(\theta)$ be $i$'s rank at $\theta$,\footnote{ 
	Formally, 
	$r_{i}(\theta) = \frac{1}{n}\left\vert\left\lbrace j\in [n]\colon \bar{u}_{j}(\theta_{-j}) > \bar{u}_{i}(\theta_{-i})\right\rbrace\right\vert + \frac{1}{n} \left\vert\left\lbrace j\in [n]\colon (\bar{u}_{j}(\theta_{-j}) = \bar{u}_{i}(\theta_{-i}))\land (i \geq j)\right\rbrace\right\vert.$
	}
and let $r_{i}^{\ast}(\theta_{-i})$ denote $i$'s robust rank, i.e. $r_{i}^{\ast}(\theta_{-i}) = \max_{\theta_{i}\in\Theta_{i}} r_{i}(\theta_{i}, \theta_{-i})$.

We define multi-unit ranking-based mechanisms as follows.
\begin{definition}[Ranking-based mechanism]
	Let $p \in (0, 1]$.
	The \emph{ranking-based mechanism $q^{p}$ with threshold $p$} is defined for all $i\in [n]$ and $\theta_{-i}\in\Theta_{-i}$ by:
	\begin{equation*}
		q^{p}_{i}(\theta_{-i}) = 
		\begin{cases}
			\min\left(1, \frac{k}{pn}\right)\quad&\text{if $r_{i}^{\ast}(\theta_{-i}) \leq p$ \text{and} $u_{i}(\theta_{-i}) \geq 0$;}
			\\
			0\quad&\text{else.}
		\end{cases}
	\end{equation*}
\end{definition}
The ranking-based mechanism $q^{p}$ is DIC: $q_{i}(\theta_{-i})$ does not depend on $i$'s report.
Further, $q^{p}$ is feasible: there are at most $pn$ agents with a robust rank of $p$, implying that the total winning probability across all agents is at most $pn \cdot \frac{k}{pn}$.

We define informational size as in the single-unit case.
\begin{definition}
For $\theta\in\Theta$, the \emph{informational size at $\theta$} is denoted $\delta(\theta)$ and is given by
\begin{equation*}
    \delta(\theta) =
    \max\limits_{i\in [n], \theta_{i}^{\prime}\in\Theta_{i}} \left\vert r_{i}(\theta_{i}, \theta_{-i}) - r_{i}(\theta_{i}^{\prime}, \theta_{-i})\right\vert.
\end{equation*}
\end{definition}
We also recall the notion of regular sequences of environments.
In the multi-unit setting, an environment specifies the number of agents, the number of goods, the type spaces, and the joint distribution of types and principal-payoffs.
\begin{definition}[Regularity]\label{def:OA:regularity}
A sequence $(n, k^{n}, \Theta^{n}, \mu^{n})_{n\in\mathbb{N}}$ of environments\footnote{To be sure, $k^{n}$ refers to the number of units in the $n$'th environment, not to $k$ raised to the power of $n$.} with associated peer values $(\bar{u}^{n})_{n\in\mathbb{N}}$ is \emph{regular} if for all $\varepsilon > 0$ there exists $p \in (0, 1]$ such that
\begin{equation*}
\lim_{n\to\infty}\mu^{n}\left\lbrace\theta\in\Theta^{n}\colon 
    \left\vert\left\lbrace j\in [n]\colon
    \bar{u}_{j}^{n}(\theta_{-j}) + \varepsilon\geq \max_{i\in [n]}\bar{u}_{i}^{n}(\theta_{-i})\right\rbrace\right\vert
    \geq p n 
    \right\rbrace = 1.
\end{equation*}
\end{definition}

The next result shows that ranking-based mechanisms are approximately optimal in regular environments with large $n$ and vanishing informational size; if the ratio $k/n$ is bounded away from $0$, then regularity can be dropped.
\begin{theorem}\label{thm:multi_unit:ranking_based}
    Let $(n, k^{n}, \Theta^{n}, \mu^{n})_{n\in\mathbb{N}}$ be a sequence of environments. Suppose the associated informational size $(\delta^{n})_{n\in\mathbb{N}}$ converges to $0$ in probability; that is, for all $d > 0$,
    \begin{equation*}
        \lim_{n\to\infty}\mu^{n}\left\lbrace\theta\in\Theta^{n}\colon \delta^{n}(\theta) > d\right\rbrace = 0.        
    \end{equation*}
    The difference between the principal's expected utility from an optimal ranking-based mechanism and an optimal DIC mechanism vanishes as $n\to\infty$ if one of the following holds:
    \begin{enumerate}
        \item The ratio $k^{n} / n$ is bounded away from $0$ along the sequence.
        \item The ratio $k^{n} / n$ vanishes as $n\to\infty$ and the sequence of environments is regular.
    \end{enumerate}
\end{theorem}
The proof for a vanishing ratio $k^{n}/n$ is almost identical to the proof of \Cref{thm:ranking_based} from the main text.
The proof simplifies if the ratio is bounded away from $0$.

\begin{proof}[Proof of \Cref{thm:multi_unit:ranking_based}]

Fixing $n$, consider the $n$'th environment.
For all $i, \ell\in [n]$ and $\theta\in\Theta^{n}$, let $\bar{u}_{i}^{n}(\theta_{-i})$ denote $i$'s peer value at $\theta_{-i}$, and let $\bar{u}^{n}(\ell, \theta)$ denote the $\ell$'th highest peer value at $\theta$.
Let $U^{n}$ denote the principal's utility (as a function of DIC mechanisms), and let $q^{n, p}$ denote the ranking-based mechanism with threshold $p\in (0, 1]$.
Let $\bar{U}^{n}$ denote the principal's utility from allocating to the $k^{n}$ highest positive peer values, i.e.
\begin{equation}\label{eq:multi_unit:ranking_based:upperbound}
    \bar{U}^{n} = \frac{1}{k^{n}}\sum_{\theta\in\Theta} \mu(\theta)\sum_{\ell\in [k^{n}]} \max\left(0, \bar{u}^{n}(\ell, \theta)\right).
\end{equation}
Every DIC mechanism yields a utility of at most $\bar{U}^{n}$.
Hence, it suffices to show that for all $\varepsilon > 0$ there exists $p\in (0, 1]$ such that for all but finitely many $n$ it holds $U^{n}(q^{n, p}) \geq \bar{U}^{n} - \varepsilon$, possibly invoking regularity of the sequence if $k^{n} / n$ vanishes as $n\to\infty$.
We may assume that the ratio $k^{n} / n$ converges to a point in $[0, 1]$.\footnote{Here, we use that a sequence of real numbers converges to $0$ if (and only if) every subsequence has a further subsequence converging to $0$. For all $n$, let $U^{\ast, n} = \sup_{p \in (0, 1]} U^{n}(q^{n, p})$ denote the supremum of the principal's utility across ranking-based mechanisms. To prove \Cref{thm:multi_unit:ranking_based}, we show that the difference of $U^{\ast, n}$ to \eqref{eq:multi_unit:ranking_based:upperbound} vanishes as $n\to\infty$. Fixing a subsequence of environments, there is a further subsequence along which $k^{n} / n$ converges to a point in $[0, 1]$ (since $k^{n} / n \in [0, 1]$ for all $n$). We prove that the difference of $U^{\ast, n}$ to \eqref{eq:multi_unit:ranking_based:upperbound} vanishes along this subsubsequence. To that end, it suffices to show that for all $\varepsilon > 0$ there exists $p\in (0, 1]$ such that for all but finitely many $n$ the utility $U^{n}(q^{n, p})$ is within $\varepsilon$ of \eqref{eq:multi_unit:ranking_based:upperbound}.}

Let $\varepsilon > 0$.
We first derive a lower bound on $U^{n}(q^{n, p})$ for fixed $n\in\mathbb{N}$ and $p\in (0, 1]$.
Denote $\alpha_{n}(p) =  \min\left(\frac{1}{k^{n}}, \frac{1}{pn}\right)$.
Then,
\begin{align}
    \nonumber
     &U^{n}(q^{n, p})
     \\
     \nonumber
     =
     &
     \frac{1}{k^{n}}
     \sum_{\theta\in\Theta^{n}}\mu^{n}(\theta) \sum_{i\in [n]} \bar{u}_{i}^{n}(\theta_{-i}) \min\left(1, \frac{k^{n}}{pn}\right) \bm{1}_{\left(r_{i}^{n, \ast}(\theta_{-i}) \leq p\right)} \bm{1}_{\left(\bar{u}_{i}^{n}(\theta_{-i}) \geq 0\right)}
     \\
     \nonumber
     \geq 
     &
     \sum_{\theta\in\Theta^{n}}\mu^{n}(\theta) \sum_{i\in [n]} \max\left(0, \bar{u}_{i}^{n}(\theta_{-i})\right) \alpha_{n}(p) \bm{1}_{\left(r_{i}^{n}(\theta) \leq p - \delta^{n}(\theta)\right)} 
     \\
     \geq
     &
     \sum_{\theta\in\Theta^{n}}\mu^{n}(\theta) \sum_{i\in [n]} \max\left(0, \bar{u}_{i}^{n}(\theta_{-i})\right) \alpha_{n}(p) \bm{1}_{\left(r_{i}^{n}(\theta) \leq p\right)} 
     \label{eq:multi_unit:ranking_based:lower_bound1}
     \\
     &
     -
     \sum_{\theta\in\Theta^{n}}\mu^{n}(\theta) \sum_{i\in [n]} \alpha_{n}(p) \left(\bm{1}_{\left(r_{i}^{n}(\theta) \leq p \right)} -  \bm{1}_{\left(r_{i}^{n}(\theta) \leq p - \delta^{n}(\theta)\right)} \right),
     \label{eq:multi_unit:ranking_based:lower_bound2}
\end{align}
where the first inequality follows from the definition of informational size, and the second inequality follows since all payoffs are in $[-1, 1]$.

Consider the sum in \eqref{eq:multi_unit:ranking_based:lower_bound2}.
Since no two agents have the same rank, it holds
\begin{align*}
    &\sum_{\theta\in\Theta^{n}}\mu^{n}(\theta) \sum_{i\in [n]} \alpha_{n}(p) \left(\bm{1}_{\left(r_{i}^{n}(\theta) \leq p - \delta^{n}(\theta)\right)} -  \bm{1}_{\left(r_{i}^{n}(\theta) \leq p\right)} \right)
    \\
    =
    &\sum_{\theta\in\Theta^{n}}\mu^{n}(\theta)\min\left(\frac{1}{k^{n}}, \frac{1}{pn}\right) \left(\lfloor np \rfloor - \lfloor n (p - \delta^{n}(\theta))\rfloor\right).
\end{align*}
Since informational size $\delta^{n}$ vanishes in probability, this sum vanishes as $n\to\infty$ for every fixed $p\in (0, 1]$.
Thus, to complete the proof it suffices to show that there is $p\in (0, 1]$ such that the sum in \eqref{eq:multi_unit:ranking_based:lower_bound1} is at least $\bar{U}^{n} - \varepsilon$ for all but finitely many $n$.
Write the sum in \eqref{eq:multi_unit:ranking_based:lower_bound1} as
\begin{align*}
    & \sum_{\theta\in\Theta^{n}}\mu^{n}(\theta) \sum_{i\in [n]} \max\left(0, \bar{u}_{i}^{n}(\theta_{-i})\right) \alpha_{n}(p) \bm{1}_{\left(r_{i}^{n}(\theta) \leq p\right)}
    \\
    =&
    \sum_{\theta\in\Theta^{n}}\mu^{n}(\theta) \sum_{\ell \in[n]} \max\left(0, \bar{u}^{n}(\ell, \theta)\right) \alpha_{n}(p) \bm{1}_{(\ell\leq pn)}
    \\
    =&
    \underbrace{\frac{1}{k^{n}}\sum_{\theta\in\Theta^{n}}\mu^{n}(\theta) \sum_{\ell \in[n]} \max\left(0, \bar{u}^{n}(\ell, \theta)\right) \bm{1}_{(\ell \leq k^{n})}}_{= \bar{U}^{n}}
    \\
    &+ \sum_{\theta\in\Theta^{n}}\mu^{n}(\theta) \sum_{\ell \in[n]} \max\left(0, \bar{u}^{n}(\ell, \theta)\right) \left(\alpha_{n}(p) \bm{1}_{(\ell \leq pn)}   - \frac{1}{k^{n}}\bm{1}_{(\ell \leq k^{n})}\right)
\end{align*}
Hence, it suffices to find $p \in (0, 1]$ such that all but finitely many $n$ satisfy
\begin{equation}
    \sum_{\theta\in\Theta^{n}}\mu^{n}(\theta) \sum_{\ell \in[n]} \max\left(0, \bar{u}^{n}(\ell, \theta)\right) \left(\alpha_{n}(p) \bm{1}_{(\ell \leq pn)}   - \frac{1}{k^{n}}\bm{1}_{(\ell \leq k^{n})}\right) \geq - \varepsilon
    .
    \label{eq:multi_unit:ranking_based:lower_bound}
\end{equation}
We distinguish two cases.
First, suppose $k^{n} / n$ vanishes as $n\to\infty$.
Regularity implies that there exists $p\in (0, 1]$ such that
\begin{equation*}
    \lim_{n\to\infty}\mu^{n}\left\lbrace\theta\in\Theta^{n}\colon 
    \left\vert\left\lbrace j\in [n]\colon
    \bar{u}_{j}^{n}(\theta_{-j}) + \varepsilon\geq \bar{u}^{n}(1, \theta)\right\rbrace\right\vert
    \geq p n
    \right\rbrace = 1.
\end{equation*}
Fix $p$ with this property.
The left side of \eqref{eq:multi_unit:ranking_based:lower_bound} is bounded below by
\begin{equation}\label{eq:multi_unit:ranking_based:vanishingratiobound}
    \sum_{\theta\in\Theta^{n}}\mu^{n}(\theta) \left(\alpha_{n}(p) \lfloor pn\rfloor \max\left(0, \bar{u}^{n}(\lfloor pn\rfloor, \theta)\right) - \max\left(0, \bar{u}^{n}(1, \theta)\right)\right)
\end{equation}
Since $k^{n}/ n$ vanishes, we have $\alpha_{n}(p) \lfloor pn\rfloor = 1$ for sufficiently large $n$ (for the fixed $p$).
Next, if $\theta$ is a type profile such that at least $pn$ agents have a peer value within $\varepsilon$ of the highest peer value $\bar{u}^{n}(1, \theta)$, then $\bar{u}^{n}(\lfloor pn\rfloor, \theta) \geq \max_{i\in [n]}\bar{u}_{i}^{n}(\theta_{-i}) - \varepsilon$.
Hence, as $n\to\infty$, the lower bound \eqref{eq:multi_unit:ranking_based:vanishingratiobound} is at most $ - \varepsilon$ all but finitely many $n$.

Second, suppose $k^{n} / n$ converges to a point in $(0, 1]$ as $n\to\infty$. 
Set $p = \lim_{n\to\infty} k^{n} / n$.
(This choice is infeasible if $k^{n}/n$ vanishes since we used $p > 0$ throughout.)
The left side of \eqref{eq:multi_unit:ranking_based:lower_bound} is bounded below as follows:
\begin{align}
    \nonumber
    &\sum_{\theta\in\Theta^{n}}\mu^{n}(\theta) \sum_{\ell \in[n]} \max\left(0, \bar{u}^{n}(\ell, \theta)\right) \left(\alpha_{n}(p) \bm{1}_{(\ell \leq pn)}   - \frac{1}{k^{n}}\bm{1}_{(\ell \leq k^{n})}\right)
    \\
    \nonumber
    =& \sum_{\theta\in\Theta^{n}}\mu^{n}(\theta) \sum_{\ell \in[n]} \max\left(0, \bar{u}^{n}(\ell, \theta)\right) \left(\alpha_{n}(p) - \frac{1}{k^{n}}\right) \bm{1}_{(\ell \leq pn)}
    \\
    \nonumber
    &+ \sum_{\theta\in\Theta^{n}}\mu^{n}(\theta) \sum_{\ell \in[n]} \max\left(0, \bar{u}^{n}(\ell, \theta)\right) \alpha_{n}(p) \left(\bm{1}_{(\ell \leq k^{n})} - \bm{1}_{(\ell \leq pn)}\right)
    \\
    \geq & \sum_{\theta\in\Theta^{n}}\mu^{n}(\theta) \lfloor pn\rfloor \left(\alpha_{n}(p) - \frac{1}{k^{n}}\right) - \sum_{\theta\in\Theta^{n}}\mu^{n}(\theta) n \alpha_{n}(p) \left(\frac{\lfloor pn\rfloor}{n} - \frac{k^{n}}{n}\right),
    \label{eq:multi_unit:ranking_based:boundedratiobound}
\end{align}
where the inequality uses that all peer values are in $[-1, 1]$ and $\alpha_{n}(p) \leq \frac{1}{k^{n}}$.
Using the choice of $p$ and the definition of $\alpha_{n}(p)$, we find $n \alpha_{n}(p) \to \frac{1}{p}$ as $n\to\infty$ (and we recall $p > 0$). The choice of $p$ further implies $\frac{\lfloor pn\rfloor}{k^{n}} \to 1$ and $\frac{\lfloor pn\rfloor}{n} - \frac{k^{n}}{n} \to 0$ as $n\to\infty$.
Thus, the lower bound in \eqref{eq:multi_unit:ranking_based:boundedratiobound} vanishes as $n\to\infty$, and hence \eqref{eq:multi_unit:ranking_based:lower_bound} is at least $-\varepsilon$ for all but finitely many $n$.

Finally, observe that in the second case---i.e. $k^{n} / n$ converges to a point in $(0, 1]$---we did not invoke regularity.
With this observation, one obtains the claim in \Cref{thm:multi_unit:ranking_based} for sequences along which the ratio $k^{n} / n $ is bounded away from $0$.
\end{proof}

\subsubsection{NP-Hardness}\label{appendix:multiunit_allocation:hardness}

We now show that the complexity result from the main text extends to multi-unit allocation problems. This implies the existence of stochastic extreme points; otherwise, an optimal deterministic mechanism can be found in time polynomial in the total number of types by the ellipsoid method (unless $P=NP$). We shall not further pursue the detailed characterizations from the main text. 

Again, it suffices to show that the following decision problem is NP-complete.

\begin{definition}[\textsc{Det}-$(n,k)$]
    The input consists of an integer $z$, finite sets $\Theta_{1}, \ldots, \Theta_{n}$ and weights $w_i: \Theta_{-i}\to \lbrace 0, 1\rbrace$ for all $i\in [n]$.
    The problem is to decide whether there is a deterministic DIC mechanism $q$ (for $k$ units and $n$ agents with respective type spaces $\Theta_{1}, \ldots, \Theta_{n}$) such that $\sum_{i, \theta} w_{i}(\theta_{-i}) q_{i}(\theta_{-i}) \geq z$.
\end{definition}

\begin{theorem}\label{thm:npcomplete_k}
	For fixed integers $n,k\geq 1$ such that $n\geq k+2$, the problem \textsc{Det}-$(n,k)$ is NP-complete.
\end{theorem}

Let $\hat G=(\hat V,\hat E)$ be a simple undirected graph. For $k\in\mathbb{N}$, we say that $S\subseteq V$ is a \emph{$k$-stable set} if the subgraph of $\hat G$ induced by $S$ contains no clique of size larger than $k$. As in the main text, we have the following graph-theoretic characterization, which we shall use to prove \Cref{thm:npcomplete_k}.

\begin{lemma}
\label{lemma:bijection_k}
	There is a bijection between deterministic DIC mechanisms and $k$-stable sets in the feasibility graph.
\end{lemma}

\begin{proof}[Proof of \Cref{thm:npcomplete_k}]
We reduce from the following decision problem \textsc{$k$-StableSet}.
The input is an integer $\hat{z}$ and a (simple undirected) graph $\hat{G}$.
The problem is to determine whether $\hat{G}$ admits a $k$-stable set of cardinality $\hat{z}$ or greater.
It follows from Theorem 4 in \citet{lewis1980node} that $k$-\textsc{StableSet} is NP-complete.\footnote{In their language, the relevant graph property is ``has no clique with more than $k$ vertices.'' This property is non-trivial (there are infinitely many graphs with and without this property) and hereditary (every induced subgraph of a graph with the property also has the property).}

Given an instance $(\hat{G},\hat{z})$ of \textsc{$k$-StableSet}, we now construct an instance of \textsc{Det}–$(n,k)$. Let $\mathcal{K}_{k+1}$ denote the collection of all cliques of $\hat{G}$ of size $k+1$. 
Let $\Theta_1=\mathcal{K}_{k+1}$, $\Theta_2=\Theta_3=\hat V$, and $\Theta_4=\ldots=\Theta_n=\{\circ\}$.  Let $G=(V,E)$ be the corresponding feasibility graph.

For each $K\in\mathcal{K}_{k+1}$, we construct a gadget $P(K)\subset V$ as follows.
Suppose the vertices of $K$ are $v_1,\ldots,v_{k+1}$. Define type profiles $\theta^1,\ldots,\theta^{k+1}$ and $\tilde\theta^1,\ldots,\tilde\theta^k$ where for all $j$,
$$
    \theta^j=(K,v_j,v_j,\circ,\ldots,\circ) \qquad 
    \tilde\theta^j=(K,v_{j},v_{j+1},\circ,\ldots,\circ).
$$
Let $P(K)$ be the subgraph induced by the vertices
$$
    \{(i,\theta_{-i}^j)\mid i=1,\ldots,k+1,\,j=1,\ldots,k+1\}\cup \{(i,\tilde\theta_{-i}^j)\mid i=2,\ldots,k+2,\,j=1,\ldots,k\}.
$$
\Cref{fig:gadget} illustrates. Let $Q(K)$ be the subgraph induced by the vertices in
\begin{equation*}
    P(K)\setminus \{(\cdot,v_j,v_j,\circ,\ldots,\circ)\mid j=1,\ldots,k+1\}.
\end{equation*}
The vertices in $P(K)\setminus Q(K)$ mirror the vertices in the clique $K$ of $\hat G$.

Assign weight $1$ to every vertex in $G$ that is contained in some gadget $P(K)$ and weight $0$ to all other vertices. Finally, set
$$
z=\hat{z}+ (2k^2-k) |\mathcal{K}_{k+1}|.
$$

\begin{figure}
\centering
\begin{tikzpicture}[scale=2.8, line cap=round, line join=round]
  \node (A) at (0,0) {$ (K,\,\cdot,\,v_1,\,\circ)$};
  \node (B) at (1,0) {$ (K,\,v_1,\,\cdot,\,\circ)$};
  \node (C) at (0.5,0.5) {$ (\cdot,\,v_1,\,v_1,\,\circ)$};
  \draw (A) -- (B) -- (C) -- (A);
  
  \node (X) at (2,0) {$ (K,\,\cdot,\,v_2,\,\circ)$};
  \node (Y) at (1.5,-0.5) {$ (K,\,v_1,\,v_2,\,\cdot)$};
  \draw (B) -- (X) -- (Y) -- (B);
  
  \node (Z) at (3,0) {$ (K,\,v_2,\,\cdot,\,\circ)$};
  \node (W) at (2.5,0.5) {$ (\cdot,\,v_2,\,v_2,\,\circ)$};
  \draw (X) -- (Z) -- (W) -- (X);
  
  \node (V) at (4,0) {$ (K,\,\cdot,\,v_3,\,\circ)$};
  \node (U) at (3.5,-0.5) {$ (K,\,v_2,\,v_3,\,\cdot)$};
  \draw (Z) -- (V) -- (U) -- (Z);
  
  \node (S) at (5,0) {$ (K,\,v_3,\,\cdot,\,\circ)$};
  \node (T) at (4.5,0.5) {$ (\cdot,\,v_3,\,v_3,\,\circ)$};
  \draw (V) -- (S) -- (T) -- (V);
  
  \draw (A) -- (B) -- (X) -- (Z) -- (V) -- (S);
\end{tikzpicture}
\caption{A representative gadget $P(K)$ for $k=2$. For $k=1$, the gadget is the induced path from the proof of \Cref{thm:npcomplete}.}
\label{fig:gadget}
\end{figure}

We make a few observations about $P(K)$ and $Q(K)$. 
\begin{itemize}
    \item For $K,K'\in\mathcal{K}_{k+1}$ such that $K\neq K'$, it holds $Q(K)\cap Q(K)=\emptyset$ and no vertex in $P(K)\setminus P(K')$ is adjacent to any vertex in $P(K')\setminus P(K)$.
    \item The total number of vertices in $P(K)$ is $2k^2+(k+1)$ (namely, $2k+1$ cliques of size $(k+1)$ that share $2k$ vertices in total). Thus, the total number of vertices in $Q(K)$ is $2k^2$. 
    \item The maximum $k$-stable sets in $Q(K)$ have size $2k^2-k$ (there are $k$ cliques of size $k+1$ in $Q(K)$). 
    There exist maximum $k$-stable sets in $Q(K)$ that select all vertices of $Q(K)$ except
    $k$ of the vertices that appear in multiple maximal cliques of $Q(K)$ (i.e., vertices where $i=2,3$).
    \item  The maximum $k$-stable sets in $P(K)$ have size $2k^2$. For any $k$ vertices in $P(K)\setminus Q(K)$, there exists a maximum $k$-stable sets in $P(K)$ consisting of these $k$ vertices plus a maximum $k$-stable set in $Q(K)$.
    \end{itemize}

We now complete the proof by showing that the constructed instance of \textsc{Det}–$(n,k)$ has a solution of value at least 
$
z=\hat{z}+ (2k^2-k) |\mathcal{K}_{k+1}|
$ 
if and only if the original graph $\hat{G}$ has a $k$–stable set of cardinality at least $\hat{z}$.

First, suppose that $\hat{G}$ admits a $k$–stable set $\hat{S}\subseteq\hat{V}$ with $|\hat{S}|\ge \hat{z}$.  
Define 
$$
S(0)=\{(\cdot,v,v,\circ,\ldots,\circ)\mid v\in \hat S\}.
$$
For each clique $K\in \mathcal{K}_{k+1}$, let $S(K)$
be a maximum $k$-stable set in $Q(K)$ such that $S(0)\cap P(K)\cup S(K)$ is a $k$-stable set in $P(K)$ (which exists by the observations above). 
Then, the $k$-stable set 
$S=S(0)\cup\bigcup_{K\in \mathcal{K}_{k+1}} S(K)$
has size $|S|\geq z=\hat z+ (2k^2-k) |\mathcal{K}_{k+1}|$.

Now suppose $G$ admits a $k$-stable set of weight $\hat z+ (2k^2-k) |\mathcal{K}_{k+1}|$ or greater. By the observations made above, $G$ admits another $k$-stable set $S$ of weight $\hat z+ (2k^2-k) |\mathcal{K}_{k+1}|$ or greater such that, for all $K\in\mathcal{K}_{k+1}$, $S$ contains at most $k$ vertices from $P(K)\setminus Q(K)$. Now define 
$$
\hat{S} = \{ v\in\hat{V} : (\cdot, v, v,\circ,\ldots\circ) \in S\}.
$$
The set $\hat{S}$ is $k$-stable in $\hat{G}$ since it selects at most $k$ vertices from every clique $K\in\mathcal{K}_{k+1}$ (of size $k+1$). Moreover, since $|\bigcup_{K\in\mathcal{K}_{k+1}} (S\cap Q(K))|\leq (2k^2-k) |\mathcal{K}_{k+1}|$, it holds $|\hat{S}|\geq \hat z=z-(2k^2-k) |\mathcal{K}_{k+1}|$.
\end{proof}

\subsection{A Classical Symmetry Condition Makes Peer Information Uninteresting}\label{appendix:symmetry_nothing_goes}
Here, we show that the principal cannot meaningfully elicit information if allocation is mandatory and the environment meets a symmetry condition.

In this part of the appendix, for every agent $i$, the value $u_{i}$ has only finitely many possible realizations.
We write $\mu(u_{1}, \ldots, u_{n}, \theta_{1}, \ldots, \theta_{n})$ for the joint probability of a profile $(u_{1}, \ldots, u_{n})$ of values and a profile $(\theta_{1}, \ldots, \theta_{n})$ of types.

Let $\Xi$ be the set of permutations of $[n]$ (i.e., bijections from $[n]$ to $[n]$).
The environment is \emph{symmetric} if all type spaces agree ($\Theta_{1} = \ldots = \Theta_{n}$) and all permutations $\xi\in \Xi$ satisfy
\begin{equation*}
    \mu(u_{1}, \ldots, u_{n}, \theta_{1}, \ldots, \theta_{n}) = \mu(u_{\xi(1)}, \ldots, u_{\xi(n)}, \theta_{\xi(1)}, \ldots, \theta_{\xi(n)}).
\end{equation*}

In \Cref{sec:conclusion}, we alluded to the following special symmetric environment: each agent's type equals their value ($\theta_{i} = u_{i}$ with probability $1$) and the distribution of values is invariant to permutations.

\begin{theorem}\label{prop:symmetry_nothing_goes}
In a symmetric environment, the principal's expected utility equals $\mathbb{E}[u_{1}]$ in every DIC mechanism that always allocates the good.
\end{theorem}
Intuitively, symmetry is too strong for peer selection since it implies that even the information of $n-2$ agents is insufficient for distinguishing between the remaining two agents, i.e. $\mathbb{E}[u_{i} \vert \theta_{-ij}] = \mathbb{E}[u_{j}\vert\theta_{-ij}]$ for all distinct $i$ and $j$ and types $\theta_{-ij}$ of the others. (This intuition is incomplete, though, since $j$'s information could also be used to determine whether $i$ wins but not whether $j$ wins.)

The proof of \Cref{prop:symmetry_nothing_goes} establishes the following stronger result.
\begin{theorem}
    If $q$ is DIC and always allocates, and if for all $i$ the allocation $q_{i}(\theta_{-i})$ is invariant with respect to permutations of $\theta_{-i}$, then $q$ is constant.
\end{theorem}
This result implies \Cref{prop:symmetry_nothing_goes} since, as we show, it is without loss to focus on such permutation-invariant mechanisms if the environment is symmetric.

\begin{proof}[Proof of \Cref{prop:symmetry_nothing_goes}]
    For a moment, let us denote each agent's allocation as a function of the entire type profile.
    Say a DIC mechanism $q$ is \emph{symmetric} if all $\xi\in\Xi$ satisfy
    \begin{equation}\label{eq:symmetric_nothing_goes:mechanism_symmetry}
        q_{i}(\theta_{1}, \ldots, \theta_{n}) = q_{\xi^{-1}(i)}(\theta_{\xi(1)}, \ldots, \theta_{\xi(n)}).
    \end{equation}
    Here, $(\theta_{\xi(1)}, \ldots, \theta_{\xi(n)})$ is the profile where the type of an agent $i$ is $\theta_{\xi(i)}$, which in turn is the type of agent $\xi(i)$ in the original profile $(\theta_{1}, \ldots, \theta_{n})$.
    Symmetry says that agent $i$'s winning probability at the permuted profile equals $\xi(i)$'s winning probability in the original profile.

    We first verify the expected result that in a symmetric environment, given an arbitrary DIC mechanism $q$ that always allocates, there is a symmetric DIC mechanisms that always allocates and that generates the same expected utility as $q$.
    Indeed, consider the mechanism $q^{\prime}$ define for all $i\in [n]$ and $(\theta_{1}, \ldots, \theta_{n})\in\Theta$ by
    \begin{equation*}
        q^{\prime}_{i}(\theta_{1}, \ldots, \theta_{n}) = \frac{1}{n!} \sum_{\xi\in\Xi} q_{\xi^{-1}(i)}(\theta_{\xi(1)}, \ldots, \theta_{\xi(n)}).
    \end{equation*}
    One may verify that $q^{\prime}$ is DIC and always allocates.
    By construction, $q^{\prime}$ is symmetric.
    We show that $q$ and $q^{\prime}$ generate the same expected utility for the principal.
    It holds:
    \begin{align*}
        U(q) 
        &= \sum_{u, \theta, i}  \mu(u_{1}, \ldots, u_{n}, \theta_{1}, \ldots, \theta_{n}) q_{i}(\theta_{1}, \ldots, \theta_{n}) u_{i}
        \\
        &= \sum_{\xi\in\Xi}\frac{1}{n!}\sum_{u, \theta, i}  \mu(u_{1}, \ldots, u_{n}, \theta_{1}, \ldots, \theta_{n}) q_{i}(\theta_{1}, \ldots, \theta_{n}) u_{i}
        \\
        &= \sum_{\xi\in\Xi}\frac{1}{n!}\sum_{u, \theta, i}  \mu(u_{\xi(1)}, \ldots, u_{\xi(n)}, \theta_{\xi(1)}, \ldots, \theta_{\xi(n)}) q_{i}(\theta_{\xi(1)}, \ldots, \theta_{\xi(n)}) u_{\xi(i)};
    \end{align*}
    here the first equality is by definition, the second is clear, and the third is a change of variables in the summation (since for each fixed permutation $\xi$ the map $(u_{1}, \ldots, u_{n}, \theta_{1}, \ldots, \theta_{n})\mapsto (u_{\xi(1)}, \ldots, u_{\xi(n)}, \theta_{\xi(1)}, \ldots, \theta_{\xi(n)})$ ranges over all profiles of values and types).
    Since the environment is symmetric, we get
    \begin{equation*}
        U(q) 
        = \sum_{\xi\in\Xi}\frac{1}{n!}\sum_{u, \theta, i}  \mu(u_{1}, \ldots, u_{n}, \theta_{1}, \ldots, \theta_{n}) q_{i}(\theta_{\xi(1)}, \ldots, \theta_{\xi(n)}) u_{\xi(i)}.
    \end{equation*}   
    By another change of variables, for each fixed $\xi, u, \theta$, we have 
    \begin{equation*}
    \sum_{i} q_{i}(\theta_{\xi(1)}, \ldots, \theta_{\xi(n)}) u_{\xi(i)} = \sum_{i} q_{\xi^{-1}(i)}(\theta_{\xi(1)}, \ldots, \theta_{\xi(n)}) u_{i}.
    \end{equation*}
    Thus
   \begin{equation*}
        U(q) 
        = \sum_{\xi\in\Xi}\frac{1}{n!}\sum_{u, \theta, i}  \mu(u_{1}, \ldots, u_{n}, \theta_{1}, \ldots, \theta_{n}) q_{\xi^{-1}(i)}(\theta_{\xi(1)}, \ldots, \theta_{\xi(n)}) u_{i}.
    \end{equation*}   
    The right side is simply the expected utility from $q^{\prime}$.

    To complete the proof, it suffices to show that every symmetric DIC mechansims that always allocates yields an expected utility of $\mathbb{E}[u_{1}]$.
    We show that every symmetric DIC mechanism $q$ that always allocates is actually constant.
    The claim then follows since, by symmetry of the environment, every constant mechanism yields $\mathbb{E}[u_{1}]$.

    For each agent $i$, we again drop $i$'s report from $i$'s allocation.
    Symmetry of $q$ implies that $q_{i}(\theta_{-i})$ is invariant to permutations of $\theta_{-i}$ (consider \eqref{eq:symmetric_nothing_goes:mechanism_symmetry} for permutations $\xi$ such $\xi(i) = i$).

    For this proof, we introduce some special notation.
    Recall that the agents share the common type space $\Theta_{1}$.
    A generic type in $\Theta_{1}$ is denoted $t$.
    A generic type profile of $n-1$ types is denoted $\bm{t}$.
    The space $T$ of such $\bm{t}$ equals the $(n-1)$-fold product of $\Theta_{1}$.
    Thus, agent $i$'s allocation when the others report $\bm{t}$ equals $q_{i}(\bm{t})$, and this allocation is invariant to permutations of $t$.
    We denote $[n-1] = \lbrace 1, \ldots, n-1\rbrace$.

    Given $\bm{t} = (t_{1}, \ldots, t_{n-1}) \in T$ and $j\in [n-1]$, we denote by $\bm{t}_{-j}$ the profile of types in $\bm{t}$ other than $t_{j}$; that is, $\bm{t}_{-j} = (t_{1}, \ldots, t_{j-1}, t_{j+1}, \ldots, t_{n-1})$.
    For a type $t^{\prime}\in\Theta_{1}$, we then denote by $(t^{\prime}, \bm{t}_{-j})$ the profile of $n-1$ types obtained from $\bm{t}$ by replacing $t_{j}$ by $t^{\prime}$; that is, $(t^{\prime}, \bm{t}_{-j}) = (t_{1}, \ldots, t_{j-1}, t^{\prime}, t_{j+1}, \ldots, t_{n-1})$.

    For the proof, it shall be useful to consider the behavior of an agent's winning probability at profiles of the form $(t^{\prime}, \bm{t}_{-j})$ as we vary $j$.
  We establish the following auxiliary lemma.
  \begin{lemma}\label{lemma:symmetry_nothing_goes:auxiliary}
    Let $i\in [n]$.
    All $t^{\prime}, t^{\prime\prime}\in \Theta_{1}$ and $\bm{t}\in T$ satisfy
    \begin{equation*}
      \sum\limits_{j=1}^{n-1}
        \left(
        q_{i}(t^{\prime}, \bm{t}_{-j})
        -
        q_{i}(t^{\prime\prime}, \bm{t}_{-j})
        \right)
      =
      0
      .
    \end{equation*}
  \end{lemma}

  \begin{proof}[Proof of \Cref{lemma:symmetry_nothing_goes:auxiliary}]
    For notational simplicity, we prove the claim for $i=n$ (later commenting on how to adapt the notation for the general case).
    Denote $\bm{t} = (t_{1}, \ldots, t_{n-1})$.

    In an intermediate step, let $j \in [n-1]$ be an agent distinct from agent $n$.
    We show
    \begin{equation}\label{eq:lemma:symmetry_nothing_goes:auxiliary:1}
        q_{n}(t^{\prime\prime}, \bm{t}_{-j}) -  q_{n}(t^{\prime}, \bm{t}_{-j}) = q_{j}(t^{\prime\prime}, \bm{t}_{-j}) - q_{j}(t^{\prime}, \bm{t}_{-j}).
    \end{equation}
    Note that $\bm{t}_{-j}$ is a profile of $n-2$ types.
    Suppose the $n-2$ agents other than $n$ and $j$ report $\bm{t}_{-j} = (t_{1}, \ldots, t_{j-1}, t_{j+1}, \ldots, t_{n-1})$.
    Let $\theta^{\prime}$ be the profile where agent $n$ reports $t^{\prime}$, agent $j$ reports $t^{\prime\prime}$, and agents other than $n$ and $j$ report $\bm{t}_{-j}$.
    Let $\theta^{\prime\prime}$ be the profile obtained from $\theta^{\prime}$ by permuting $n$'s and $j$'s reports.
    Since $\theta^{\prime}$ and $\theta^{\prime\prime}$ differ only in a permutation of $n$'s and $j$'s reports, the allocation of agents other than $n$ and $j$ is the same across $\theta^{\prime}$ and $\theta^{\prime\prime}$.
    Since the object is always allocated, the sum of $n$'s and $j$'s allocation is also the same across $\theta^{\prime}$ and $\theta^{\prime\prime}$.
    Thus:
    \begin{equation}\label{eq:lemma:symmetry_nothing_goes:auxiliary:2}
        q_{n}(t^{\prime\prime}, \bm{t}_{-j}) + q_{j}(t^{\prime}, \bm{t}_{-j}) = q_{n}(t^{\prime}, \bm{t}_{-j}) + q_{j}(t^{\prime\prime}, \bm{t}_{-j});
    \end{equation}
    the left side is the sum of $n$'s and $j$'s allocation at $\theta^{\prime}$, the right side the sum at $\theta^{\prime\prime}$.
    Rearranging \eqref{eq:lemma:symmetry_nothing_goes:auxiliary:2} yields \eqref{eq:lemma:symmetry_nothing_goes:auxiliary:1}.

    Next, we sum \eqref{eq:lemma:symmetry_nothing_goes:auxiliary:1} across all $j\in [n]$ distinct from $i$.
    Recall again that $\bm{t} = (t_{1}, \ldots, t_{n-1})$ is a profile of $n-1$ types.
    Consider the profile where agents $1$ to $n-1$ report $\bm{t}$, and agent $n$ reports $t^{\prime}$.
    At this profile, for all $j$ distinct from $n$, agent $j$'s allocation equals $q_{j}(t^{\prime}, \bm{t}_{-j})$, while $n$'s allocation is $q_{n}(\bm{t})$.
    Since the object is allocated, we have $\sum_{j\neq i} q_{j}(t^{\prime}, \bm{t}_{-j}) = 1 - q_{n}(\bm{t})$.
    By a similar argument, we have $\sum_{j\neq i} q_{j}(t^{\prime\prime}, \bm{t}_{-j}) = 1 - q_{n}(\bm{t})$.
    Thus summing \eqref{eq:lemma:symmetry_nothing_goes:auxiliary:1}  across $j\in [n-1]$ yields 
    $\sum_{j\in [n-1]}
        (
        q_{n}(t^{\prime}, \bm{t}_{-j})
        -
        q_{n}(t^{\prime\prime}, \bm{t}_{-j})
        ) = 0$, as desired.

    For the general case (where $i$ not need equal $n$), the argument is analogous. One should now think of $\bm{t} = (t_{1}, \ldots, t_{n-1})$ as the reports of all agents other than $i$ (for some fixed assignment of these agents to $t_{1}, \ldots, t_{n-1}$). The profile $\bm{t}_{-j}$ should be thought of as the profile $\bm{t}$ except $j$'s type.
    \end{proof}

    We next use \Cref{lemma:symmetry_nothing_goes:auxiliary} to complete the proof.
    Let $i\in [n]$ be arbitary.
    We show $i$'s winning probability is constant in the reports of others.
    To that end, let us fix an arbitrary type $t^{\ast}\in T$.
    For all $k \in \lbrace 0, \ldots, n-1\rbrace$, let $T_{k}$ denote the subset of profiles in $T$ where exactly $k$-many entries are distinct from $t^{\ast}$.
    Let $p_{i}$ denote $i$'s winning probability when all other agents report $t^{\ast}$.
    We will show via induction over $k$ that $i$'s winning probability is equal to $p_{i}$ whenever the others report a profile in $T_{k}$.
    This completes the proof since $T = \cup_{k=0}^{n-1} T_{k}$.

    \emph{Base case $k = 0$}. Immediate from the definitions of $p_{i}$ and $T_{0}$.

    \emph{Induction step.}
    Let $k \geq 1$.
    Suppose all $\hat{\bm{t}}\in \cup_{\ell=0}^{k-1} T_{\ell}$ satisfy $q_{i}(\hat{\bm{t}}) = p_{i}$.
    Letting $\bm{t} \in T_{k}$ be arbitrary, we show $q_{i}(\bm{t}) = p_{i}$.

    Denote $\bm{t} = (t_{1}, \ldots, t_{n})$.
    Since $q_{i}(\bm{t})$ is invariant to permutations of $\bm{t}$, we may assume that exactly the first $k$ entries of $\bm{t}$ are distinct from $t^{\ast}$, while all other entries equal $t^{\ast}$.
    That is, $\bm{t} = (t_{1}, \ldots, t_{k}, t^{\ast}, \ldots, t^{\ast})$.

    Let $\tilde{\bm{t}} = (t_{1}, \ldots, t_{k-1}, t^{\ast}, \ldots, t^{\ast})$ be the profile obtained from $\bm{t}$ by replacing $t_{k}$ by $t^{\ast}$.
    We now invoke \Cref{lemma:symmetry_nothing_goes:auxiliary} to infer
    \begin{equation}\label{eq:strong_anonymity:permutation_sum}
      \sum\limits_{j=1}^{n-1}
        q_{i}(t_{k}, \tilde{\bm{t}}_{-j})
      =
      \sum\limits_{j=1}^{n-1}
        q_{i}(t^{\ast}, \tilde{\bm{t}}_{-j})
      .
    \end{equation}
    
    Consider the profiles appearing in the sum on the left of \eqref{eq:strong_anonymity:permutation_sum} as $j$ varies from $1$ to $n-1$.
    \begin{enumerate}
      \item Let $j \leq k-1$. Since exactly the first $k-1$ entries of $\tilde{\bm{t}}$ are distinct from $t^{\ast}$, it follows that $(t_{k}, \tilde{\bm{t}}_{-j})$
      is another profile where exactly $k-1$ entries differ from $t^{\ast}$.
      Hence the induction hypothesis implies $q_{i}(t_{k}, \tilde{\bm{t}}_{-j}) = p_{i}$.
      \item Let $j > k-1$. In the profile $(t_{k}, \tilde{\bm{t}}_{-j})$, the first $k-1$ entries are $t_{1}, \ldots, t_{k-1}$, the $j$'th entry is $t_{k}$, and all remaining entries are $t^{\ast}$.
      Hence, $(t_{k}, \tilde{\bm{t}}_{-j})$ is a permutation of $\bm{t}$. 
      Hence, $q_{i}(t_{k}, \tilde{\bm{t}}_{-j}) = q_{i}(\bm{t})$.
    \end{enumerate}
    Hence the sum on the left-hand side of \eqref{eq:strong_anonymity:permutation_sum} is given by
    $\sum_{j\in[n-1]}
      q_{i}(t, \tilde{\bm{t}}_{-j})
      =
      (k - 1)p_{i} + (n - k) q_{i}(\bm{t})$

    Now consider the sum on the right-hand side of \eqref{eq:strong_anonymity:permutation_sum}.
    For all $j$, the profile $(t^{\ast}, \tilde{\bm{t}}_{-j})$ contains at most $(k-1)$-many entries different from $t^{\ast}$ since $\bm{t}$ contains $k$ entries different from $t^{\ast}$.
    By the induction hypothesis, therefore, the sum on the right-hand side of \eqref{eq:strong_anonymity:permutation_sum} equals $(n-1)p_{i}$.

    Equation \eqref{eq:strong_anonymity:permutation_sum} thus simplifies to $(k - 1)p_{i} + (n - k) q_{i}(\bm{t}) = (n-1) p_{i}$.
    Equivalently, $(n - k)(q_{i}(\bm{t}) - p_{i}) = 0$.
    Since $k \leq n-1$, we conclude $q_{i}(\bm{t}) = p_{i}$, as promised.
\end{proof}

\end{document}